\documentclass[12pt,a4paper]{article}

\usepackage[english]{babel}
\usepackage{epsfig}%
\usepackage{longtable}
\usepackage{multirow}
\usepackage[numbers, authoryear, comma, longnamesfirst, sectionbib]{natbib}  
\usepackage{amssymb}
\usepackage{amsmath}
\usepackage{graphicx}
\usepackage{verbatim}
\usepackage{lscape}
\usepackage{amsthm}
\usepackage{rotate}
\usepackage{rotating}
\usepackage{pdflscape}
\usepackage{booktabs}
\usepackage{subfigure}
\usepackage{setspace}
\usepackage{placeins}
\usepackage{amsfonts}
\usepackage{booktabs}
\usepackage{footnote}	 
\usepackage{float}
\usepackage[font=small,labelfont=bf]{caption}
\usepackage{fullpage}
\usepackage{adjustbox}
\usepackage{pdflscape}
\usepackage{caption}
\usepackage{hyperref}
\usepackage{fancyref}                                                   
\usepackage{xr}

\makeatletter
\newcommand*{\addFileDependency}[1]{%
  \typeout{(#1)}%
  \@addtofilelist{#1}%
  \IfFileExists{#1}{}{\typeout{No file #1.}}%
}
\makeatother

\newcommand*{\myexternaldocument}[1]{%
  \externaldocument{#1}%
  \addFileDependency{#1.tex}%
  \addFileDependency{#1.aux}%
}

\myexternaldocument{online_appendix}
\usepackage{xcolor}    
\usepackage[flushleft]{threeparttable}   
\hypersetup{
    colorlinks,
    linkcolor={red!50!black},
    citecolor={blue!50!black},
    urlcolor={blue!80!black}
}

\usepackage{lettrine} 
\usepackage{paralist} 
\usepackage{fancyhdr} 
\usepackage{titlesec} 
\usepackage{appendix} 
\usepackage[bottom]{footmisc} 
\usepackage{array} 
\usepackage{color}    
\usepackage{colortbl} 
\usepackage{libertine}
\usepackage[libertine]{newtxmath}
\usepackage[a4paper,hmargin={0.85in,0.85in},vmargin=0.9in]{geometry}
\usepackage{epstopdf}
\usepackage{dsfont}
\usepackage{mathtools}
\usepackage{setspace}
\usepackage{graphicx} 
\usepackage{placeins}
\usepackage{lscape}

\graphicspath{{./Plot/}}

\newtheorem{theorem}{Theorem}

\newtheorem{proposition}[theorem]{Proposition}

\theoremstyle{definition}

          \def\E{\mbox{E}}     
  
     \def\Var{\mbox{Var}}

\newcommand{\addifdefined}[1]{%
  \ifdefined#1
    \expandafter\@firstofone
  \else
    \expandafter\@
  \fi}

\begin{document}

\title{Illiquidity at Risk\thanks{
We thank the participants at the 2nd Italian Conference on Economic Statistics (University of Florence, 2024), the ICEEE conference (University of Palermo, 2025), the SoFiE conference (ESSEC Business School, 2025), the CREST seminar (ENSAE Paris, 2026), and the QFFE conference (Marseille, 2026). Paolo Santucci de Magistris also acknowledges the research support of the European Union’s Next Generation EU program through the Italian PRIN 2022 - M4C2, Investment 1.1 - “Monitoring Risks in Financial Markets” (Codice Cineca: 2022NEL482 - CUP: I53D23003410008).}}
\author{Demetrio Lacava\thanks{University of Messina, Department of Economics, Piazza Pugliatti, 1, 98122 Messina, Italy. E-mail: dlacava@unime.it}
\and
Paolo Santucci de Magistris \thanks{Luiss University, Department of Economics and Financial Markets, Viale Romania 32, 00197
Roma, Italy. E-mail: sdemagistris@luiss.it.}\hspace{1.5mm} \thanks{Corresponding author.} }

\date{}
\maketitle

\begin{abstract}

Market efficiency relies fundamentally on stable liquidity. Consequently, forecasting liquidity dynamics is a priority for both investors and regulators. We introduce a new tail-risk metric, Illiquidity-at-Risk (IlliQaR), designed to quantify the magnitude of extreme liquidity dry-ups. Relying upon the realized Amihud (a precise illiquidity measurement derived from high-frequency data as the ratio of realized volatility to trading volume) we assess the predictive power of various linear and non-linear econometric models, with a specific focus on the impact of discontinuous jump components. Accounting for these jumps is essential for achieving accurate probability coverage and better IlliQaR predictions during periods of systemic stress, where standard continuous models systematically underestimate the severity of liquidity evaporation.  Our empirical analysis, encompassing the S\&P 500 index and a cross-section of 25 large U.S. equities, demonstrates that incorporating jumps significantly improves forecasts of illiquidity. Our results suggest that individual stock IlliQaR violations often cluster during periods of S\&P 500 liquidity stress. This indicates that \textit{Illiquidity at Risk} is not just a localized concern but a systemic one, where the main index acts as a leading indicator for extreme dry-ups in individual stock liquidity.

\end{abstract}

\vspace{0.3cm}

\emph{J.E.L. classification}: C15, F31, G12, G15

\spacing{1.55}
\newpage

\section{Introduction}

Financial markets rely on two pillars for efficient functioning: price discovery and liquidity. While volatility has long been the primary metric for assessing risk and price discovery, liquidity, i.e. the ability to trade substantial quantities of an asset quickly and at low cost, is equally critical. In many contexts, liquidity is the invisible infrastructure of the market; when it functions well, it is taken for granted, but its disappearance can trigger systemic crises. Conceptually, liquidity risk differs from price risk. While price risk (or volatility) refers to the uncertainty regarding the future value of an asset, liquidity risk refers to the uncertainty regarding the ability to realize that value. It is the risk that an investor cannot convert a financial asset into cash (or vice versa) without incurring a prohibitive cost or significantly moving the price. Conceptually, liquidity is an intrinsic and unobservable market characteristic; its sudden disappearance can transform idiosyncratic shocks into systemic crises, establishing market liquidity as a fundamental risk factor in asset pricing, see among others \cite{pastor2003liquidity} and \cite{acharya2005asset}.

While the existing literature has focused extensively on modeling the level of illiquidity, market participants are often more concerned with the tails, i.e. the sudden ``evaporation'' of liquidity that characterizes flash crashes and financial panics. To address the challenge of quantifying this dimension of liquidity risk, this paper introduces a novel contribution: Illiquidity-at-Risk (IlliQaR). Much like Value-at-Risk (VaR) in market risk management, IlliQaR is designed to capture the tail risk of market liquidity, answering the question: 'What is the maximum level of illiquidity expected over a given horizon at a specific confidence level?'.  The reliability of IlliQaR as a risk metric rests on two critical pillars. First, the use of high-frequency data to construct a highly precise measurement of illiquidity allows for the filtration of observation noise, which is shown to mask the true tail behavior when employing lower-frequency proxies. Second, empirical evidence demonstrates that a robust IlliQaR framework must explicitly account for discontinuous illiquidity bursts. As these bursts are the primary drivers of severe liquidity shocks, failure to model them results in a significant underestimation of risk during periods of systemic stress.

We construct IlliQaR by modeling the conditional density of illiquidity, paying particular attention to the role of discontinuous jumps in liquidity dynamics. We argue that failing to account for these jumps leads to a severe underestimation of liquidity risk, leaving portfolios exposed to unexpected transaction costs exactly when trading is most urgent. The inclusion of a jump component is not merely an econometric convenience, but is heavily rooted in market microstructure theory. Sudden evaporation of liquidity—or illiquidity jumps—can be driven by the sudden binding of dealers' capital constraints, see \cite{brunnermeier2009market}, the synchronized withdrawal of algorithmic `phantom liquidity' during periods of stress or flash crashes (\citealp{kirilenko2017flash}), and discrete shocks to adverse selection risk following information arrivals, as in the seminal work of \cite{glosten1985bid}.

The necessity of such a precise risk metric brings us to the fundamental problem of measurement. Before one can model the risk of illiquidity, one must define and measure the underlying latent variable. The academic literature proposes a taxonomy of illiquidity based on three main dimensions: tightness, depth, and resiliency. Tightness refers to the cost of a round-trip transaction and is typically approximated by the bid-ask spread. Following the pioneering work of \cite{Roll1}, numerous studies including \cite{hasbrouck2009trading}, \cite{CorwinSchultz}, and \cite{Abdi2017} have focused on estimating the effective spread from low-frequency data. While bid-ask tightness provides a snapshot of transaction costs, it represents only a single dimension of market liquidity. A market can exhibit narrow spreads yet lack the necessary depth to absorb significant order flow without precipitating adverse price movements.  Depth is visible in the limit order book as the volume available at the best bid and ask, but it is harder to reconstruct historically. Finally, and perhaps most importantly for stability, is resiliency: the speed at which prices recover to their equilibrium level after a large trade causes a dislocation.

Measuring resiliency and the price impact of trades requires a more sophisticated approach than simply observing spreads. \cite{kyle1985continuous} famously introduced ``lambda'' to proxy for the elasticity of prices, but it was the contribution of \cite{Amihud2002illiquidity} that provided the most enduring and practical measure: the ratio of absolute returns to trading volume. While the Amihud measure provides an intuitive proxy for price impact by scaling absolute returns by trading volume, daily proxies often introduce substantial noise, necessitating the use of higher-frequency realized measures for greater accuracy. In this paper, we emphasize the importance of high-precision measurement to disentangle genuine liquidity information from measurement error. We therefore rely on the \textit{realized Amihud} measure first proposed by \cite{RanaldoSantucci2020} and refined in \cite{lacava2023realized}. By aggregating high-frequency intraday data, this measure provides a ``noise-reduced" signal of daily illiquidity. Precision here is not merely a statistical luxury; it is a prerequisite for risk management. If the measurement of illiquidity is contaminated by noise, any risk metric derived from it (like IlliQaR) will yield false alarms or, worse, fail to detect stress scenarios.

Building on this high-precision proxy, we evaluate a number of econometric models to forecast illiquidity and estimate IlliQaR. We employ both linear and non-linear specifications, in particular the Multiplicative Error Model (MEM) framework of \cite{Engle:2002}, which is uniquely suited for non-negative processes like illiquidity. To account for the well-known persistence of liquidity (the fact that illiquid days tend to cluster together), we incorporate the Heterogeneous Autoregressive (HAR) structure of \cite{Corsi2009}. Crucially, we augment these models with a ``jump" component, utilizing the MEM-J specification of \cite{Caporin:Rossi:DeMagistris:2017}. This is motivated by the empirical observation that liquidity does not always deteriorate gradually; it often ``jumps" abruptly in response to news or market shocks. We investigate whether these jumps are merely transient noise or if they carry information about future risks. By explicitly modeling the probability of these jumps, we aim to correct the distributional assumptions that underpin standard risk models, particularly addressing the positive skewness and fat tails characteristic of liquidity distributions.

Our empirical analysis yields several robust findings that extend our understanding of illiquidity dynamics. We analyze the daily realized Amihud series for the S\&P 500 and a cross-section of 25 major individual U.S. equities. While the S\&P 500 index serves as a barometer for systemic liquidity risk, the transmission of these dry-ups to individual constituents is non-uniform. By analyzing 25 large-cap equities, we bridge the gap between market-wide IlliQaR and idiosyncratic liquidity failures, examining whether the jump dynamics observed at the index level are a fundamental property of the underlying assets or a result of aggregation.

First, consistent with \cite{lacava2023realized}, we confirm that illiquidity is a long-memory process characterized by strong clustering. This aligns with the findings of \cite{hafner2023dynamic,hafner2025permanent}, who attribute the observed persistence of illiquidity (as measured by the range Amihud proxy introduced by \citealp{lacava2023realized}) to a slowly varying trend component that is potentially subject to structural breaks. Furthermore, we demonstrate that illiquidity dynamics are also characterized by significant jump activity. 
Second, and perhaps most consequentially, we find that identifying and modeling these jumps is essential for short-term forecasting. In the context of 1-step-ahead forecasts, the MEM-J specification significantly outperforms models that assume purely continuous dynamics. This superiority is driven by the model's ability to react rapidly to the onset of liquidity stress. Conversely, over medium-to-long horizons, the importance of the jump component fades as illiquidity mean-reverts; in these cases, simpler asymmetric specifications perform adequately. 
Third, regarding the estimation of IlliQaR, we find that standard models fail to provide adequate coverage for extreme events, tending to underestimate the frequency of severe illiquidity spikes. Our proposed models, which account for both the discontinuous illiquidity jump component and the specific error distribution, successfully correct this bias and provide accurate probability coverage in the tails. Notably, further improvements in IlliQaR accuracy are achieved by disentangling illiquidity jumps from price jumps—specifically, by utilizing bipower variation rather than realized volatility. Taken together, these results confirm that IlliQaR is a viable and indispensable tool for modern risk management.

Finally, we investigate the economic determinants of IlliQaR violations to understand why liquidity dries up. By regressing the probability of an extreme illiquidity event against various market variables, we find a strong asymmetric effect, analogous to the leverage effect in volatility: liquidity is significantly more likely to evaporate during market downturns than during rallies. Furthermore, we document a strong positive relationship between IlliQaR violations and the VIX index, as well as measures of economic policy uncertainty. Interestingly, risk aversion plays a distinct role; as investors become more risk-averse, the probability of a liquidity crisis increases, likely due to a withdrawal of market-making capital. These relationships hold both in-sample and in out-of-sample forecasting tests, validating the economic rationale behind our risk metric.

The remainder of the paper is organized as follows. Section \ref{sec:IlliQaR} formally defines the concept of Illiquidity-at-Risk, introduces the realized Amihud and the baseline econometric frameworks, specifically the Multiplicative Error Model (MEM) and the Heterogeneous Autoregressive (HAR) specifications. Section \ref{sec:jump_model} extends this framework by incorporating a jump component (MEM-J) and outlines the identification mechanism for conditional jumps. Section \ref{sec:empirical_analysis} is devoted to the empirical application: Section \ref{sec:dataset} provides descriptive statistics for the S\&P 500 index; Section \ref{sec:results} discusses the in-sample estimation results and the significance of the jump parameters; and Section \ref{sec:forecast} evaluates the out-of-sample forecasting performance. Section \ref{sec:illiquidity_determinats} focuses on the evaluation of the IlliQaR metric via coverage tests and analyzes the economic drivers of liquidity stress. Section \ref{sec:individual} extends the analysis by assessing IlliQaR on individual US stocks. Finally, Section \ref{sec:conclusion} concludes. Additional empirical findings and robustness checks are provided in the Supplementary Document.

\section{Illiquidity-at-Risk}\label{sec:IlliQaR}

We introduce the concept of Illiquidity-at-Risk. Analogous to the VaR for large negative returns, IlliQaR is defined as the threshold level, $\text{IlliQaR}_t(p)$, that the realized illiquidity measure, $\text{Illiq}_t$, is expected to exceed with probability $p$. Formally
\begin{equation}\label{eq:IlliQaR}
	\text{Pr}(\text{Illiq}_t>\text{IlliQaR}_t(p)|\mathcal{F}_{t-1})=p, \quad t=1,2,\ldots,T,
\end{equation} 
where $\mathcal{F}_{t-1}$ is the available information set. The evaluation of the probability in  \eqref{eq:IlliQaR} requires several ingredients.  In particular, IlliQaR relies on three methodological pillars: the derivation of a precise illiquidity measurement to proxy the unobservable true process, the implementation of a dynamic predictive model to account for time-series dependencies, and the use of a flexible econometric distribution to accurately model tail illiquidity events.

\subsection{Measuring Illiquidity}\label{sec:meas}

As a measurement of illiquidity we consider the \textit{realized Amihud}, introduced by \cite{RanaldoSantucci2020}, which refines the illiquidity proxy proposed by \cite{Amihud2002illiquidity}. Realized illiquidity of a given security is defined as the ratio of two observable quantites referring respectively to realized volatility and volume computed on a given unit interval (e.g. a day, a week, a month). In particular,  assuming to split a unit period of time into $M$ subperiods, the realized Amihud on a given interval of unit-length is defined as
\begin{equation}
	\text{Illiq}=\frac{RPV}{\nu} \label{eq:realized_Amihud},
\end{equation}
where $RPV=\sum_{i=1}^{M}|r_i|$ denotes the realized power variation \citep{barndorff2003realized} of order one (or the realized absolute variation) with $r_i$ being the log-return on the $i$-th subperiod. Similarly, $\nu=\sum_{i=1}^{M}\nu_i$ denotes the trading volume, which is obtained as the sum of the volume generated on each sub-interval interval $i=1,\ldots M$. \cite{lacava2023realized} provide a comprehensive theoretical derivation of the properties of the realized Amihud in measuring \textit{stochastic illiquidity} by assuming that the market of a given security is populated by traders with different reservation prices. In particular, \cite{lacava2023realized} show that $\text{Illiq}$ converges to the true illiquidity signal aggregated on a given interval of unit length as $M \to \infty$. Following the framework outlined in \cite{RanaldoSantucci2020}, the theoretical quantity measured by the realized Amihud is the \textit{integrated illiquidity}, defined as  $\text{Illiq}^* = \frac{1}{\int_{0}^{1} \ell(s) ds}$, where $\ell(s)$ measures the instantaneous sensitivity of the number of trades to new information, representing a dimension of illiquidity related to market depth.
\cite{lacava2023realized}  show that the precision of the realized Amihud in measuring daily integrated illiquidity based on intradaily returns sampled at 5-minutes intervals is several times more efficient than that obtained with daily Amihud. In particular, the daily Amihud of \cite{Amihud2002illiquidity} is obtained as a special case of realized Amihud, and it is obtained when $M=1$, i.e. using one observation on returns per day.

\subsection{Linear and non-linear models for illiquidity }\label{sec:lin-nonlin}

We consider several econometric specifications designed to model the dynamic evolution of illiquidity and of its quantiles, with the goal of providing a precise prediction of IlliQaR. By computing an illiquidity measure over disjoint daily intervals, we obtain a time series of illiquidity measurements, $\{\text{Illiq}_t\}_{t=1}^{T}$, where $T$ denotes the sample size. Given that illiquidity is a non-negative, persistent, and heavy-tailed process, its dynamic and distributional features can be effectively analyzed using the econometric framework adopted in the realized variance literature, following the seminal contribution of \cite{andersen2003modeling}.

First, we consider a linear benchmark, the Heterogeneous Autoregressive (HAR) model of \cite{Corsi2009}.  The model is defined as
    \begin{equation}
    	\text{Illiq}_t=\omega + \alpha_1 \text{Illiq}_{t-1} +\alpha_2 \text{Illiq}_{t-1:t-5} + \alpha_3 \text{Illiq}_{t-1:t-21} + u_t, \quad u_t|\mathcal{F}_{t-1} \sim N(0,\sigma^2_u),
    	\label{eq:har_lin}
    \end{equation}
where  $\text{Illiq}_{t-1:t-5}$ ($\text{Illiq}_{t-1:t-21}$) denote the weekly (monthly) realized Amihud, computed as the rolling average over the last 5 (21) days, and $\mathcal{F}_{t-1}$ is the information set available at time $t-1$. This specification is particularly well-suited for modeling the long-range dependence structure of illiquidity, through a simple linear additive structure of daily, weekly, and monthly components. Recent studies by \cite{hafner2023dynamic, hafner2025permanent} highlight the importance of disentangling slow-moving trends from fast-moving innovations in illiquidity. The HAR specification can be viewed as a reduced-form of this component-based approach; it approximates the multi-scale dynamics of illiquidity through a cascade of autoregressive terms that represent varying speeds of adjustment. It follows that for the Gaussian HAR model, IlliQaR is
\[
\text{IlliQaR}_t^{\text{HAR}}(p)=\omega + \alpha_1 \text{Illiq}_{t-1} +\alpha_2 \text{Illiq}_{t-1:t-5} + \alpha_3 \text{Illiq}_{t-1:t-21}+\Phi^{-1}(1-p)\sigma_u,
\]
where $\Phi^{-1}(1-p)$ denotes the inverse cumulative distribution function (CDF) of the standard Gaussian random variable evaluated at $1-p$ (as we consider exceedance over the right tail).

 We also extend the analysis to a class of non-linear specifications specifically designed to accommodate non-negative stochastic processes, namely the class of Multiplicative Error models (MEM), as introduced by \cite{Engle:2002} and revised by \cite{Engle:Gallo:2006} to account for an asymmetric response to the sign of past news (i.e. the Asymmetric MEM - AMEM).  The AMEM for $\text{Illiq}_t$ is specified as 
\begin{equation}
	\begin{array}{l}
\text{Illiq}_{t}=\mu _{t}\epsilon _{t }, ~ \qquad \epsilon _{t}|\mathcal{F}_{t-1}\sim \Gamma(\vartheta ,\frac{1}{\vartheta }),\\
\mu _{t}=\omega +\alpha_1 \text{Illiq}_{t-1}+\beta_1 \mu _{t-1}+\gamma
D_{t-1}\text{Illiq}_{t-1}, \quad t=1,2,\ldots,T,
\end{array}\label{eq:amem}
\end{equation}
where  $D_{t-1}$ a dummy variable taking value of 1 if the return of the considered asset is negative, 0 otherwise. Since $\mu_t$ follows the GARCH(1,1) dynamic process of \citep{Bollerslev:1986}, the usual parameter constraints for positiveness ($\omega,\alpha_1,\beta_1,\gamma>0$) and stationarity ($\alpha_1 + \beta_1 +\gamma/2>0$) are imposed. The conditional density of the error term is
\begin{equation*}
\begin{array}{l}
	f(\epsilon_t|\mathcal{F}_{t-1})=\frac{1}{\Gamma(\vartheta)}\vartheta^\vartheta\epsilon_t^{\vartheta-1}e^{-\vartheta\epsilon_t},\qquad \epsilon_t>0,
\end{array} \label{eq:density}
\end{equation*}
which is the one of a Gamma distribution with shape parameter $\vartheta>0$ and scale $\frac{1}{\vartheta}$. This means that $\epsilon_t$ is a random variable with unit mean and variance equal to $\frac{1}{\vartheta}$, denoted as $\epsilon_t\sim \Gamma(1,\vartheta)$ in the mean-shape representation. It follows that $\E(\text{Illiq}_t|\mathcal{F}_{t-1})=\mu_t$ and $\Var(\text{Illiq}_t|\mathcal{F}_{t-1})=\mu^2_t/\vartheta$. 
The conditional density of $\text{Illiq}_{t}$ is therefore available in closed form, thus allowing us to estimate the model parameters by maximum likelihood (ML). The asymptotic properties of the ML estimator of the unknown coefficients are discussed in \cite{Engle:2002} and \cite{Engle:Gallo:2006}. By recalling the quasi-ML principle, they show that the ML estimator is consistent and efficient, irrespectively of the appropriateness of the distribution for the error term.  We also consider the HAR specification for $\mu_t$ (see \citealp{Gallo:Otranto:2015} and \citealp{Caporin:Rossi:DeMagistris:2017}), that is
    \begin{equation}
    	\mu_t=\omega + \alpha_1 \text{Illiq}_{t-1} +\alpha_2 \text{Illiq}_{t-1:t-5} + \alpha_3 \text{Illiq}_{t-1:t-21} +\beta \mu_{t-1} + \gamma  D_{t-1} \text{Illiq}_{t-1}.
    	\label{eq:har}
    \end{equation}
 We call this model G-AMEM-HAR. The G-AMEM-HAR nests all the other specifications for $\mu_t$. For instance, the AMEM is obtained by imposing $\alpha_2=\alpha_3=0$, while with $\beta=0$ the model reduces to the AMEM-HAR. Under the various MEM specifications, the IlliQaR is
 \begin{equation}\label{eq:illiqar_mem}
 \text{IlliQaR}^\text{MEM}_t(p)= \frac{\mu_t}{\vartheta}\gamma^{-1}(\vartheta,\Gamma(\vartheta)(1-p)),
 \end{equation}
 where $\gamma^{-1}(\cdot)$ is the inverse of the lower incomplete Gamma function.

\subsection{Illiquidity jumps}\label{sec:jump_model}

Although the adoption of the Gamma distribution as in the baseline MEM model is justified in view of its robustness \citep[see][]{Cipollini:Engle:Gallo:2013}, other distributions are widely considered in the current literature: \cite{Engle:2002} suggests to use the exponential distribution, while mixture models are provided by \cite{Lanne:2006}. However, as shown by \cite{Caporin:Rossi:DeMagistris:2017}, often these distributional assumptions do not ensure an adequate coverage of the probability of tail events of volatility.  We therefore consider a variant of the MEM specification that features conditional jumps dynamics in illiquidity, as originally proposed by \cite{Caporin:Rossi:DeMagistris:2017}. In such a model, $\text{Illiq}_t$ is the result of the product of three elements
\begin{equation}
	\begin{array}{l}
        \text{Illiq}_t=\mu_tZ_t\epsilon_t,
	\end{array}\label{eq:mem-j}
\end{equation}
where $\mu_t$ and $\epsilon_t$ are defined as in \eqref{eq:amem}, while $Z_t$ is the jump component, assumed to be independent of $\epsilon_t$. The jump term is modeled as a compound Poisson random variable, where the expected number of jump arrivals at time $t$ ($N_t$) is governed by a Poisson random variable with a time-varying intensity $\kappa_t$: when $N_t>0$, the jump size is given by the sum of independent Gamma random variables, $Y_t|\mathcal{F}_{t-1}\sim \Gamma(d_{\kappa_t},\zeta)$ in mean-shape representation, corresponding to a scale given by $\frac{d_{\kappa_t}}{\zeta}$.  In particular, $Z_t=d_{\kappa_t}$ if $N_t=0$ and $Z_t=\sum_{j=1}^{N_t}Y_{j,t}$ for $N_t>0$, where $d_{\kappa_t}=(e^{-\kappa_t}+\kappa_t)^{-1}$ ensures $E[Z_t\epsilon_t|\mathcal{F}_{t-1}]=1$.
As a consequence of Poisson distributed jumps, the probability of observing a $j\geq 0$ jumps, conditioning on $\mathcal{F}_{t-1}$, is 
 \begin{equation*}
 		\text{Pr}(N_t=j|\mathcal{F}_{t-1})=\frac{e^{-\kappa_t}\kappa^j_t}{j!},\qquad j=0,1,2,\dots. 
 	\label{eq:poisson}
 \end{equation*}
Similarly to \cite{Chan:Maheu:2002}, we allow for a time-varying dynamics of the jump size, by specifying an autoregressive structure for $\kappa_t$ as 
\begin{equation*}
	\kappa_t=\phi_1+\phi_2\kappa_{t-1}+\phi_3\xi_{t-1},
	\label{eq:jump_intensity}
\end{equation*}
where the jumps innovations are defined as
\begin{equation*}
	\xi_t=E(N_t|\mathcal{F}_t)-E(N_t|\mathcal{F}_{t-1})=\sum_{j=0}^{\infty}j \text{Pr}(N_t=j|\mathcal{F}_t)-\kappa_t.
\end{equation*}
Imposing $\phi_1>0$ and $1>\phi_2>\phi_3>0$ is sufficient to ensure the positiveness and stationarity of $\kappa_t$, see \cite{Chan:Maheu:2002}. In particular, \cite{Maheu:McCurdy:2004} and \cite{maheu2013jumps} successfully applied this specification to study conditional jumps in stock returns, while \cite{Caporin:Rossi:DeMagistris:2016} focused the analysis on realized variance. Furthermore, acknowledging that $N_t$ cannot be directly observed in $\mathcal{F}_t$, $\xi_t$ can be interpreted as the forecast of $N_t$ as new information becomes available. Finally, the filtered probability needed to compute $E(N_t|\mathcal{F}_t)$ is obtained via the Bayes rule as
\begin{equation}
\text{Pr}(N_t=j|\mathcal{F}_t)=\frac{f(\text{Illiq}_t|N_t=j,\mathcal{F}_{t-1})\text{Pr}(N_t=j|\mathcal{F}_{t-1})}{f(\text{Illiq}_t|\mathcal{F}_{t-1})},
\label{eq:bayes_rule}
\end{equation}
where the density of $\text{Illiq}_t$ conditional on $N_t=j$ and $\mathcal{F}_{t-1}$ is
\begin{equation*}
	\begin{large}
f_{\text{MEM-J}}(\text{Illiq}_t|N_t=j,\mathcal{F}_{t-1})=
\begin{cases}
	\frac{1}{\text{Illiq}_t}\left(\frac{\vartheta \text{Illiq}_t}{d_{\kappa_t} \mu_t}\right)^{\vartheta}~ \frac{e^{\left(\frac{-\vartheta \text{Illiq}_t}{d_{\kappa_t} \mu_t}\right)}}{\Gamma(\vartheta)}, \qquad \qquad \qquad \qquad \qquad \quad ~ N_t=0\\
	\frac{2}{\text{Illiq}_t}\left(\frac{\text{Illiq}_t}{\mu_t}\frac{\vartheta \zeta}{d_{\kappa_t}}\right)\left(\frac{j\zeta+\vartheta}{2}\right)\frac{1}{\Gamma(j\zeta)\Gamma(\vartheta)}\mathds{K}_{j\zeta-\vartheta}\left(2\sqrt{\frac{\text{Illiq}_t}{\mu_t}\frac{\vartheta \zeta}{d_{\kappa_t}}}\right),\quad N_t=j>0,
\end{cases}
\end{large}
\end{equation*}
where $\mathds{K}(\cdot)$ is the modified Bessel function of second kind, while the denominator in Eq. \eqref{eq:bayes_rule} is given by
\begin{equation*}
	f_{\text{MEM-J}}(\text{Illiq}_t|\mathcal{F}_{t-1})=e^{-\kappa_t}f_{\text{MEM-J}}(\text{Illiq}_t|N_t=0,\mathcal{F}_{t-1})+\sum_{j=1}^{\infty}\frac{e^{-\kappa_t} \kappa_t^j}{j!} f_{\text{MEM-J}}(\text{Illiq}_t|N_t=j,\mathcal{F}_{t-1}).
\end{equation*}
The conditional density, which is therefore a mixture of Gamma and Kappa distributions, is available in closed form, allowing for model parameters estimation by ML. The following proposition derives the analytical expression for the IlliQaR under the MEM-J specification.
\begin{proposition}\label{theo1}
The IlliQaR for the MEM-J model is obtained as the value $\text{IlliQaR}_t(p)$ such that
\begin{equation}
\text{IlliQaR}^\text{MEM-J}_t(p): \qquad p=1-F_{\text{MEM-J}}(\text{IlliQaR}_t^\text{MEM-J}(p)|\mathcal{F}_{t-1}),
\end{equation}
where $F_{\text{MEM-J}}(\cdot|\mathcal{F}_{t-1})$
denotes the conditional CDF of the MEM-J, that is
\begin{equation*}
	F_{\text{MEM-J}}(\text{Illiq}_t|\mathcal{F}_{t-1})=e^{-\kappa_t}F_{\text{MEM-J}}(\text{Illiq}_t|N_t=0,\mathcal{F}_{t-1})+\sum_{j=1}^{\infty}\frac{e^{-\kappa_t} \kappa_t^j}{j!} F_{\text{MEM-J}}(\text{Illiq}_t|N_t=j,\mathcal{F}_{t-1}),
\end{equation*}
where $F_{\text{MEM-J}}(\text{Illiq}_t|N_t=0,\mathcal{F}_{t-1})=\frac{1}{\Gamma(\vartheta)}\int_{0}^{\frac{\vartheta \text{Illiq}_t}{d_{\kappa_t}\mu_t}}s^{\vartheta-1}e^{-s}ds$
is the CDF of a Gamma-distributed random variable with shape $\vartheta$ and scale $\frac{d_{\kappa_t} \mu_t}{\vartheta}$ and
\[
F_{\text{MEM-J}}(\text{Illiq}_t|N_t=j,\mathcal{F}_{t-1})=\frac{2^{2-j\zeta-\vartheta}}{\Gamma(j\zeta)\Gamma(\vartheta)}\int_{0}^{2\sqrt{\frac{\zeta\vartheta\text{Illiq}_t}{d_{\kappa_t}\mu_t}}}s^{j\zeta+\vartheta-1}\mathds{K}_{|j\zeta-\vartheta|}\left(s\right)ds,
\]
is the CDF of a Kappa distributed random variable. 
\end{proposition}
The CDF of the MEM-J is an essential element for the calculation of IlliQaR as well as for testing the adequacy of this specification as illustrated in Section \ref{sec:testing}.

\subsection{Density Forecast Evaluation: The Berkowitz Test} 
\label{sec:testing}

The evaluation of the IlliQaR is based on the Berkowitz test \citep{Berkowitz:2001} for the adequacy of the density function with the realization of the dependent variable. Considering the flexibility of such a test, it can be applied to evaluate the fit of the full density as well as to specific quantile. In our application, we consider three quantiles, corresponding to the 1, 5 and 10\% of the Normal distribution.  Importantly, the test is applied on the conditional density function (CDF) of $\text{Illiq}_t$ as expressed in
\begin{equation}
	y_t=F(\text{Illiq}_t|\mathcal{F}_{t-1})=\int_{0}^{\text{Illiq}_t} f(x|\mathcal{F}_{t-1})dx.
	\label{eq:cdf}
\end{equation}
Taking as an example the linear HAR model, the CDF is that of a Normal distribution. As for the MEM specifications, $F_{\text{MEM}}(\text{Illiq}_t|\mathcal{F}_{t-1})$ is given by the Gamma CDF. Finally, for the jump models in \eqref{eq:cdf}, the CDF is built as a mixture of a Gamma and a Kappa CDFs, as in Proposition \ref{theo1}. 

The test is constructed under the null hypothesis that the model is correctly specified: if this is the case, the empirical CDF follows an uniform distribution, i.e. $y_t\sim U(0,1)$. The Berkowitz is therefore derived as a LR test for the comparison of the empirical CDF, $y_t$, and a new truncated variable, computed starting from the inverse of the standard normal distribution of $y_t$ itself. In other words, let the $s_t$ be defined as
\begin{equation*}
	s_t=\Phi^{-1}(y_t),
\end{equation*} 
we compute the new truncated variable, $s^*_t$ as
\begin{equation}
s^*_t=
	\begin{cases}
	\text{IlliQaR}_t(p) &\qquad \text{if} \quad s_t\leq \text{IlliQaR}_t(p)\\
	s_t \qquad          &\qquad  \text{if} \quad  s_t> \text{IlliQaR}_t(p).
		\end{cases}
\end{equation}
The (right) tail coverage test can be derived using the LR principle based on
the censored normal density of $s^*_t$. Under the null of correct tail
coverage the test statistic is distributed as $\chi^2(2)$ as it corresponds
to a test for comparing the mean and variance estimated from
the truncated likelihood of $s^*_t$ to those expected under correct
coverage, i.e. zero mean and unit variance. 

\section{Empirical Analysis}\label{sec:empirical_analysis}

We conduct an empirical assessment of IlliQaR by examining a sample of daily illiquidity measures of the S\&P 500 from January 3, 2005 until October 15, 2021, and of 25 US stocks from January 3, 2012 to January 10, 2024. 

\subsection{Data}\label{sec:dataset}
Our empirical analysis focuses on the assessment of daily time series of the realized Amihud of the U.S. stock market, as represented by the S\&P 500 index. In analogy with \cite{RanaldoSantucci2020} and \cite{lacava2023realized}, daily illiquidity is computed as the ratio between the daily realized volatility (RV), which is the square root of the realized variance $RV=\sqrt{\sum_{i=1}^{I}r_i^2}$ \citep[computed using returns sampled at 5-minutes intervals, see][among many others]{liu2015does}, and the daily volume.\footnote{In practice, realized power variation (RPV) and realized variance (RV) exhibit an extremely high correlation, with an $R^2$ exceeding 99\%. Consequently, substituting RPV with RV in \eqref{eq:realized_Amihud} has a negligible impact on the resulting illiquidity series.  As an alternative to return-based proxies, \cite{lacava2023realized} propose a low-frequency yet efficient estimator based on the daily price range—the difference between the daily high and low log-prices. \cite{hafner2023dynamic, hafner2025permanent} have successfully employed this estimator for illiquidity prediction.}   Our data for RV are sourced from the Oxford-Man Institute, while daily trading volume is obtained from Datastream.
	\begin{figure}[h!]
 \centering
	\begin{subfigure}
{\includegraphics[height=9cm,width=16cm]{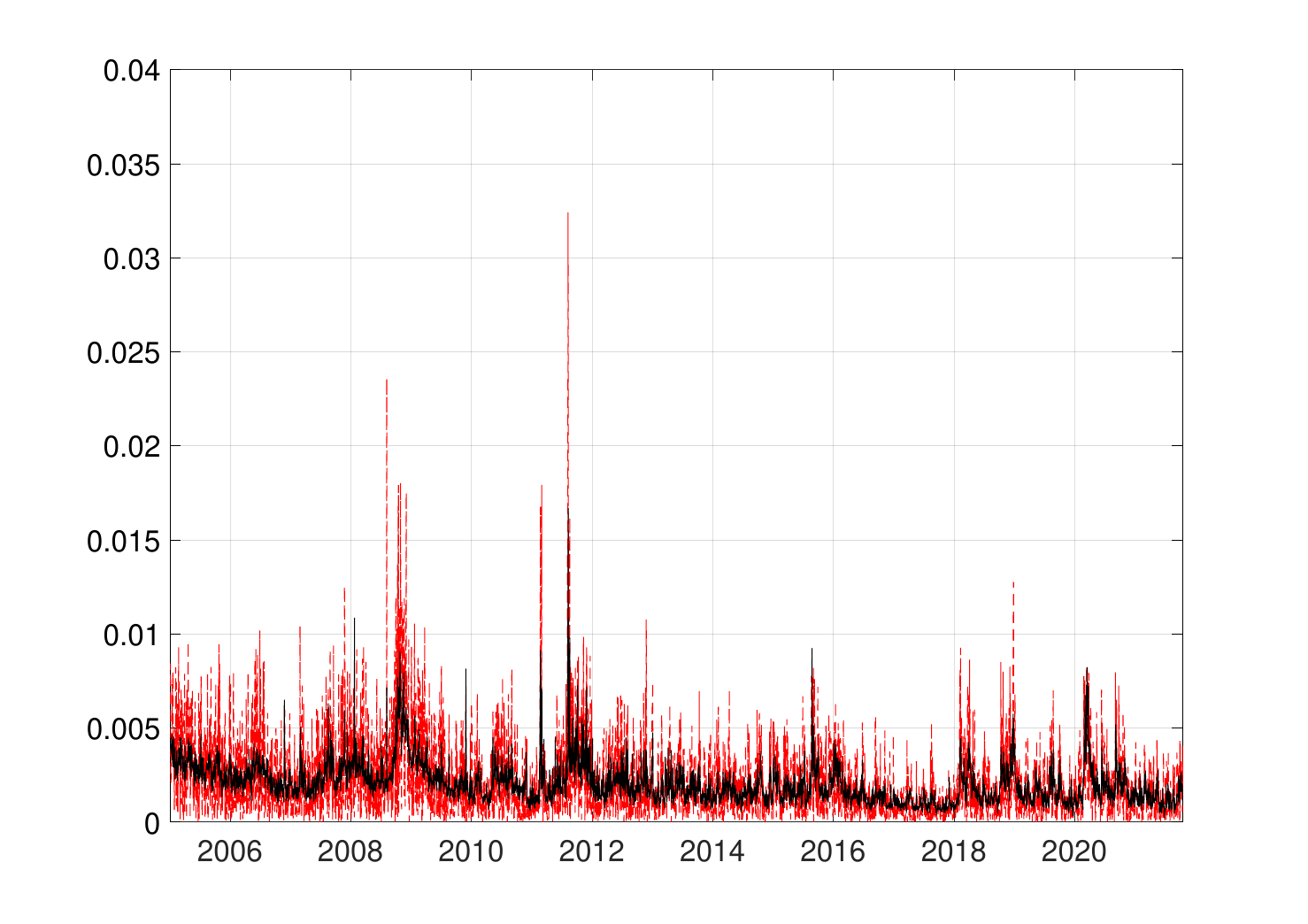}}
	\end{subfigure}
	\caption{Realized (black solid line) and daily Amihud (red dashed line) series for the S\&P 500.	\label{fig:amihud}}
\end{figure}

Figure \ref{fig:amihud} displays the time series of the realized Amihud (in black) for the S\&P 500. The persistence feature of $\text{Illiq}_t$ is evident at first sight, with long periods of low illiquidity followed by long periods of high illiquidity. Within the volatility literature, this is a well-established phenomenon known as {\it volatility clustering} (although in our case it is more appropriate to define it as {\it illiquidity clustering}).  Altogether, this gives us a justification for considering well-established volatility models for predicting illiquidity. Furthermore, the series exhibits frequent spikes, particularly during the early portion of the sample which coincides with periods of financial turbulence. Indeed, elevated illiquidity levels have characterized much of the past fifteen years, driven by a succession of systemic shocks following the 2008 Global Financial Crisis, including the European Sovereign Debt Crisis and the COVID-19 pandemic. 
Figure \ref{fig:amihud} also reports the daily Amihud (in red). The comparison with the realized Amihud signals two important features of these illiquidity measurements. First, realized Amihud and daily Amihud follow similar dynamic patterns. Second, the latter is much noisier than the former, thus confirming the efficiency gains in measuring illiquidity using high-frequency prices.

Table \ref{tab:stats} reports the sample statistics of realized Amihud, RV, trading volume, daily Amihud, and absolute returns for the S\&P 500. The realized Amihud displays kurtosis in excess compared to the reference value of the Normal distribution, as a consequence of extremely large realizations. These extreme realizations, typically originating from sharp illiquidity episodes, make the realized Amihud positively skewed, thus calling for a proper modeling framework. For instance, the MEM-J is expected to be able to assign the correct probability to these events. Indeed, due to the presence of measurement error, the traditional daily Amihud measure exhibits higher volatility, as indicated by its standard deviation, compared to the realized Amihud. Furthermore, the daily Amihud displays lower kurtosis, suggesting it is less effective at isolating extreme illiquidity realizations due to the noisy measurement of the latter. In the subsequent sections, we evaluate which econometric specifications best capture the dynamic and probabilistic properties of illiquidity to provide robust forecasts of future liquidity conditions.
\begin{table}[h!]
\centering
\setlength{\tabcolsep}{8pt}
\renewcommand{\arraystretch}{1.2}
\begin{tabular}{lccccc}
    \toprule
    &  RV & $|r|$ & Volume & $\text{Realized Amihud}$ & $\text{Daily Amihud}$  \\
    \midrule
    Mean           & 0.0080 & 0.0086 & 3.8015 & 0.0020 & 0.0018  \\
    Median         & 0.0061 & 0.0055 & 3.6412 & 0.0018 & 0.0012  \\
    St. Deviation  & 0.0065 & 0.0105 & 0.0065 & 0.0011 & 0.0018 \\
    Skewness       & 3.6558 & 3.5439 & 1.1037 & 2.4919 & 2.7596  \\
    Kurtosis       & 24.151 & 24.142 & 5.628 & 18.183 & 20.632 \\
    \midrule
    N. obs        & 4216   &        &        &        &        \\
    \bottomrule
\end{tabular}
\caption{S\&P 500 descriptive statistics. Realized Amihud and daily Amihud are scaled by a factor of 1.000E9, absolute returns are scaled by a factor of $\sqrt{\pi/2}$ while volume are divided by a factor of 1.000E8. \label{tab:stats}}
\end{table}

\subsection{Estimation results}\label{sec:results}
The estimation of the various MEM specifications presented in Section \ref{sec:IlliQaR} is performed on the \textit{full-sample} period, which, in the case of the S\&P 500, covers the years between January 3, 2005 to October 15, 2021. Panel a) of Table \ref{tab:results} shows the estimated coefficients for the S\&P 500 index. To model the dynamics of $\mu_t$, we consider a range of specifications including the linear (HAR) and non-linear (MEM) as discussed in Section \ref{sec:lin-nonlin}. Specifically, we consider the MEM and AMEM with both a GARCH(1,1) and HAR structures, alongside several jump-augmented specifications, namely the AMEM-J, AMEM(1,2)-J, G-AMEM-HAR-J, MEM-HAR-J, and AMEM-HAR-J. 
\begin{sidewaystable}
	\begin{adjustbox}{max width=0.88\linewidth,center}
	\begin{tabular}{lccccccccccccc|ccc}
    \vspace{0.4cm}
		Panel a)& \multicolumn{16}{c}{\Large\textbf{Parameter Estimates}}\\
        \vspace{0.4cm}
  &  \multicolumn{13}{c}{\large{Realized Amihud}} & \multicolumn{3}{c}{\large{Daily Amihud}}\\
& (I)      & (II)      & (III)       & (IV)      & (V)      & (VI)      & (VII)      & (VIII)      & (IX)         & (X)                 & (XI)         & (XII)   & (XIII) & (II)& (IV)  & (XI)    \\
\midrule
$\omega$      & $0.0002^a$                   & $0.0002^a$                   & $0.0001^a$                   & $0.0001^a$                   & $0.0001^a$                   & $0.0001^a$                   & $0.0001^a$                   & $0.0001^a$                   & $0.0001^a$                   & $0.0001^a$                   & $0.0001^a$                   & $0.0001^a$                   & $0.0001^a$  & $0.0004^a$                 & $0.0001^a$                 & $0.0001^a$                 \\
              & (0.0000)                     & (0.0000)                     & (0.0000)                     & (0.0000)                     & (0.0000)                     & (0.0000)                     & (0.0000)                     & (0.0000)                     & (0.0000)                     & (0.0000)                     & (0.0000)                     & (0.0000)                     & (0.0000)    & (0.0001)                   & (0.0000)                   & (0.0000)                   \\[2mm]
$\alpha_d$    & $0.3814^a$                   & $0.2802^a$                  & $0.4055^a$                   & $0.3077^a$                   & $0.3720^a$                   & $0.2741^a$                   & $0.2909^a$                   & $0.4092^a$                   & $0.3022^a$                   & $0.3183^a$                   & $0.2771^a$                   & $0.3686^a$                   & $0.2603^a$  & $-0.1314^a$                & $0.0056$                   & $0.0000$                   \\
              & (0.0623)                     & (0.0597)                     & (0.0234)                     & (0.0243)                     & (0.0259)                     & (0.0288)                     & (0.0260)                     & (0.0196)                     & (0.0186)                     & (0.0180)                     & (0.0201)                     & (0.0221)                     & (0.0246)    & (0.0296)                   & (0.0106)                   & (0.0278)                   \\[2mm]
$\alpha_w$    & $0.3599^a$                   & $0.4018^a$                   &                              &                              & $0.4045^a$                   & $0.4462^a$                   & $0.0731^b$                   &                              &                              &                              & $0.0771^b$                   & $0.4174^a$                   & $0.4686^a$  & $0.3656^a$                 &                            &                  $0.0000$          \\
              & (0.0848)                     & (0.0824)                     &                              &                              & (0.0338)                     & (0.0341)                     & (0.0373)                     &                              &                              &                              & (0.0360)                     & (0.0330)                     & (0.0344)    & (0.0541)                   &                            &                 (0.0258)           \\[2mm]
$\alpha_m$    & $0.1772^a$                   & $0.1784^a$                   &                              &                              & $0.1617^a$                   & $0.1662^a$                   & $0.0692^a$                   &                              &                              &                              & $0.0972^a$                   & $0.1676^a$                   & $0.1732^a$  & $0.4737^a$                 &                            &                           $0.0432^a$ \\
              & (0.0437)                     & (0.0415)                     &                              &                              & (0.0255)                     & (0.0254)                     & (0.0162)                     &                              &                              &                              & (0.0162)                     & (0.0275)                     & (0.0299)    & (0.0766)                   &                            &          (0.0132)                  \\[2mm]
$\beta_1$     &                              &                              & $0.5571^a$                   & $0.6112^a$                   &                              &                              & $0.4481^a$                   & $0.5613^a$                   & $0.6230^a$                   & $0.4724^a$                   & $0.4649^a$                   &                              &             &                            & $0.8961^a$                 & $0.8411^a$                 \\
              &                              &                              & (0.0260)                     & (0.0260)                     &                              &                              & (0.0430)                     & (0.0212)                     & (0.0199)                     & (0.0376)                     & (0.0386)                     &                              &             &                            & (0.0100)                   & (0.0195)                   \\[2mm]
$\beta_2$     &                              &                              &                              &                              &                              &                              &                              &                              &                              & $0.1318^a$                   &                              &                              &             &                            &                            &                            \\
              &                              &                              &                              &                              &                              &                              &                              &                              &                              & (0.0332)                    &                              &                              &             &                            &                            &                            \\[2mm]
$\gamma$      &                              & $0.1134^a$                   &                              & $0.0892^a$                   &                              & $0.0980^a$                   & $0.1105^a$                   &                              & $0.0948^a$                   & $0.1018^a$                   & $0.1175^a$  &                              & $0.1054^a$  & $0.1479^a$                 & $0.1421^a$                 & $0.1644^a$                 \\
              &                              & (0.0166)                     &                              & (0.0084)                     &                              & (0.0101)                     & (0.0101)                     &                              & (0.0067)                     & (0.0071)                     & (0.0082)                     &                              & (0.0092)    & (0.0317)                   & (0.0166)                   & (0.0223)                   \\[2mm]
$\vartheta$      &                              &                              & $12.4790^a$                  & $12.9010^a$                  & $12.4745^a$                  & $12.8165^a$                  & $13.1089^a$                  & $17.9103^a$                  & $19.1291^a$                  & $19.2810^a$                  & $19.7753^a$                  & $18.1070^a$                  & $19.2893^a$ &                            & $1.2266^a$                 & $21.6723^a$                 \\
              &                              &                              & (0.3845)                     & (0.4033)                     & (0.3688)                     & (0.3615)                     & (0.4239)                     & (0.5097)                     & (0.7224)                     & (0.5722)                     & (0.5661)                     & (0.0.5981)                   & (1.8122)    &                            & (0.0260)                   & (0.1421)                   \\[2mm]
$\zeta$       &                              &                              &                              &                              &                              &                              &                              & $17.3756^a$                  & $17.8975^a$                  & $17.9585^a$                  & $18.4562^a$                  & $16.9308^a$                  & $18.1812^a$ &                            &                            & $0.7757^a$                 \\
              &                              &                              &                              &                              &                              &                              &                              & (1.9085)                     & (0.5399)                     & (0.4641)                     & (0.3884)                     & (0.8604)                     & (2.8700)    &                            &                            & (0.0395)                   \\[2mm]
$\phi_1$      &                              &                              &                              &                              &                              &                              &                              & $0.0139^b$                   & $0.0119^a$                   & $0.0111^a$                   & $0.0076$                     & $0.0099$                     & $0.0099$    &                            &                            & $0.0344^b$                   \\
              &                              &                              &                              &                              &                              &                              &                              & (0.0056)                     & (0.0040)                     & (0.0040)                     & (0.0055)                     & (0.0078)                     & (0.0079)    &                            &                            & (0.0135)                   \\[2mm]
$\phi_2$      &                              &                              &                              &                              &                              &                              &                              & $0.9422^a$                   & $0.9560^a$                   & $0.9588^a$                   & $0.9712^a$                   & $0.9572^a$                   & $0.9617^a$  &                            &                            & $0.9896^a$                 \\
              &                              &                              &                              &                              &                              &                              &                              & (0.0222)                     & (0.0148)                     & (0.0148)                     & (0.0211)                     & (0.0346)                     & (0.0334)    &                            &                            & (0.0041)                   \\[2mm]
$\phi_3$      &                              &                              &                              &                              &                              &                              &                              & $0.2181^a$                   & $0.2139^a$                   & $0.2011^a$                   & $0.1420^a$                   & $0.1548^a$                   & $0.1491^b$  &                            &                            & $0.0224^c$                   \\
              &                              &                              &                              &                              &                              &                              &                              & (0.0364)                     & (0.0329)                     & (0.0311)                     & (0.0491)                     & (0.0542)                     & (0.0658)    &                            &                            & (0.0111)                   \\[2mm]
		\midrule
LogLik & 24753.09 & 24830.34 & 25759.31 & 25830.86 & 25758.53 & 25816.78 & 25865.27 & 25922.29 & 26013.86 & 26021.59 & 26052.75 & 25920.96 & 25995.17    & 21160.20  & 22861.33 & 22877.67 \\
\bottomrule
     \\
		Panel b)& \multicolumn{16}{c}{\Large\textbf{Ljung-Box test}}\\  
		\multicolumn{16}{l}{Residuals: $\epsilon_t$}\\[2mm]
LB 1          & 0.1015   & 0.0033   & 0.3999   & 0.3658   & 0.0321   & 0.0024   & 0.4420   & 0.4614   & 0.2808   & 0.8426   & 0.7770   & 0.0435   & 0.0045      & 0.3440 & 0.0063 & 0.0007\\[2mm]
LB 5          & 0.0000   & 0.0000   & 0.0163   & 0.0387   & 0.0000   & 0.0000   & 0.4748   & 0.0043   & 0.0129   & 0.0679   & 0.5702   & 0.0000   & 0.0000      & 0.0052 & 0.0805 & 0.0252 \\[2mm]
LB 10         & 0.0000   & 0.0000   & 0.0011   & 0.0054   & 0.0000   & 0.0000   & 0.3228   & 0.0001   & 0.0003   & 0.0004   & 0.3018   & 0.0000   & 0.0000      & 0.0131 & 0.1600 & 0.0526 \\[2mm]
      	\midrule

	\end{tabular}
	\end{adjustbox}
	\caption{Estimates for S\&P 500. Panel a): Estimated coefficients (robust standard errors in parenthesis); Panel b):  $p$-value of the Ljung-Box statistics. Sample period: January 3, 2005 - October 15, 2021. The superscripts a, b and, c denote significant coefficients at $1\%$, $5\%$ and, $10\%$ level, respectively. The estimated models are (I) HAR, (II) AHAR, (III) MEM, (IV) AMEM, (V) MEM-HAR, (VI) AMEM-HAR, (VII) G-AMEM-HAR, (VIII) MEM-J, (IX) AMEM-J, (X) AMEM(2,1)-J, (XI) G-AMEM-HAR-J, (XII) MEM-HAR-J, (XIII) AMEM-HAR-J.\label{tab:results}}
\end{sidewaystable}

The estimation of all model parameters is performed by ML and reported in Panel a) of Table \ref{tab:results}. The coefficients governing the conditional mean $\mu_t$ are all highly significant (at a 1\% significance level), with an average persistence of 95\% among the estimated models. As for $\hat\alpha_d$ coefficient, which summarizes the impact of news on market illiquidty, we find an average value of 0.33, with a peak at 0.41 for the MEM-J (model VII). As for the coefficient $\hat\gamma$, which measures the impact of bad news (in the form of negative returns) on illiquidity, it enters the model with a positive sign, pointing at a more heavily reaction of illiquidity against negative returns rather than against positive returns. This is also confirmed by the fact that, when the asymmetric effect is included in the model, $\hat\alpha_d$ reduces by 22\% on average, meaning that a large part of the impact of the price news on illiquidity is due to negative returns. We interpret this result as evidence in favor of the idea that illiquidity is strongly related to uncertainty among investors and that it causes a larger part of illiquidity persistence. 
This translates into estimated values for $\hat\beta$ (0.53, on average) that are typically below the usual value found for volatility \citep[see][Ch. 1]{Bauwens:Hafner:Laurent:2012}. Finally, as expected, $\hat\vartheta$ (which measures the inverse of the variance of the error term) is higher for the more sophisticated MEM-J models, where a significant part of the variability of illiquidity is explained by the conditional jump component. This result emphasizes the need to adopt a model that is able to assign the correct probability to extremely large realizations of the realized Amihud measure.  As for the specifications with a jump component, the process $\kappa_t$ for the jump intensity in the model is found to be a very persistent process ($\hat\phi_2$ is above 0.94 in all MEM-J specifications employing the realized Amihud), meaning that, similarly to the continuous part of illiquidity, also the number of jump arrivals strongly depends on its past realizations. The unconditional mean of the jump process, $\frac{\phi_1}{(1-\phi_2)}$, is between 0.23 and 0.27, corresponding to a jump every 4 days.

For comparison, we estimate the AHAR, AMEM, and G-AMEM-HAR-J specifications using the daily Amihud series (see the final three columns of Table \ref{tab:results}). The estimated coefficients for $\mu_t$ indicate that the daily series requires substantially more smoothing than its realized counterpart due to its more erratic nature. Specifically, the AMEM results yield a lower $\hat{\alpha}_d$ of $0.006$ (versus $0.306$ for the realized series) and a higher $\hat{\beta}_1$ of $0.896$ (versus $0.611$). Furthermore, the $\hat{\vartheta}$ estimate of $1.227$ is nearly an order of magnitude smaller than the $12.901$ obtained using realized Amihud. This disparity implies that the error term variance is approximately 11 times larger when the daily Amihud is used as the illiquidity proxy. These results highlight the significant precision gains achieved by utilizing high-frequency price data, which enhances model fit, improves explicability, and allows for a clearer separation between stochastic components of the model. As an additional illustration of the difficulty in disentangling noise from stochastic components, when estimating the G-AMEM-HAR-J on the daily Amihud series (last column of Table \ref{tab:results}), we find that $\hat{\phi}_1 \approx 0.03$ and $\hat{\phi}_2 \approx 0.99$. This leads to an unconditional jump arrival rate of nearly three jumps per day, signaling that liquidity jumps cannot be effectively isolated from the innovation component. Consequently, the log-likelihood remains essentially unchanged relative to a MEM without jumps, while the jump process variance is 25 times higher than that of the model estimated with the realized Amihud. This indicates that the jump component is poorly identified; the surge in variance primarily absorbs measurement noise rather than isolating genuine illiquidity jumps.

Panel b) of Table \ref{tab:results} presents diagnostic tests for all models, reporting $p$-values for the Ljung-Box test at 1, 5, and 10 lags to assess the properties of residuals. As far as models without jump are concerned, i.e. models (I)-(VII), we reject the null of no serial correlation of residuals at a 5\% significance level, with the only exception of the G-AMEM-HAR models (model VII) and for the HAR (I), the MEM (III), and the AMEM (IV), for the first lag. Results do not change if we consider jump models, i.e. models (VIII)-(XIII). In detail, we fail to reject the null hypothesis for the  G-AMEM-HAR-J (model XI), which is the only model able to correctly account for the persistent nature of the illiquidity series. A similar performance is observed for the AMEM(1,2)-J (model X), for which we do not reject the null at a 5\% level for lag 1 and 5, while residuals serial correlation is detected for lag 10. Once again, we interpret this result as an evidence in favor of the view that illiquidity needs to be modeled through a model that is able to assign the correct probability to extreme realizations. In Section \ref{sec:forecast}, models are compared by evaluating their out-of-sample forecasting capability by means of the model confidence set procedure.

\subsection{Forecasting Illiquidity}\label{sec:forecast}

This section presents the results of our forecasting analysis, which evaluates the relative performance of the candidate econometric models in predicting realized illiquidity. Accurate illiquidity forecasts are of great importance to both market makers and investors. For market makers, high liquidity facilitates the enforcement of no-arbitrage conditions, thereby ensuring price efficiency. For investors, precise forecasts are essential given the critical role that liquidity risk plays in optimal portfolio construction and risk management.
To evaluate the predictive power of the candidate models, we conduct an out-of-sample forecasting exercise across three distinct horizons: $h=1, 5,$ and $22$ days, corresponding to daily, weekly, and monthly frequencies. The out-of-sample period spans from October 16, 2015, to October 15, 2021. Forecasts are generated using coefficients estimated over the in-sample window (January 3, 2005, to October 15, 2015).  The in-sample parameter estimates, as reported in Table \ref*{tab:results_insample} in the Supplementary Document, are highly stable and do not exhibit significant deviations from the full-sample results, as shown in Table \ref{tab:results}.

To statistically evaluate the relative predictive accuracy of the candidate models, we employ the model confidence set (MCS) procedure developed by \cite{Hansen:Lunde:Nason:2011}. This procedure identifies the superior set of models that exhibit statistically indistinguishable forecasting performance at a given significance level. Following \cite{Patton:2011}, we utilize the QLIKE loss function, as it is robust to the choice of the proxy for the latent variable. The MCS is constructed using a 10\% significance level, and our results identify the models that are not outperformed by any others in the set.
\begin{table}[h!]
	\centering
    \setlength{\tabcolsep}{3pt}
\renewcommand{\arraystretch}{1.1}
	\begin{adjustbox}{max width=0.95\linewidth,center}
	\begin{tabular}{lcllcllc}
		\multicolumn{2}{c}{1-step   ahead}                                      &                      & \multicolumn{2}{c}{5-step ahead}                                        &                      & \multicolumn{2}{c}{22-step ahead}                            \\[0.5mm]
		\midrule
		\multicolumn{1}{c}{Model}                 & \multicolumn{1}{c}{p-value} & \multicolumn{1}{c}{} & \multicolumn{1}{c}{Model}                 & \multicolumn{1}{c}{p-value} & \multicolumn{1}{c}{} & \multicolumn{1}{c}{Model}      & \multicolumn{1}{c}{p-value} \\[0.5mm]
MEM                                     & 0.0012                  &  & AMEM-J                       & 0.0000                        &  & AMEM-J                & 0.0000 \\
AMEM                                    & 0.0027                  &  & AMEM                              & 0.0000                        &  & AMEM(2,1)-J                & 0.0000 \\
MEM-HAR                                 & 0.0036                  &  & MEM-J                             & 0.0009                        &  & MEM-J                      & 0.0000 \\
HAR                                     & 0.0080                  &  & AMEM(2,1)-J                       & 0.0015                        &  & AMEM                       & 0.0000 \\
MEM-J                                   & 0.0080                  &  & MEM                               & 0.0024                        &  & G-AMEM-HAR-J           & 0.0000 \\
AMEM-J                             & 0.0080                  &  & G-AMEM-HAR-J                  & 0.0073                        &  & MEM                        & 0.0000 \\
AMEM-HAR                                & 0.0145                  &  & G-AMEM-HAR                        & 0.0203                        &  & MEM-HAR-J                  & 0.0000 \\
MEM-HAR-J                               & 0.0145                  &  & HAR                               & 0.0717                        &  & AMEM-HAR-J                 & 0.0000 \\
AMEM(2,1)-J                             & 0.0145                  &  & MEM-HAR-J                         & 0.0717                        &  & G-AMEM-HAR                 & 0.0016 \\
AHAR                                    & 0.0488                  &  & MEM-HAR                           & 0.0717                        &  & AMEM-HAR                   & 0.0144 \\
AMEM-HAR-J                              & 0.0488                  &  & {\textbf{AHAR$^*$}}       & {\textbf{0.7437$^*$}} &  & MEM-HAR                    & 0.0211 \\
G-AMEM-HAR                              & 0.0488                  &  & {\textbf{ AMEM-HAR-J$^*$}} & {\textbf{0.7437$^*$}} &  & AHAR                       & 0.0211 \\
{\textbf{G-AMEM-HAR-J$^*$}} & {$\mathbf{1.0000^*}$ } &  & \multicolumn{1}{l}{{\textbf{AMEM-HAR$^*$}}}               & $\mathbf{1.0000^*}$& & { \textbf{HAR$^*$}} & $\mathbf{1.0000^*}$ \\
		\bottomrule
	\end{tabular}
	\end{adjustbox}
		\caption{Forecast analysis for S\&P 500. Model Confidence Set for the out-of-sample forecasting performance (for 1, 5 and 22 step-ahead). Significance level 10\% (best set of models in bold and identified by an asterisk). Loss function: QLike. Estimation period: January 3, 2005  - October 15, 2015. Out-of-sample period: October 16, 2015 - October 15, 2021.\label{tab:mcs} }
\end{table}

 Table \ref{tab:mcs} summarizes the results of the MCS procedure (best set of models in bold and identified by an asterisk).  Our results underscore the superior forecasting performance of models incorporating jump dynamics, particularly at the short-term (1-step-ahead) horizon. Specifically, for the S\&P 500, the G-AMEM-HAR-J emerges as the unique best-performing model for the 1-step-ahead horizon, with no other specifications included in the MCS. However, as the forecasting horizon extends to the medium (5-step) and long term (22-step), the importance of the jump component diminishes. At these horizons, the AMEM-HAR and the linear HAR emerge as the top-performing models, respectively. As the forecasting horizon increases, the exclusion of jump models from the superior set of models highlights a shift in dynamics: jumps are critical for identifying `flash' shocks, but less relevant for long-term prediction because illiquidity eventually reverts to its historical average. 

\section{Backtesting Illiquidity-at-Risk}\label{sec:illiquidity_determinats}

In this section, we assess the ability of the considered models to provide an accurate characterization of illiquidity risk. This assessment centers on the models' capacity to provide adequate coverage for the probability of extreme illiquidity spikes as measured by the Illiquidity-at-Risk (IlliQaR) metric defined in Section \ref{sec:IlliQaR}.
\begin{figure}[h!]
		\includegraphics[height=9cm,width=16cm]{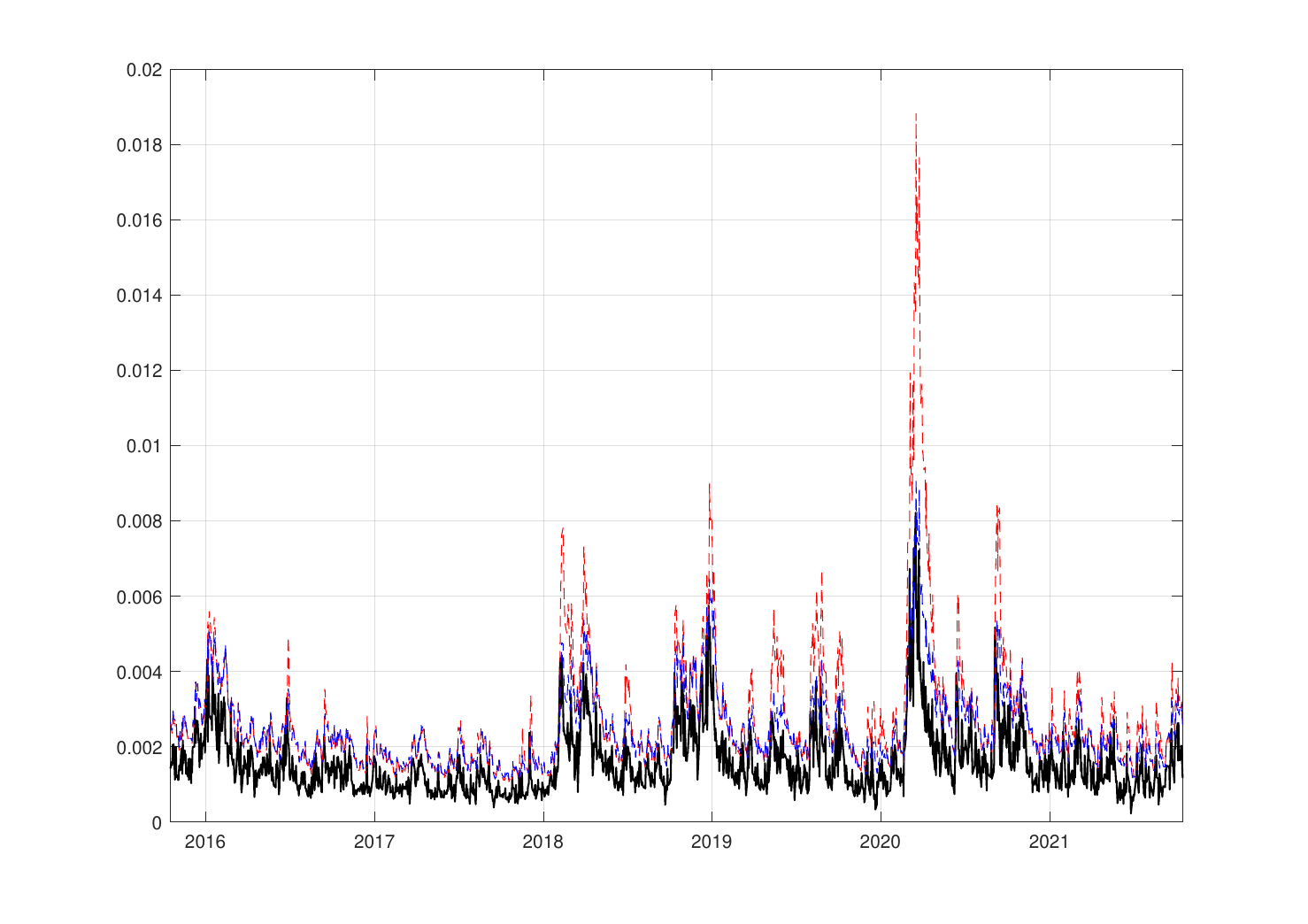}
	\caption{Realized Amihud series (black solid line) and 95\% out-of-sample IlliQaR forecasts from the G-AMEM-HAR-J (red dashed line) and G-AMEM-HAR (blue dashed line) models. \label{fig:IaR}}
\end{figure}

Figure \ref{fig:IaR} illustrates the out-of-sample 5\% IlliQaR for two competing specifications. The first specification, the G-AMEM-HAR-J model (red dashed line), explicitly accounts for discontinuous dynamics through the inclusion of jump components; the expression for the IlliQaR in this framework is derived from Proposition \ref{theo1}. The second is the G-AMEM-HAR baseline specification (blue dashed line), which relies on a standard Gamma MEM framework with the corresponding IlliQaR expression provided in \eqref{eq:illiqar_mem}. While baseline IlliQaR trajectories for both specifications are similar during calm periods, the jump-augmented model is significantly more responsive to new information. Consequently, the G-AMEM-HAR-J specification assigns higher probabilities to tail events, especially during market turmoil, with high illiquidity and increased systemic risk.

\begin{table}[h!]
    \centering
    \small
    \renewcommand{\arraystretch}{1.2}
    \begin{tabular}{lcccclccc}
        \toprule
        & \multicolumn{3}{c}{\textbf{No-Jump Models}} & & & \multicolumn{3}{c}{\textbf{Jump Models}} \\
        \cmidrule(lr){2-4} \cmidrule(lr){7-9}
        & 1\% & 5\% & 10\% & & & 1\% & 5\% & 10\% \\
        \midrule
        \multicolumn{9}{c}{\textit{Panel A: Full Sample}} \\
        \midrule
        HAR         & 0.0000 & 0.0000 & 0.0000 & & MEMJ         & 0.1826 & 0.4070 & 0.2392 \\
        AHAR        & 0.0000 & 0.0000 & 0.0000 & & AMEM-J        & 0.4627 & 0.0073 & 0.7580 \\
        MEM         & 0.0000 & 0.0000 & 0.0000 & & AMEM(2,1)-J  & 0.4708 & 0.0017 & 0.4868 \\
        AMEM        & 0.0000 & 0.0000 & 0.0000 & & G-AMEM-HAR-J & 0.3896 & 0.0183 & 0.4171 \\
        MEM-HAR     & 0.0000 & 0.0000 & 0.0000 & & MEMJ HAR     & 0.1133 & 0.0113 & 0.0494 \\
        AMEM-HAR    & 0.0000 & 0.0000 & 0.0000 & & MEMJ AHAR    & 0.0406 & 0.0038 & 0.1455 \\
        G-AMEM-HAR  & 0.0000 & 0.0000 & 0.0000 & &              &        &        &        \\
        \midrule
        \multicolumn{9}{c}{\textit{Panel B: Out-of-Sample}} \\
        \midrule
        HAR         & 0.0000 & 0.0000 & 0.0000 & & MEMJ         & 0.0095 & 0.0010 & 0.0012 \\
        AHAR        & 0.0000 & 0.0000 & 0.0000 & & AMEM-J        & 0.0155 & 0.0001 & 0.0005 \\
        MEM         & 0.0462 & 0.0001 & 0.0000 & & AMEM(2,1)-J  & 0.0160 & 0.0001 & 0.0008 \\
        AMEM        & 0.0118 & 0.0000 & 0.0000 & & G-AMEM-HAR-J & 0.0170 & 0.0012 & 0.0111 \\
        MEM-HAR     & 0.0033 & 0.0008 & 0.0000 & & MEMJ HAR     & 0.0090 & 0.0033 & 0.0073 \\
        AMEM-HAR    & 0.0013 & 0.0000 & 0.0000 & & MEMJ AHAR    & 0.0031 & 0.0025 & 0.0199 \\
        G-AMEM-HAR  & 0.0004 & 0.0000 & 0.0000 & &              &        &        &        \\
        \bottomrule
    \end{tabular}
    \caption{Table reports the p-values of the Berkowitz test for the S\&P 500 full sample and out-of-sample IlliQaR($p$) computed at $p=1\%, 5\%, 10\%$.}
    \label{tab:berkowitz_full_out}
\end{table}

The top panel of Table \ref{tab:berkowitz_full_out} reports the $p$-value of the Berkowitz test for the full-sample IlliQaR computed at the 1\%, 5\% and 10\% of the Gaussian distribution. The results reveal a clear hierarchy: regardless of the specification chosen for the conditional mean $\mu_t$, jump-augmented models consistently outperform their continuous-path counterparts. Across nearly all quantiles, jump-augmented specifications successfully characterize upper-tail dynamics, yielding $p$-values that consistently exceed the 0.10 significance threshold (marginal exceptions are limited to the 5\% quantile). We interpret this as robust evidence that realized illiquidity is driven by a stochastic process with a significant jump component; failing to account for these discontinuities leads to a systematic mispricing of tail risk.

These limitations are further corroborated by the out-of-sample results presented in the bottom panel of Table \ref{tab:berkowitz_full_out}. While no-jump models are universally rejected across most specifications, largely due to their inability to account for sudden and discontinuous liquidity evaporation, the jump-augmented specifications maintain significantly higher $p$-values, particularly at the 1\% level. This disparity confirms that modeling discrete shocks within the MEM-J framework is essential for achieving accurate probability coverage. By explicitly identifying and incorporating these jump components, the model successfully captures the non-linear dynamics and heavy-tailed distributions characteristic of extreme market stress. However, it is important to note that the superior performance of jump-augmented models is conditional upon the precision of the underlying illiquidity metric. While the MEM-J framework is theoretically superior for IlliQaR forecasting, its predictive gains are highly sensitive to the granularity of the input data; without precise measurement, the identification of genuine liquidity shocks is easily confounded by noise. 
\begin{table}[h!] 
    \centering
    \small
    \renewcommand{\arraystretch}{1.2}
    \begin{tabular}{lcccclccc}
        \toprule
        & \multicolumn{3}{c}{\textbf{Full Sample}} & & & \multicolumn{3}{c}{\textbf{Out-of-Sample}} \\
        \cmidrule(lr){2-4} \cmidrule(lr){7-9}
        & 1\% & 5\% & 10\% & & & 1\% & 5\% & 10\% \\
        \midrule
        AHAR  & 0.000 & 0.000 & 0.000 & & AHAR  & 0.000 & 0.000 & 0.000 \\
        AMEM  & 0.000 & 0.000 & 0.000 & & AMEM  & 0.000 & 0.000 & 0.000 \\
        JUMPS & 0.000 & 0.000 & 0.000 & & JUMPS & 0.000 & 0.000 & 0.000 \\
        \bottomrule
    \end{tabular}
    \caption{Table reports the p-values of the Berkowitz test for the full sample and out-of-sample IlliQaR($p$) for $p=1\%, 5\%, 10\%$ for the daily Amihud. JUMPS denotes the G-AMEM-HAR-J model. \label{tab:berkowitz_daily}}
\end{table}

The necessity of a precise illiquidity measurement for analyzing tail risk is further reinforced by a comparison with the daily Amihud measure, which proves to be a significantly less efficient proxy for daily illiquidity than the realized Amihud. As reported in Table \ref{tab:berkowitz_daily}, the failure to reject the null hypothesis of the Berkowitz test suggests that the inherent noise in low-frequency measures effectively masks the true tail behavior of the process. Consequently, the daily Amihud is rendered unsuitable for providing reliable risk coverage, as it lacks the granularity required to distinguish between transitory fluctuations and genuine liquidity shocks.

\subsection{Price Jumps}

The results displayed in Tables \ref{tab:berkowitz_full_out} and \ref{tab:berkowitz_daily} point to an improvement in the quality of the IlliQaR coverage when illiquidity jumps are accounted for in the modeling, while simultaneously measuring illiquidity very precisely using the realized Amihud proxy. However, even accounting for liquidity jumps and precisely measuring illiquidity is not completely sufficient to achieve perfect coverage of extreme illiquidity tail events. As shown in Table \ref{tab:berkowitz_full_out}, during the out-of-sample period, even jump models frequently fail to provide correct coverage of the tails. 

We argue that this is due to the potential interference of price jumps in the measurement of illiquidity. Specifically, when major public news hits the market (e.g., central bank announcements or earnings releases), reservation prices across all traders shift in the same direction to establish a new equilibrium. This common informational shock generates significant price volatility with minimal corresponding trading volume, as volume is primarily driven by trader disagreement (the investor-specific diffusive component), see among others \cite{boudt2014intraday} or \cite{scaillet2020high} in the context of the Bitcoin market. Consequently, standard realized volatility (RV) aggregates both the continuous diffusive component and discrete price jumps, causing artificial spikes in the realized Amihud measure that do not reflect baseline market illiquidity.

To isolate the pure diffusive component that genuinely drives trading volume and to assess how price jumps affect the adequacy of the \textsf{IlliQaR} framework, we replace RV with jump-robust proxies. Specifically, we replicate the analysis using the bipower variation \citep[BPV,][]{barndorff2004power} to explicitly purge the impact of information jumps (price jumps) on illiquidity measurement.
Since BPV is robust to price jumps by construction, the ratio $\text{Illiq}^C=BPV/\nu$ isolates the continuous component of illiquidity, filtering out the contribution of discontinuous price movements to the realized price impact. This robustness check allows us to disentangle two distinct sources of tail risk: genuine illiquidity jumps driven by trading frictions and the sudden withdrawal of liquidity providers, and price discontinuities that mechanically inflate the realized Amihud through the volatility numerator.
\begin{table}[h!]
    \centering
    \small
    \renewcommand{\arraystretch}{1.2}
    \begin{tabular}{lcccclccc}
        \toprule
        & \multicolumn{3}{c}{\textbf{No-Jump Models}} & & & \multicolumn{3}{c}{\textbf{Jump Models}} \\
        \cmidrule(lr){2-4} \cmidrule(lr){7-9}
        & 1\% & 5\% & 10\% & & & 1\% & 5\% & 10\% \\
        \midrule
        \multicolumn{9}{c}{\textit{Panel A: Full Sample}} \\
        \midrule
HAR        & 0.0000 & 0.0000 & 0.0000 &  & MEM-J            & 0.4473 & 0.3585 & 0.9001 \\
AHAR       & 0.0000 & 0.0000 & 0.0000 &  & AMEM-J      & 0.2670 & 0.2158 & 0.7771 \\
MEM        & 0.0000 & 0.0000 & 0.0000 &  & AMEM(2,1)-J      & 0.4659 & 0.5577 & 0.6699 \\
AMEM       & 0.0000 & 0.0000 & 0.0000 &  & G-AMEM-HAR-J & 0.6554 & 0.4369 & 0.4151 \\
MEM-HAR    & 0.0000 & 0.0000 & 0.0000 &  & MEM-HAR-J        & 0.1530 & 0.3582 & 0.1490 \\
AMEM-HAR   & 0.0000 & 0.0000 & 0.0000 &  & AMEM-HAR-J       & 0.5359 & 0.3383 & 0.1342 \\
G-AMEM-HAR & 0.0000 & 0.0000 & 0.0000 &  &                  &        &        & \\
        \midrule
        \multicolumn{9}{c}{\textit{Panel B: Out-of-Sample}} \\
        \midrule
HAR        & 0.0000 & 0.0000 & 0.0000 &  & MEM-J            & 0.2303 & 0.0055 & 0.8925 \\
AHAR       & 0.0000 & 0.0000 & 0.0000 &  & AMEM-J      & 0.1392 & 0.0692 & 0.6029 \\
MEM        & 0.0000 & 0.0000 & 0.0000 &  & AMEM(2,1)-J      & 0.1400 & 0.0794 & 0.4231 \\
AMEM       & 0.0000 & 0.0000 & 0.0000 &  & G-AMEM-HAR-J & 0.2360 & 0.0480 & 0.5584 \\
MEM-HAR    & 0.0000 & 0.0000 & 0.0000 &  & MEM-HAR-J        & 0.3583 & 0.0190 & 0.2455 \\
AMEM-HAR   & 0.0000 & 0.0000 & 0.0000 &  & AMEM-HAR-J       & 0.1402 & 0.1171 & 0.4409 \\
G-AMEM-HAR & 0.0000 & 0.0000 & 0.0000 &  &                  &        &        &  \\
        \bottomrule
    \end{tabular}
    \caption{Table reports the p-values of the Berkowitz test for the S\&P 500 full sample and out-of-sample IlliQaR($p$) computed at $p=1\%, 5\%, 10\%$, based on the jump robust realized Amihud (Illiq$_t^C$).}
    \label{tab:berkowitz_full_out_bv}
\end{table}
Table \ref{tab:berkowitz_full_out_bv} reports the $p$-values of the Berkowitz test\footnote{Due to space constraints, the results for model estimation, forecast evaluation, and the determinants of Illiquidity at Risk are available upon request.} for the jump-robust IlliQaR computed at the 1\%, 5\%, and 10\% quantiles for the S\&P 500, for both the full sample and the out-of-sample period. The results confirm the same qualitative hierarchy documented in Table \ref{tab:berkowitz_full_out}: models without a jump component are systematically rejected across all quantiles, while jump-augmented specifications provide adequate coverage. Notably, the $p$-values obtained under the jump-robust measure are generally higher than those obtained with the realized Amihud,  in both the full sample and the out-of-sample period. For instance, at the 1\% quantile, the MEM-J $p$-value rises from 0.183 to 0.447, and the AMEM-HAR-J from 0.041 to 0.536. This improvement suggests that discontinuous price movements introduce additional dispersion into the tail of the illiquidity distribution that the MEM-J framework only partially accommodates. When this source of variation is filtered out at the measurement stage, the model's task of accurately characterizing extreme illiquidity events becomes marginally easier. This improvement is also confirmed in the out-of-sample analysis, where a $p$-value lower than 1\% is detected only for the MEM-J at the 5\% quantile.

\subsection{Illiquidity at Risk and Its Determinants}\label{sec:determinants}

Finally, we investigate the economic determinants of IlliQaR violations to identify the factors driving extreme illiquidity episodes. We consider a Logit model for the binary dependent variable, $I_t$, which equals 1 if $\text{Illiq}_t>\text{IlliQaR}_t(p)$ and 0 otherwise. Building on our finding that illiquidity dynamics are closely linked to market uncertainty, we consider the following logit specification
\begin{equation}Pr(\text{I}_t=1 | \mathbf{x}_{t-1}) = \frac{\exp(\delta_0 + \mathbf{x}_{t-1}'\delta)}{1 + \exp(\delta_0 + \mathbf{x}_{t-1}'\delta)},
\label{eq:logit}
\end{equation}
where $\delta$ is a vector of coefficients, and $\mathbf{x}_{t-1}$ is a vector including several macro-financial variables, which we use to explore the determinants of tail liquidity risk. The Cboe Volatility Index (VIX) serves as a proxy for forward-looking market sentiment and implied volatility. The Economic Policy Uncertainty (EPU) index of \cite{Baker:Bloom:Davis:2016} accounts for broader macroeconomic and political risk. The TED spread is a proxy for systemic stress within the interbank lending market, specifically to account for fluctuations in funding liquidity. Furthermore, we introduce a binary indicator, $D$, which equals 1 for negative returns. This allows the model to isolate the asymmetric sensitivity of illiquidity to downward price innovations, a phenomenon often observed during market panics.

Following the methodology of \cite{Bekaert2013}, we  also decompose the VIX index into two distinct components: market uncertainty and risk aversion. As a preliminary step, we estimate a linear model by regressing S\&P 500 realized volatility ($\text{RV}_t$) on its lagged values and the lagged VIX
\begin{equation*}
\text{RV}_t = \beta_0 + \beta_1 \text{RV}_{t-1} + \beta_2 \text{VIX}_{t-1} + \epsilon_t.\end{equation*}
The fitted values from this regression serve as our proxy for uncertainty, defined as $\text{Uncertainty}_t := \widehat{E(\text{RV}_t | \mathcal{F}_{t-1})}$, leading to
$$\text{Uncertainty}_t = \hat{\beta}_0 + \hat{\beta}_1 \text{RV}_{t-1} + \hat{\beta}_2 \text{VIX}_{t-1}.$$
The residual difference between the squared VIX and this uncertainty measure represents our proxy for risk aversion
$$\text{Risk Aversion}_t = \text{VIX}^2_t - \text{Uncertainty}_t.$$
We then replace the VIX with these two components, resulting in the following logit specification
\begin{equation}\text{Pr}(\text{I}_t=1 | \mathbf{x}_{t-1}) = \frac{\exp(\delta_0 + \delta_1 \text{Uncertainty}_{t-1} + \delta_2 \text{Risk Aversion}_{t-1} + \delta_3 \text{EPU}_{t-1} + \delta_4 \text{TED}_{t-1} + \delta_5 D_{t-1})}{1 + \exp(\delta_0 + \delta_1 \text{Unc}_{t-1} + \delta_2 \text{Risk Aversion}_{t-1} + \delta_3 \text{EPU}_{t-1} + \delta_4 \text{TED}_{t-1} + \delta_5 D_{t-1})}, \label{eq:logit_2}\end{equation}

Panel (a) of Table \ref{tab:logit_horiz} reports the estimation results for the full sample (columns 1–2) and the out-of-sample analysis (columns 3–4). In the full sample, the only significant predictor is the dummy variable ($D_{t-1}$), with a marginal effect of 0.016. This suggests that the probability of IlliQaR violations increases following negative returns. During the out-of-sample period, macro-financial risk factors, specifically the VIX, EPU, and TED spread, play a significant role in predicting unexpected illiquidity bursts (the LR test suggest a joint significant role of these factors). This shift may be attributed to the 2015–2021 period, which was characterized by increased political uncertainty and potential shifts in market structure and news reactivity. Recognizing that market participants may react to information without delay, Panel (b) presents results using contemporaneous rather than lagged regressors. In this specification, the dummy variable remains significant, with a marginal effect of approximately 0.05. Furthermore, the VIX becomes highly significant, particularly regarding its association with IlliQaR violations in the out-of-sample period. The decomposition of the VIX reveals a notable distinction: while the uncertainty coefficient is negative and large in magnitude, risk aversion is positive and significant across both the full sample and out-of-sample analyses. This provides a valuable insight for investors, suggesting they can better anticipate peaks in tail risk exceeding the threshold IlliQaR by monitoring specific observable signals, namely risk aversion, rather than the aggregate VIX alone. The TED spread is negatively and significantly associated with IlliQaR violations. 
\begin{table}[h!]
	\centering
    \setlength{\tabcolsep}{3pt}
\renewcommand{\arraystretch}{1.1}
	\begin{adjustbox}{max width=\linewidth,center}
	\begin{tabular}{lccccccccc}
			\toprule
			& \multicolumn{4}{c}{Panel a) - Lagged Regressors} & & \multicolumn{4}{c}{Panel b) - Contemporaneous Regressors} \\
			\cmidrule{2-5} \cmidrule{7-10}
			& \multicolumn{2}{c}{Full sample} & \multicolumn{2}{c}{Out of sample} & & \multicolumn{2}{c}{Full sample} & \multicolumn{2}{c}{Out of sample} \\
			\cmidrule{2-3} \cmidrule{4-5} \cmidrule{7-8} \cmidrule{9-10}
			 & \multicolumn{4}{c}{Parameter Estimates} & & \multicolumn{4}{c}{Parameter Estimates} \\
			\midrule
			Constant & $-3.2479^a$ & $-3.0703^a$ & $-4.1858^a$ & $-3.6973^a$ & & $-4.0424^a$ & $-3.1838^a$ & $-5.0866^a$ & $-3.1700^a$ \\
			         & (0.1501)    & (0.2506)    & (0.2822)    & (0.3223)    & & (0.1729)    & (0.2774)    & (0.4397)    & (0.5902)    \\
			VIX    & $0.0042$    &      ----       & $0.0654^a$  &       ----      &  & $0.0131^b$  &     ----        & $0.1116^a$  &      ----       \\
			         & (0.0067)    &   ----          & (0.0182)    &      ----       &  & (0.0065)    &      ----       & (0.0175)    &     ----        \\
			         & $[0.0002]$  &   ----          & $[0.0021]$  &   ----          & & $[0.0005]$  &     ----        & $[0.0015]$  &     ----        \\
			Uncert.  &             & $-28.4122$  & ----            & $46.4553$   & &   ----          & $-152.9560^a$ &    ----        & $-99.4808$  \\
			         &       ----      & (47.2839)   &  ----           & (54.3246)   &   &    ----         & (52.0721)   &   ----          & (89.7807)   \\
			         &      ----       & $[-1.3664]$ &   ----          & $[1.5490]$  &  &  ----            & $[-6.1645]$ &          ----   & $[-1.3724]$ \\
			Risk Av. &     ----        & $0.0001$    & ----            & $0.0001$    &  &   ----          & $0.0012^a$  &       ----      & $0.0021^a$  \\
			         &       ----      & (0.0004)    & ----            & (0.0004)    &  & ----            & (0.0004)    &  ----           & (0.0008)    \\
			         &      ----       & $[0.0000]$  & ----            & $[0.0000]$  & &    ----         & $[0.0001]$  &     ----        & $[0.0000]$  \\
			EPU    & $0.0008$    & $0.0013^c$  & $-0.0037^b$ & $-0.0015$   & & $0.0013^c$  & $0.0022^a$  & $-0.0048^b$ & $-0.0029$   \\
			         & (0.0007)    & (0.0007)    & (0.0018)    & (0.0015)    & & (0.0008)    & (0.0007)    & (0.0021)    & (0.0020)    \\
			         & $[0.0000]$  & $[0.0001]$  & $[-0.0001]$ & $[-0.0001]$ & & $[0.0001]$  & $[0.0001]$  & $[-0.0001]$ & $[0.0000]$  \\
			TED    & $0.0023$    & $0.0026$    & $0.0433^b$  & $0.0420^b$  & & $0.0024$    & $0.0025$    & $-2.9462^a$ & $-3.4556^a$ \\
			         & (0.0149)    & (0.0137)    & (0.0184)    & (0.0193)    & & (0.0170)    & (0.0140)    & (0.7220)    & (0.8579)    \\
			         & $[0.0001]$  & $[0.0001]$  & $[0.0014]$  & $[0.0014]$  & & $[0.0001]$  & $[0.0001]$  & $[-0.0391]$ & $[-0.0477]$ \\
			D      & $0.3220^b$  & $0.3153^b$  & $0.1928$    & $0.3487$    & & $1.1996^a$  & $1.1104^a$  & $1.5621^a$  & $1.5442^a$  \\
			         & (0.1402)    & (0.1426)    & (0.2854)    & (0.2799)    & & (0.1560)    & (0.1551)    & (0.3710)    & (0.3783)    \\
			         & $[0.0158]$  & $[0.0154]$  & $[0.0062]$  & $[0.0119]$  & & $[0.0544]$  & $[0.0487]$  & $[0.0245]$  & $[0.0251]$  \\
			\midrule
			 & \multicolumn{4}{c}{Model Fit} & & \multicolumn{4}{c}{Model Fit} \\
			\midrule
			Success rate & 94.80\% & 94.80\% & 96.20\% & 96.30\% & & 94.80\% & 94.90\% & 96.40\% & 96.50\% \\
			LR p-value   & 0.1062  & 0.1389  & 0.0006  & 0.0507  & & 0.0000  & 0.0000  & 0.0000  & 0.0000  \\
			\bottomrule			
	\end{tabular}
	\end{adjustbox}
	\caption{Logit model. The table displays logit estimation results for IlliQaR violations. Panel (a) reports coefficients for lagged regressors, while Panel (b) focuses on contemporaneous specifications. Robust standard errors in parenthesis. Average marginal effects in square brackets. The superscripts a, b and c denote significance at 1\%, 5\% and 10\% levels, respectively. Success rate denotes the percentage of correctly classified observations. LR p-value refers to the p-value of the likelihood ratio test for the joint significance of the regressors, where, under the null hypothesis, the model is $\text{Pr}(I_t=1 | \mathbf{x}_{t-1}) = \frac{\exp(\delta_0) }{1 + \exp(\delta_0)}$.\label{tab:logit_horiz}}
\end{table}

Finally, Table \ref*{tab:logit_ma_quantile} in the Supplementary Document reports results from the logit model where the explanatory variables (VIX, Uncertainty, Risk Aversion, EPU, and TED) are defined as deviations from their 22-day moving averages, restricted to observations above the 95th percentile (i.e., focusing on tail events within the regressors also). Intuitively, periods of elevated volatility and macroeconomic uncertainty are likely associated with higher margin requirements, which, in turn, can impair market liquidity. These results largely confirm the findings from Table \ref{tab:logit_horiz}, with one notable exception: once we isolate extreme realizations, volatility uncertainty, which captures the predictable component of the VIX, is no longer significant in the contemporaneous regression. In contrast, the risk aversion component remains highly significant. This suggests that IlliQaR violations are not driven by high volatility per se, but specifically by unexpected spikes in risk aversion. This indicates that market liquidity is most vulnerable when volatility realizations reflect a sudden shift in investor sentiment rather than just heightened fundamental uncertainty.

\section{Individual Stocks}\label{sec:individual}

To assess the robustness of our findings beyond the aggregate market index, we extend the empirical analysis to a cross-section of 25 major individual U.S. equities. This expansion allows us to investigate whether the predictive power of jump dynamics in illiquidity forecasting remains persistent in the presence of idiosyncratic risk, which may interact differently with the market liquidity shocks, as illustrated  by \cite{pastor2003liquidity}. The sample, which covers the years between January 3, 2012 until January 10, 2024 is split into an in-sample period from January 3, 2012 to February 4, 2020 and an out-of-sample period from February 5, 2020 to January 10, 2024.

\subsection{Parameter Estimates}

\begin{landscape}
    \begin{figure}[htbp!]
	\centering
	\subfigure[$\omega$]    {\includegraphics[width=0.35\textwidth,height=5cm]{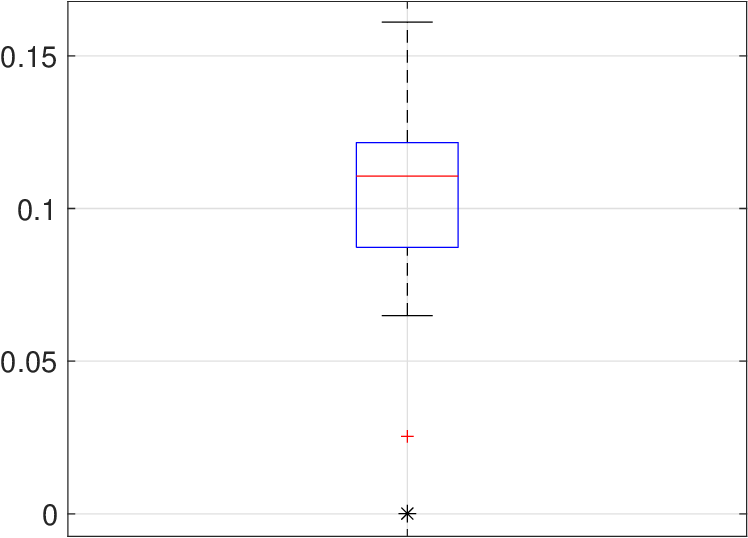}}
	\subfigure[$\alpha_1$]    {\includegraphics[width=0.35\textwidth,height=5cm]{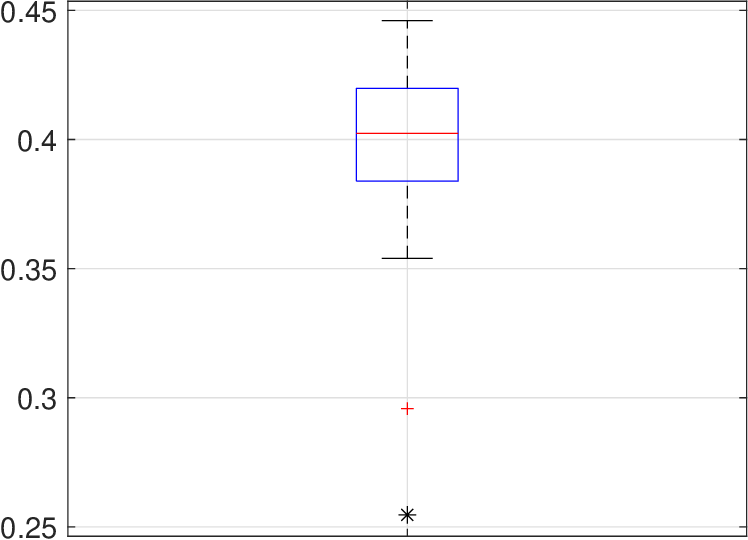}}
    \subfigure[$\alpha_2$]    {\includegraphics[width=0.35\textwidth,height=5cm]{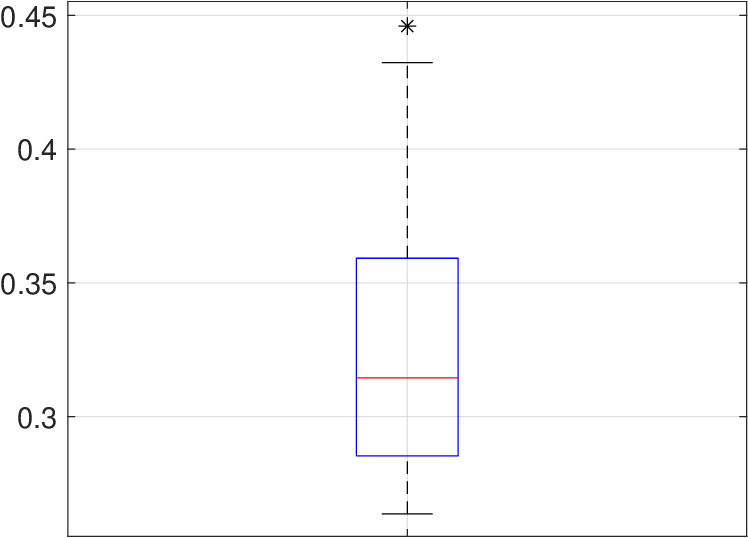}}
	\subfigure[$\alpha_3$]    {\includegraphics[width=0.35\textwidth,height=5cm]{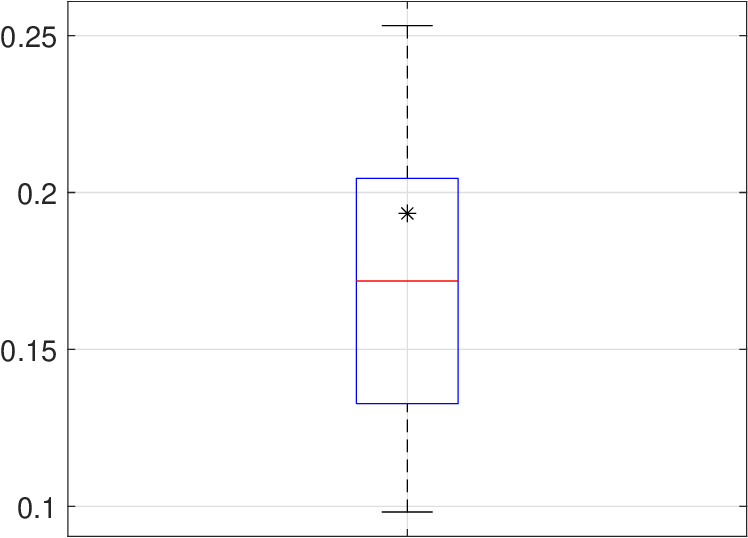}}\\
   	\subfigure[$\gamma$]    {\includegraphics[width=0.35\textwidth,height=5cm]{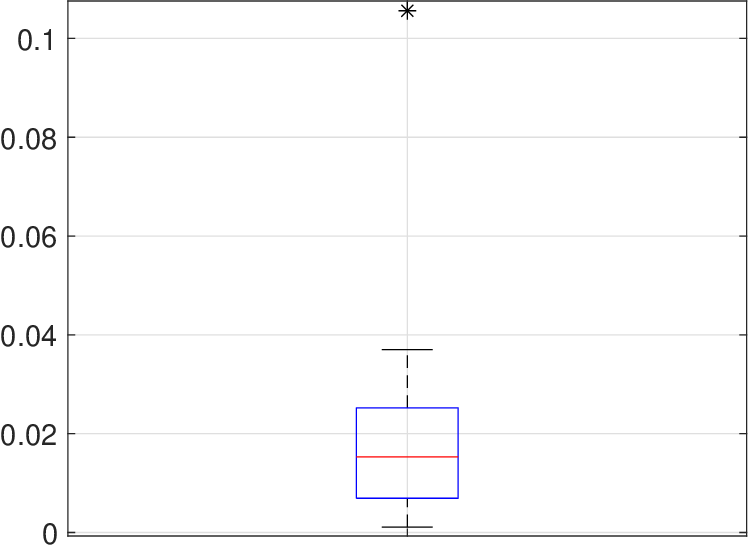}}
	\subfigure[$\vartheta$]    {\includegraphics[width=0.35\textwidth,height=5cm]{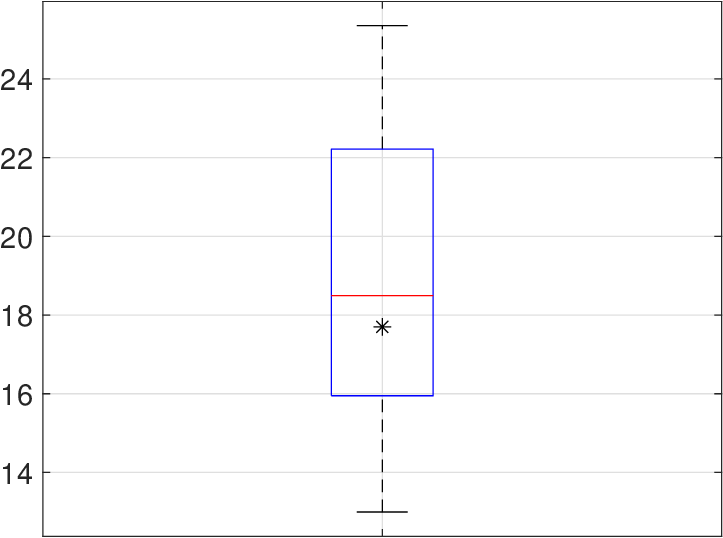}}
    \subfigure[$\zeta$]    {\includegraphics[width=0.35\textwidth,height=5cm]{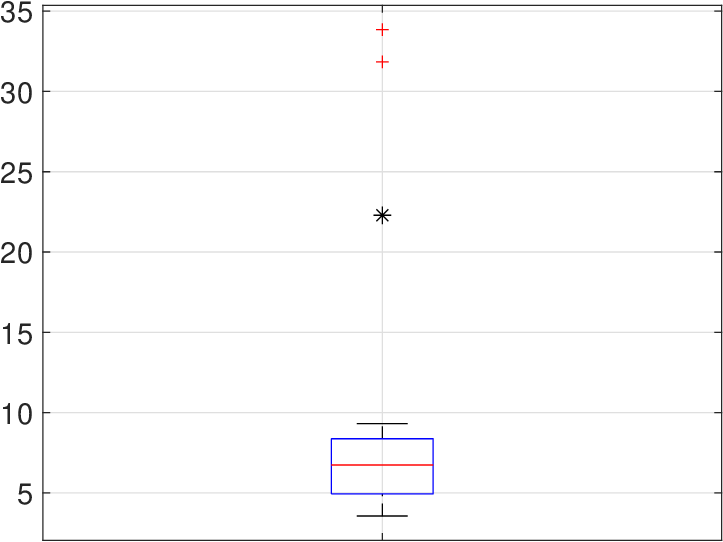}}
	\subfigure[$\kappa$]    {\includegraphics[width=0.35\textwidth,height=5cm]{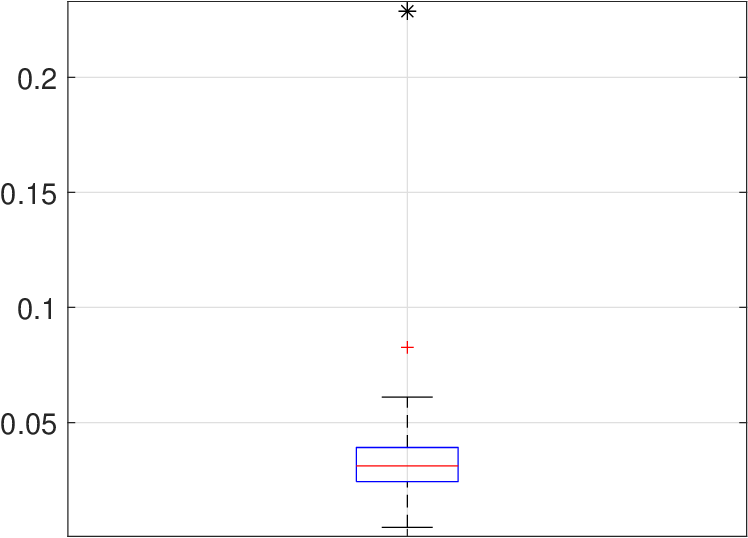}}
    
	\caption{Cross-sectional distribution of estimated coefficients across the 25 Dow Jones constituent stocks. Panels report the boxplot of each parameter of the AMEM-HAR-J across the 25 stocks. The black asterisk is the parameter estimate obtained on the S\&P 500 index.}\label{fig:boxplot_stocks}
\end{figure}
\end{landscape}

Figure \ref{fig:boxplot_stocks} reports the box plot of the cross-sectional (full-sample) parameter estimates of the AMEM-HAR-J model for the 25 stocks under investigation.\footnote{Due to space constraints, we report the complete estimation results in the Supplementary Document. Specifically, Section \ref*{app:appendix_fullsample_individual_stocks} of the Supplementary Document provides parameter estimates for three representative specifications: the AHAR, AMEM-HAR, and AMEM-HAR-J models, considering a constant jump intensity specification ($\kappa_t=\kappa$). Furthermore, MCS results are reported in Tables \ref*{tab:mcs_1_individual}--\ref*{tab:mcs_22_individual} (Section \ref*{app:individual}), while the results of the Berkowitz test are in Sections \ref*{App:berkowitz_individual_full} and \ref*{App:berkowitz_individual_out} of the Supplementary Document.}

Across all 25 equities, the intercept $\omega$ ranges between 0.07 and 0.015, much larger values than those of the S\&P 500, signaling that the average illiquidity levels of individual stocks are significantly higher than that of the overall US market. As for the HAR components ($\alpha_d$, $\alpha_w$, and $\alpha_m$), they are generally positive and statistically significant at the 1\% level. This confirms that idiosyncratic illiquidity (much like market illiquidity) is characterized by strong persistence and long-memory behavior, with the sum of the three coefficients close to 1. However, there is a slight structural difference compared with the index: $\alpha_d$ is larger for individual stocks, whereas $\alpha_w$ is smaller. Finally, the asymmetry parameter ($\gamma$), whenever significant, is positive, confirming that the ``leverage effect"—where liquidity deteriorates more severely following negative returns—holds at the individual stock level. Compared with the S\&P 500 index, the strength of this leverage effect appears to be weaker. The distributional parameters governing the continuous process are remarkably stable across the cross-section. Under the AMEM-HAR-J specification, the shape parameter of the Gamma distribution ($\vartheta$) consistently falls between 13 and 25, closely mirroring the dispersion levels observed for the S\&P 500.

Instead, the most notable differences relate to the jump intensity and distribution parameters for individual stocks. Specifically, the jump size distribution parameter $\zeta$ across the 25 equities ranges between 4 and 9, substantially lower than the index-level estimate—while the jump intensity coefficients range between 1\% and 5\%, compared to around 22\% for the index. This signals that liquidity jumps occur less frequently for individual stocks than for the broad market, but they carry the potential to trigger larger liquidity dry-ups. Consequently, while market liquidity jumps reflect sustained, systemic stress episodes, single-stock liquidity evaporation is often driven by transient, idiosyncratic shocks that rapidly dissipate despite generating extreme short-term liquidity dry-ups.

\subsection{Forecasting Analysis}

\begin{figure}[h!]
	\centering
    \subfigure[1-step ahead] 
    {\includegraphics[width=0.49\textwidth]{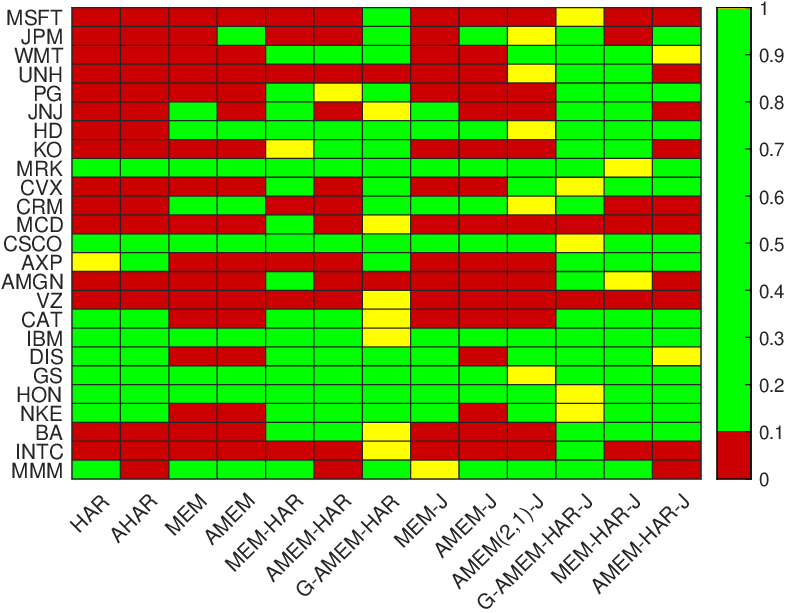}}
	\subfigure[5-step ahead] {\includegraphics[width=0.49\textwidth]{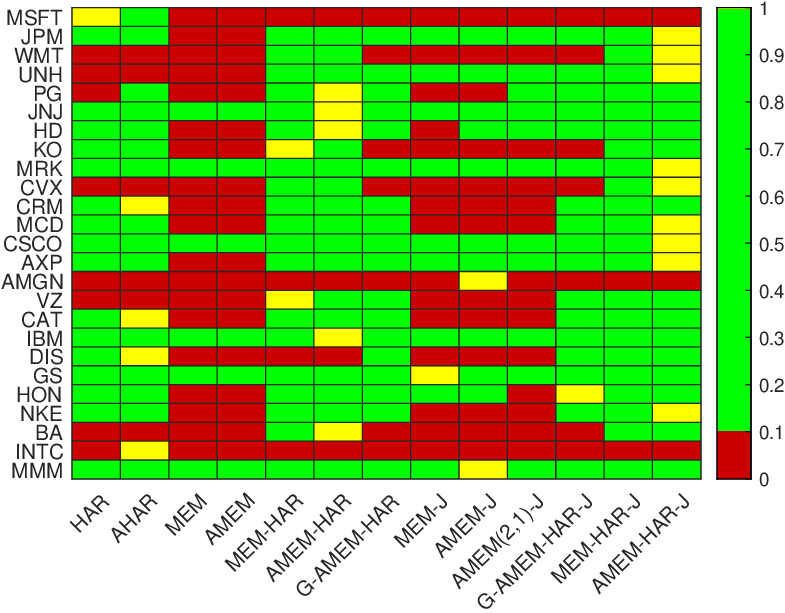}}\\
    \subfigure[22-step ahead] 
    {\includegraphics[width=0.49\textwidth]{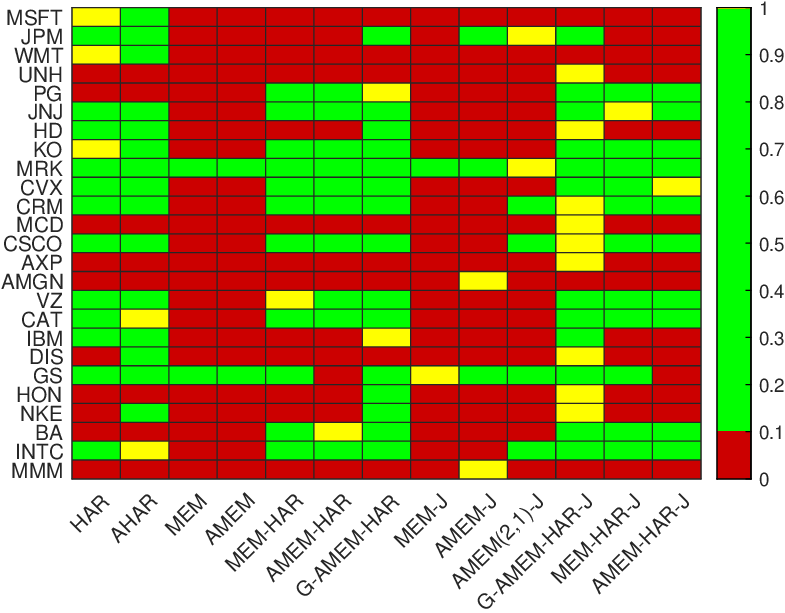}}
	
    \caption{Forecast analysis for Individual stocks. Model Confidence Set for the 1-, 5-, and 22-step ahead out-of-sample
forecasting performance. Significance level 10\%. The best model is highlighted in yellow; models included in the Model Confidence Set are shown in green; models excluded from the Model Confidence Set are shown in red.
Loss function: QLike. Estimation period: January 3, 2012 - February 4, 2020. Out-of-sample period: February
5, 2020 - January 10, 2024. Stocks are presented in descending order of market capitalization.}\label{fig:mcs_stocks}
\end{figure}

The forecasting results for individual stocks largely corroborate the findings for the S\&P 500, albeit with the degree of heterogeneity expected at a disaggregated level. Figure \ref{fig:mcs_stocks} summarizes the MCS analysis for the 25 stocks under consideration.\footnote{For a complete overview of the MCS results, see Tables \ref*{tab:mcs_1_individual}--\ref*{tab:mcs_22_individual} of the Supplementary Document.}

At the short-term horizon ($h=1$) (Panel a), models incorporating jump dynamics consistently populate the MCS. Specifications such as the G-AMEM-HAR-J and the AMEM(2,1)-J are frequently identified as top performers. For instance, for most stocks, the MCS procedure excludes simple linear HAR and standard MEM specifications, favoring instead jump-augmented models. This confirms that for individual assets, daily liquidity is characterized by abrupt, discontinuous changes that require explicit inclusion of jumps to achieve accurate one-step-ahead predictions. Consistent with the index-level analysis, the contribution of the jump component decays as the forecasting horizon extends. At the weekly and monthly horizons ($h=5$ and $22$, Panels b and c), the superior performance of the linear HAR model reaffirms its ability to capture the long-memory behavior of illiquidity through a parsimonious cascade of heterogeneous components. However, across all horizons, the G-AMEM-HAR-J is included in the MCS for most stocks, underscoring the need for accurate modeling of both the long-memory dynamics of illiquidity and its discontinuous, jumpy behavior.

\subsection{IlliQaR Analysis}

\begin{figure}[h!]
	\centering
	\subfigure[1\% - Out-of-sample]    {\includegraphics[width=0.49\textwidth]{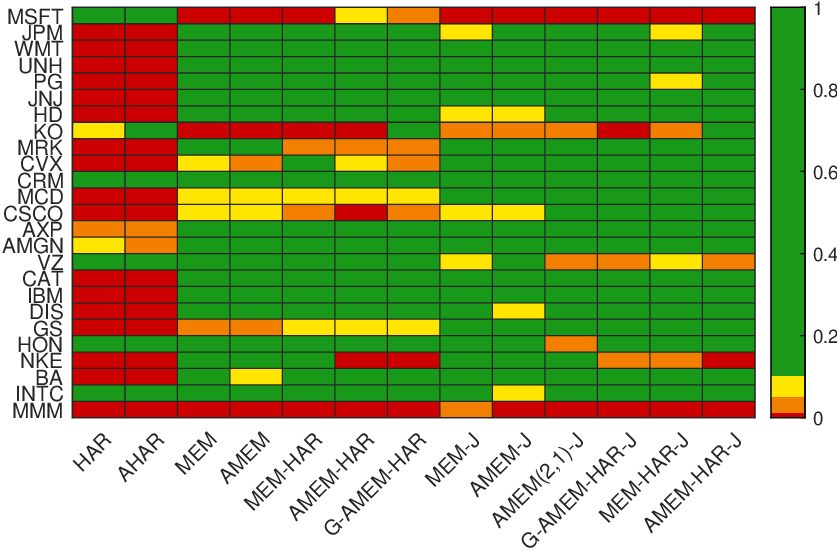}}
	\subfigure[5\% - Out-of-sample]    {\includegraphics[width=0.49\textwidth]{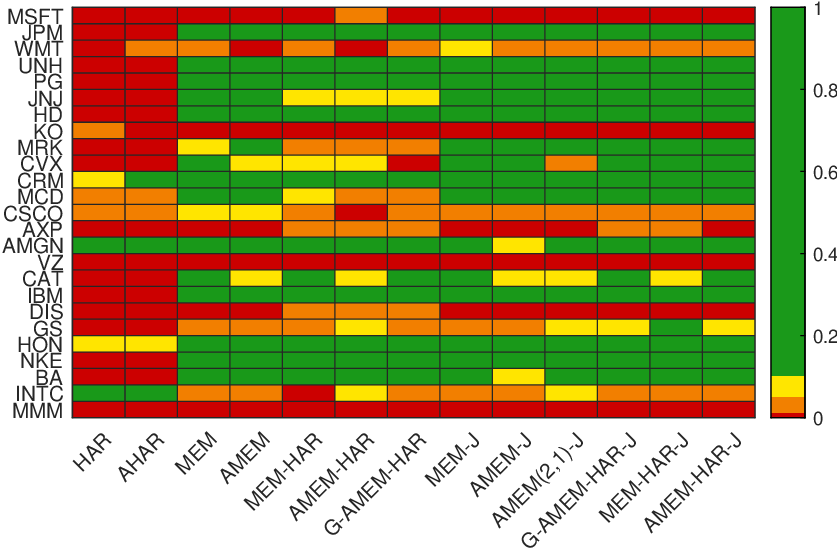}}
    	\subfigure[10\% - Out-of-sample]    {\includegraphics[width=0.49\textwidth]{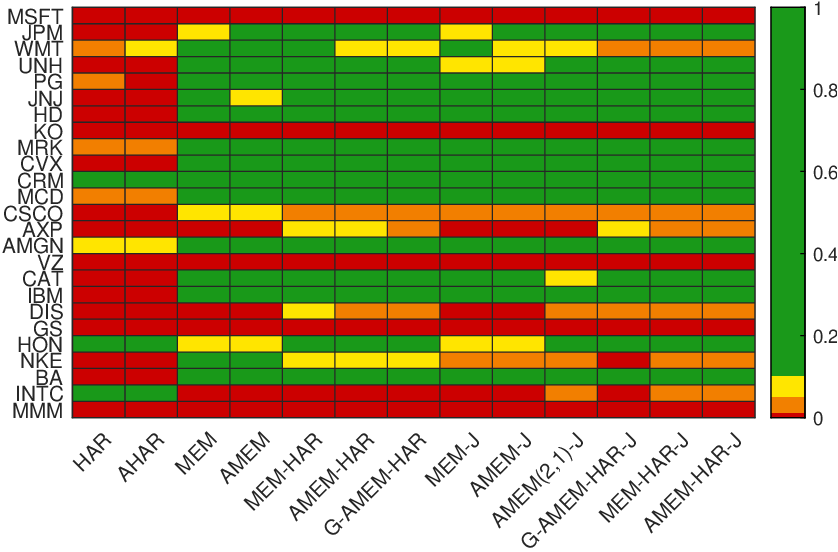}}
   
	\caption{Heatmaps of Berkowitz test p-values for the 1\% IlliQaR across 25 individual U.S. equities. Higher p-values (green) indicate adequate tail coverage; lower p-values (red) indicate rejection of the null hypothesis of correct specification.}\label{fig:pval_berkowitz_stocks_1}
\end{figure}

The necessity of jump models becomes even more apparent when evaluating the IlliQaR metric. Figure \ref{fig:pval_berkowitz_stocks_1} uses a heatmap to summarize the results of the Berkowitz test for the out-of-sample periods across the 25 stocks under analysis.\footnote{See Tables \ref*{tab:density_forecast_full_1_individual}--\ref*{tab:density_forecast_oos_10_individual} in Sections \ref*{App:berkowitz_individual_full} and \ref*{App:berkowitz_individual_out} of the Supplementary Document for the $p$-values of the Berkowitz test for density forecasts.} Standard linear models (HAR and AHAR) and continuous MEM specifications frequently fail to provide adequate coverage for 1\% tail events, with $p$-values often dropping to 0.0000 and leading to a rejection of the null hypothesis of correct specification. This indicates a systematic underestimation of the probability of extreme liquidity dry-ups. In contrast, jump-diffusion specifications (MEM-J class) outperform baseline models in accurately tracking the tail properties of realized illiquidity, particularly during periods of high market stress. For the majority of the analyzed stocks, jump-augmented models fail to reject the null hypothesis. This result is economically significant: it implies that ignoring the jump component in individual stocks exposes investors to a ``tail risk" that is far greater than what standard Gamma-distributed models predict.

In summary, the analysis of individual stocks confirms that IlliQaR is a jump-driven phenomenon. The ``evaporation" of liquidity is not merely an index-level occurrence; it is inherently discontinuous at the single-stock level as well. Consequently, risk management systems relying solely on continuous approximations of illiquidity are likely ill-equipped to handle periods of market stress. Our results suggest that individual stock IlliQaR violations often cluster during periods of S\&P 500 liquidity stress. This indicates that Illiquidity at Risk is a systemic concern rather than a localized one, with the main index acting as a leading indicator for extreme liquidity dry-ups in individual stocks.

\section{Conclusions}
\label{sec:conclusion}

This paper provides a comprehensive investigation into the dynamics of stock market illiquidity and the extent to which their level and tail behavior can be accurately predicted.  Our primary contribution is the introduction of Illiquidity-at-Risk (IlliQaR), a novel risk metric designed to quantify the magnitude of extreme liquidity dry-ups. Our analysis utilizes a refinement of the classic Illiq proxy \citep{Amihud2002illiquidity}, the realized Amihud measure \citep{RanaldoSantucci2020}, which has been demonstrated to provide a precise assessment of illiquidity \citep{lacava2023realized}. 

We employ a number of linear and non-linear econometric specifications designed to capture the defining stylized facts of illiquidity, namely positive skewness,  excess kurtosis and long-range dependence. Furthermore, we evaluate the specific contribution of illiquidity jumps in shaping the distributional properties of the realized Amihud and the resulting IlliQaR estimates. Several key findings emerge from our empirical analysis of the S\&P 500 and 25 individual US stocks. First, we find that illiquidity is a highly persistent process that is more closely associated with market uncertainty than with risk aversion. Second, jumps constitute a significant component of illiquidity risk. Accurately assigning probabilities to extreme realizations of the realized Amihud is of fundamental importance, particularly for risk management and real-time monitoring of market stress. Regarding the forecasting horizon, our results reveal a clear distinction between short-term and long-term dynamics. For one-step-ahead forecasts, the AMEM-J specification—which explicitly accounts for abrupt, discontinuous jumps—emerges as the superior model. Conversely, at medium-to-long-term horizons, illiquidity tends toward its unconditional mean and the forecast trajectories become smoother. In these instances, the parsimonious HAR-type models, which effectively capture the pseudo long-memory properties of the series, provide better predictive performance. Finally, our investigation into the determinants of IlliQaR violations indicates that the probability of observing an extreme illiquidity event is state-dependent. It responds significantly to both negative lagged returns and shifts in market risk aversion, suggesting that investors can anticipate peaks in tail risk by monitoring these observable variables.

\bibliography{Bibliography}

\begin{appendix}

\section{Proof of Proposition \ref{theo1}}\label{app:proof}
The conditional density of $\text{Illiq}_t$ is defined as an infinite mixture:\begin{equation*}f_{\text{MEM-J}}(\text{Illiq}_t|\mathcal{F}_{t-1}) = e^{-\kappa_t} f_G(\text{Illiq}_t|N_t=0,\mathcal{F}_{t-1}) + \sum_{j=1}^\infty \frac{e^{-\kappa_t}\kappa_t^j}{j!} f_K(\text{Illiq}_t|N_t=j,\mathcal{F}_{t-1}),
\end{equation*}
where $f_G$ and $f_K$ are the conditional Gamma (no jumps) and Kappa (jumps occur) densities, respectively. By interchanging the integral and summation, the conditional CDF simplifies to
\begin{equation}
F_{\text{MEM-J}}(\text{Illiq}_t|\mathcal{F}_{t-1}) = e^{-\kappa_t} F_G(\text{Illiq}_t|N_t=0,\mathcal{F}_{t-1}) + \sum_{j=1}^\infty \frac{e^{-\kappa_t}\kappa_t^j}{j!} F_K(\text{Illiq}_t|N_t=j,\mathcal{F}_{t-1}),
\end{equation}
where $F_G(\cdot)$ and $F_K(\cdot)$ denote the respective Gamma and Kappa CDFs. In particular,
\[
F_G(\text{Illiq}_t|N_t=0,\mathcal{F}_{t-1})=\frac{1}{\Gamma(\vartheta)}\int_{0}^{\frac{\vartheta \text{Illiq}_t}{d_{\kappa_t}\mu_t}}s^{\vartheta-1}e^{-s}ds,
\]
is the cumulative distribution function of a Gamma-distributed random variable with shape $\vartheta$ and scale $\frac{d_{\kappa_t} \mu_t}{\vartheta}$.

As for the the K-distribution CDF, $F_K(y;\mu, \vartheta_1, \vartheta_2)$, expressed as a function of mean and the shape parameters is given by
\begin{equation}
F_K(y;\cdot) = \frac{2^{2-\vartheta_1-\vartheta_2}}{\Gamma(\vartheta_1)\Gamma(\vartheta_2)} \int_0^{2\sqrt{\vartheta_1\vartheta_2 y/\mu}} t^{\vartheta_1+\vartheta_2-1} K_{|\vartheta_1-\vartheta_2|}(t) dt,\end{equation}where $K_a(\cdot)$ is the modified Bessel function of the second kind. For the jump components $K(\text{Illiq}_t|N_t=j)$, we set $y = \text{Illiq}_t$, $\mu = \mu_t j d_{\kappa_t}$, $\vartheta_1 = j\zeta$, and $\vartheta_2 = \vartheta$.

\setcounter{section}{1}

\clearpage
\begin{landscape} 
	\begin{table}[!h]
			\section*{Supplementary Document}
		{\footnotesize\section{Additional Empirical Results}\label{app:all_results}}
		{\footnotesize\subsection{S\&P 500 -- In-sample estimation}\label{app:appendix_in_sample_est}} 
		\setlength{\tabcolsep}{3pt}
		\renewcommand{\arraystretch}{0.8}
		\begin{adjustbox}{max width=0.75\linewidth,center}
			\begin{tabular}{lccccccccccccc|ccc}
				\vspace{0.4cm}
				Panel a)& \multicolumn{16}{c}{\Large\textbf{Parameter Estimates}}\\
				\vspace{0.4cm}
				&  \multicolumn{13}{c}{\large{Realized Amihud}} & \multicolumn{3}{c}{\large{Daily Amihud}}\\
				& (I)      & (II)      & (III)       & (IV)      & (V)      & (VI)      & (VII)      & (VIII)      & (IX)         & (X)                 & (XI)         & (XII)   & (XIII) & (II)& (IV)  & (XI)    \\
				\midrule
				$\omega$      & $0.0002^a$                   & $0.0002^a$                   & $0.0001^a$                   & $0.0001^a$                   & $0.0002^a$                   & $0.0002^a$                   & $0.0001^a$                   & $0.0001^a$                   & $0.0001^a$                   & $0.0001^a$                   & $0.0001^a$                   & $0.0002^a$                   & $0.0002^a$  & $0.0005^a$                 & $0.0001^a$                 & $0.0001^a$                 \\
				& (0.0001)                     & (0.0001)                     & (0.0000)                     & (0.0000)                     & (0.0000)                     & (0.0000)                     & (0.0000)                     & (0.0000)                     & (0.0000)                     & (0.0000)                     & (0.0000)                     & (0.0000)                     & (0.0000)    & (0.0001)                   & (0.0000)                   & (0.0000)                   \\[2mm]
				$\alpha_d$      & $0.3698^a$                   & $0.2690^a$                   & $0.3704^a$                   & $0.2638^a$                   & $0.3342^a$                   & $0.2318^a$                   & $0.2401^a$                   & $0.3786^a$                   & $0.2684^a$                   & $0.2816^a$                   & $0.2308^a$                   & $0.3316^a$                   & $0.2239^a$  & $-0.1545^a$                & $0.0000$                   & $0.0000$                   \\
				& (0.0770)                     & (0.0726)                     & (0.0329)                     & (0.0321)                     & (0.0367)                     & (0.0401)                     & (0.0372)                     & (0.0246)                     & (0.0224)                     & (0.0221)                     & (0.0266)                     & (0.0302)                     & (0.0328)    & (0.0336)                   & (0.0153)                   & (0.0361)                   \\[2mm]
				$\alpha_w$    & $0.3078^a$                   & $0.3481^a$                   &                              &                              & $0.3986^a$                   & $0.4397^a$                   & $0.0719^c$                   &                              &                              &                              & $0.1169^a$                   & $0.4529^a$                   & $0.5024^a$  & $0.3573^a$                 &                            &            $0.0000$                \\
				& (0.0885)                     & (0.0837)                     &                              &                              & (0.0454)                     & (0.0461)                     & (0.0433)                     &                              &                              &                              & (0.0396)                     & (0.0386)                     & (0.0373)    & (0.0693)                   &                            &        (0.0346)                    \\[2mm]
				$\alpha_m$    & $0.2278^a$                   & $0.2254^a$                   &                              &                              & $0.1665^a$                   & $0.1755^a$                   & $0.0934^a$                   &                              &                              &                              & $0.0846^a$                   & $0.1448^a$                   & $0.1537^a$  & $0.4678^a$                 &                            &           $0.0072$                 \\
				& (0.0426)                     & (0.0386)                     &                              &                              & (0.0374)                     & (0.0376)                     & (0.0221)                     &                              &                              &                              & (0.0186)                     & (0.0309)                     & (0.0316)    & (0.1044)                   &                            &              (0.0143)              \\[2mm]
				$\beta_1$     &                              &                              & $0.5748^a$                   & $0.6350^a$                   &                              &                              & $0.4813^a$                   & $0.5829^a$                   & $0.6484^a$                   & $0.5237^a$                   & $0.4709^a$                   &                              &             &                            & $0.8998^a$                 & $0.8913^a$                 \\
				&                              &                              & (0.0387)                     & (0.0372)                     &                              &                              & (0.0522)                     & (0.0280)                     & (0.0254)                     & (0.0449)                     & (0.0435)                     &                              &             &                            & (0.0147)                   & (0.0194)                   \\[2mm]
				$\beta_2$     &                              &                              &                              &                              &                              &                              &                              &                              &                              & $0.1089^a$                   &                              &                              &             &                            &                            &                            \\
				&                              &                              &                              &                              &                              &                              &                              &                              &                              & (0.0389)                     &                              &                              &             &                            &                            &                            \\[2mm]
				$\gamma$      &                              & $0.1226^a$                   &                              & $0.0979^a$                   &                              & $0.1054^a$                   & $0.1208^a$                   &                              & $0.0959^a$                   & $0.1018^a$                   & $0.1236^a$                     &                              & $0.1088^a$  & $0.1475^a$                 & $0.1344^a$                 & $0.1359^a$                 \\
				&                              & (0.0199)                     &                              & (0.0108)                     &                              & (0.0126)                     & (0.0132)                     &                              & (0.0078)                     & (0.0081)                     & (0.0097)                     &                              & (0.0103)    & (0.0383)                   & (0.0194)                   & (0.0259)                   \\[2mm]
				$\theta$      &                              &                              & $12.7424^a$                  & $13.3025^a$                  & $12.7148^a$                  & $13.1352^a$                  & $13.4936^a$                  & $21.1261^a$                  & $22.8732^a$                  & $22.9825^a$                  & $23.2534^a$                  & $21.2748^a$                  & $22.4339^a$ &                            & $1.2431^a$                 & $25.5677^a$                 \\
				&                              &                              & (0.5007)                     & (0.5650)                     & (0.5321)                     & (0.5670)                     & (0.5910)                     & (0.7235)                     & (1.2788)                     & (0.9606)                     & (0.7172)                     & (0.6650)                     & (0.7315)    &                            & (0.0338)                   & (1.2613)                   \\[2mm]
				$\zeta$       &                              &                              &                              &                              &                              &                              &                              & $16.7566^a$                  & $17.2338^a$                  & $17.4640^a$                  & $19.2278^a$                  & $16.1917^a$                  & $18.8551^a$ &                            &                            & $0.7549^a$                 \\
				&                              &                              &                              &                              &                              &                              &                              & (1.6643)                     & (2.1565)                     & (0.8680)                     & (0.8819)                     & (2.0630)                     & (0.5785)    &                            &                            & (0.0491)                   \\[2mm]
				$\phi_1$      &                              &                              &                              &                              &                              &                              &                              & $0.0204^a$                   & $0.0157^a$                   & $0.0154^a$                   & $0.0152^b$                   & $0.0265^b$                   & $0.0219^b$  &                            &                            & $1.0868$                   \\
				&                              &                              &                              &                              &                              &                              &                              & (0.0071)                     & (0.0052)                     & (0.0052)                     & (0.0068)                     & (0.0115)                     & (0.0090)    &                            &                            & (1.5806)                   \\[2mm]
				$\phi_2$      &                              &                              &                              &                              &                              &                              &                              & $0.9390^a$                   & $0.9586^a$                   & $0.9594^a$                   & $0.9571^a$                   & $0.9125^a$                   & $0.9344^a$  &                            &                            & $0.6808$                 \\
				&                              &                              &                              &                              &                              &                              &                              & (0.0213)                     & (0.0146)                     & (0.0149)                     & (0.0204)                     & (0.0386)                     & (0.0279)    &                            &                            & (0.4637)                    \\[2mm]
				$\phi_3$      &                              &                              &                              &                              &                              &                              &                              & $0.3475^a$                   & $0.3091^a$                   & $0.3032^a$                   & $0.2672^a$                   & $0.3501^a$                   & $0.3140^a$  &                            &                            & $0.0000$                   \\
				&                              &                              &                              &                              &                              &                              &                              & (0.0452)                     & (0.0374)                     & (0.0383)                     & (0.0458)                     & (0.0681)                     & (0.0587)    &                            &                            & (0.0305)                   \\[2mm]
				\midrule
				LogLik & 15653.31 & 15711.38 & 16191.50 & 16250.95 & 16188.51 & 16233.45 & 16270.62 & 16373.17 & 16446.18 & 16449.65 & 16469.25 & 16368.65 & 16426.82    & 13315.30  & 14274.54                   & 14283.68 \\
				\bottomrule
				\\
				Panel b)& \multicolumn{16}{c}{\Large\textbf{Ljung-Box test}}\\  
				
				\multicolumn{16}{l}{Residuals: $\epsilon_t$}\\[2mm]
				LB 1          & 0.2071   & 0.0154   & 0.9715   & 0.8884   & 0.0379   & 0.0048   & 0.4126   & 0.8461   & 0.8794   & 0.5294   & 0.4512   & 0.0233   & 0.0029      & 0.4852 & 0.0000 & 0.0000 \\
				LB 5          & 0.0000   & 0.0000   & 0.0574   & 0.0910   & 0.0000   & 0.0000   & 0.1985   & 0.0127   & 0.0318   & 0.0600   & 0.1625   & 0.0000   & 0.0000      & 0.1425 & 0.0006 & 0.0005 \\
				LB 10         & 0.0001   & 0.0000   & 0.0646   & 0.0996   & 0.0003   & 0.0000   & 0.2433   & 0.0051   & 0.0070   & 0.0067   & 0.1112   & 0.0001   & 0.0000      & 0.2922 & 0.0096 & 0.0086 \\
				\midrule
				
			\end{tabular}
		\end{adjustbox}
		\caption{Estimates for S\&P 500. Panel a): Estimated coefficients (robust standard errors in parenthesis); Panel b):  $p$-value of the Ljung-Box statistics. In-sample period: January 3, 2005 - October 15, 2015. The superscripts a, b and, c denote significant coefficients at $1\%$, $5\%$ and, $10\%$ level, respectively. The estimated models are (I) HAR, (II) AHAR, (III) MEM, (IV) AMEM, (V) MEM-HAR, (VI) AMEM-HAR, (VII) G-AMEM-HAR, (VIII) MEM-J, (IX) AMEM-J, (X) AMEM(2,1)-J, (XI) G-AMEM-HAR-J, (XII) MEM-HAR-J, (XIII) AMEM-HAR-J.
			\label{tab:results_insample}}
	\end{table}
\end{landscape}

\clearpage
\subsection{Determinants of Illiquidity at Risk}

\begin{table}[h!]
	\centering
	\setlength{\tabcolsep}{3pt}
	\renewcommand{\arraystretch}{1.1}
	\begin{adjustbox}{max width=\linewidth,center}
		\begin{tabular}{lccccccccc}
			\toprule
			& \multicolumn{4}{c}{Panel a) - Lagged Regressors} & & \multicolumn{4}{c}{Panel b) - Contemporaneous Regressors} \\
			\cmidrule{2-5} \cmidrule{7-10}
			& \multicolumn{2}{c}{Full sample} & \multicolumn{2}{c}{Out of sample} & & \multicolumn{2}{c}{Full sample} & \multicolumn{2}{c}{Out of sample} \\
			\cmidrule{2-3} \cmidrule{4-5} \cmidrule{7-8} \cmidrule{9-10}
			& \multicolumn{4}{c}{Parameter Estimates} & & \multicolumn{4}{c}{Parameter Estimates} \\
			\midrule
			Constant   & $-3.883^a$  & $-3.0887^a$ & $-3.4705^a$ & $-3.5067^a$  &  & $-3.6905^a$ & $-3.6656^a$ & $-4.5020^a$ & $-4.4153^a$  \\
			& (0.1026)    & (0.1033)    & (0.2005)    & (0.2076)     &  & (0.1337)    & (0.1321)    & (0.3363)    & (0.3342)     \\
			Tail $VIX$      & $0.0641^a$  & ----        & $0.0960^a$  & ----         &  & $0.0813^a$  & ----        & $0.1803^a$  & ----         \\
			& (0.0164)    & ----        & (0.0336)    & ----         &  & (0.0138)    & ----        & (0.0035)    & ----         \\
			& $[0.0031]$  & ----        & $[0.0032]$  & ----         &  & $[0.0033]$  & ----        & $[0.0018]$  & ----         \\
			Tail $Uncer$    & ----        & 29.1362     & ----        & $101.1290^c$ &  & ----        & -46.4346    & ----        & -48.3881     \\
			& ----        & (35.9231)   & ----        & (54.0673)    &  & ----        & (46.1503)   & ----        & (82.6789)    \\
			& ----        & $[1.4059]$  & ----        & $[3.3363]$   &  & ----        & $[-1.9285]$ & ----        & $[-0.3196]$  \\
			Tail $Risk Av.$ & ----        & 0.0003      & ----        & 0.0003       &  & ----        & $0.0009^a$  & ----        & $.0029^a$    \\
			& ----        & (0.0003)    & ----        & (0.0004)     &  & ----        & (0.0003)    & ----        & (0.0008)     \\
			& ----        & $[0.0000]$  & ----        & $[0.0000]$   &  & ----        & $[0.0000]$  & ----        & $[0.0000]$   \\
			Tail $EPU$      & -0.0010     & -0.0004     & -0.0055     & -0.0076      &  & 0.0017      & $0.0031^b$  & -0.0005     & -0.0008      \\
			& (0.0019)    & (0.0018)    & (0.0053)    & (0.0071)     &  & (0.0016)    & (0.0015)    & (0.0056)    & (0.0069)     \\
			& $[-0.0001]$ & $[0.0000]$  & $[-0.0002]$ & $[-0.0003]$  &  & $[0.0001]$  & $[0.0001]$  & $[0.0000]$  & $[0.0000]$   \\
			Tail $TED$      & 0.0057      & 0.0058      & $0.0435^b$  & $0.0441^b$   &  & 0.0072      & 0.0072      & $-6.6167^c$ & $-10.0677^c$ \\
			& (0.0129)    & (0.0127)    & (0.0196)    & (0.0196)     &  & (0.0138)    & (0.0135)    & (3.4615)    & (5.6288)     \\
			& $[0.0003]$  & $[0.0003]$  & $[0.0014]$  & $[0.0015]$   &  & $[0.0003]$  & $[0.0003]$  & $[-0.0653]$ & $[-0.0665]$  \\
			$D$        & $0.2699^c$  & $0.3124^b$  & 0.2373      & 0.3652       &  & $1.1554^a$  & $1.1719^a$  & $1.6027^a$  & $1.6119^a$   \\
			& (0.1419)    & (0.1405)    & (0.2863)    & (0.2782)     &  & (0.1574)    & (0.1553)    & (0.3643)    & (0.3658)     \\
			& $[0.0131]$  & $[0.0153]$  & $[0.0079]$  & $[0.0123]$   &  & $[0.0517]$  & $[0.0532]$  & $[0.0189]$  & $[0.0128]$ \\
			\midrule
			& \multicolumn{4}{c}{Model Fit} & & \multicolumn{4}{c}{Model Fit} \\
			\midrule
			Success Rate &  94.80\% &94.80\% &96.10\% &96.20\% & & 94.80\% &94.80\% &96.30\% &96.40\% \\
			LR p-value   &  0.0008 & 0.0603 & 0.0013 & 0.0109 & & 0.0000 & 0.0000 & 0.0000 & 0.0000  \\
			\bottomrule			
		\end{tabular}
	\end{adjustbox}
	\caption{Logit model. The table displays logit estimation results for IlliQaR violations. Panel (a) reports coefficients for lagged regressors, while Panel (b) focuses on contemporaneous specifications. Robust standard errors in parenthesis. Average marginal effects in square brackets. The explanatory variables (VIX, Uncertainty, Risk Aversion, EPU, and TED) are defined as deviations from their 22-day moving averages, considering only values exceeding the 95th percentile. The superscripts a, b and c denote significance at 1\%, 5\% and 10\% levels, respectively. Success rate denotes the percentage of correctly classified observations. LR p-value refers to the p-value of the likelihood ratio test for the joint significance of the regressors, where, under the null hypothesis, the model is $\text{Pr}(I_t=1 | \mathbf{x}_{t-1}) = \frac{\exp(\delta_0) }{1 + \exp(\delta_0)}$.\label{tab:logit_ma_quantile}}
\end{table}

\clearpage

\section{Parameter Estimates for Individual Stocks }\label{app:appendix_fullsample_individual_stocks} 
\subsection{AHAR Model}
\begin{table}[!ht]
	\centering
	\setlength{\tabcolsep}{3pt} 
	\renewcommand{\arraystretch}{1.0}
	\begin{adjustbox}{max width=1\linewidth,center}
		\begin{tabular}{lccccccccccccc}
			& \multicolumn{13}{c}{\textbf{Parameter Estimates ( MSFT to MCD)}} \\
			[1.2mm]
			& MSFT & JPM & WMT & UNH & PG & JNJ & HD & KO & MRK & CVX & CRM & MCD & \\ 
			[0.8mm]
			$\omega$   & $0.1100^{a}$ & $0.0954^{a}$ & $0.1438^{a}$ & $0.1497^{a}$ & $0.1461^{a}$ & $0.1607^{a}$ & $0.1321^{a}$ & $0.1359^{a}$ & $0.1349^{a}$ & $0.0814^{a}$ & $0.1134^{a}$ & $0.1580^{a}$ & \\
			& (0.0176) & (0.0165) & (0.0221) & (0.0249) & (0.0260) & (0.0230) & (0.0202) & (0.0235) & (0.0200) & (0.0167) & (0.0171) & (0.0274) & \\
			$\alpha_d$ & $0.3756^{a}$ & $0.3900^{a}$ & $0.3830^{a}$ & $0.4063^{a}$ & $0.3847^{a}$ & $0.3727^{a}$ & $0.3549^{a}$ & $0.3956^{a}$ & $0.3352^{a}$ & $0.2970^{a}$ & $0.3353^{a}$ & $0.3439^{a}$ & \\
			& (0.0255) & (0.0314) & (0.0240) & (0.0238) & (0.0274) & (0.0267) & (0.0258) & (0.0307) & (0.0235) & (0.0397) & (0.0251) & (0.0283) & \\
			$\alpha_w$ & $0.3615^{a}$ & $0.3357^{a}$ & $0.3627^{a}$ & $0.2735^{a}$ & $0.4006^{a}$ & $0.3924^{a}$ & $0.3999^{a}$ & $0.2699^{a}$ & $0.3617^{a}$ & $0.4650^{a}$ & $0.3742^{a}$ & $0.3866^{a}$ & \\
			& (0.0401) & (0.0395) & (0.0369) & (0.0541) & (0.0698) & (0.0616) & (0.0414) & (0.0390) & (0.0994) & (0.0538) & (0.0380) & (0.0515) & \\
			$\alpha_m$ & $0.1504^{a}$ & $0.1730^{a}$ & $0.1126^{a}$ & $0.1782^{a}$ & $0.0695$ & $0.0711$ & $0.1207^{a}$ & $0.2015^{a}$ & $0.1755^{b}$ & $0.1521^{b}$ & $0.1761^{a}$ & $0.1122^{a}$ & \\
			& (0.0377) & (0.0469) & (0.0330) & (0.0501) & (0.0586) & (0.0475) & (0.0365) & (0.0368) & (0.0886) & (0.0600) & (0.0322) & (0.0353) & \\
			\midrule
			LogLik     & 230.21 & 227.72 & -11.31 & -474.48 & 81.28 & -5.66 & -221.41 & 81.75 & 12.80 & -39.60 & -886.00 & -200.69 & \\
			LB 10 ($\hat\epsilon_t$)      & 0.0000 & 0.0068 & 0.0278 & 0.3492 & 0.1710 & 0.0000 & 0.0041 & 0.0012 & 0.2225 & 0.0000 & 0.0000 & 0.4469 & \\
			\midrule[1.5pt] 
			\midrule
			\\
			& \multicolumn{13}{c}{\textbf{Parameter Estimates (CSCO to MMM)}} \\
			[1.2mm]
			& CSCO & AXP & AMGN & VZ & CAT & IBM & DIS & GS & HON & NKE & BA & INTC & MMM \\
			[0.8mm]
			$\omega$   & $0.1301^{a}$ & $0.1259^{a}$ & $0.1513^{a}$ & $0.1055^{a}$ & $0.1179^{a}$ & $0.1340^{a}$ & $0.0890^{a}$ & $0.1321^{a}$ & $0.1096^{a}$ & $0.0995^{a}$ & $0.0583^{a}$ & $0.1215^{a}$ & $0.1338^{a}$ \\
			& (0.0177) & (0.0181) & (0.0274) & (0.0200) & (0.0201) & (0.0221) & (0.0185) & (0.0208) & (0.0180) & (0.0158) & (0.0151) & (0.0187) & (0.0196) \\
			$\alpha_d$ & $0.3947^{a}$ & $0.4001^{a}$ & $0.3129^{a}$ & $0.4125^{a}$ & $0.4094^{a}$ & $0.4187^{a}$ & $0.4022^{a}$ & $0.3513^{a}$ & $0.3746^{a}$ & $0.3773^{a}$ & $0.3372^{a}$ & $0.3962^{a}$ & $0.4235^{a}$ \\
			& (0.0224) & (0.0257) & (0.0233) & (0.0226) & (0.0297) & (0.0388) & (0.0264) & (0.0257) & (0.0232) & (0.0298) & (0.0308) & (0.0322) & (0.0291) \\
			$\alpha_w$ & $0.3446^{a}$ & $0.2665^{a}$ & $0.3396^{a}$ & $0.2808^{a}$ & $0.3190^{a}$ & $0.3850^{a}$ & $0.3290^{a}$ & $0.3864^{a}$ & $0.3082^{a}$ & $0.3060^{a}$ & $0.3371^{a}$ & $0.3357^{a}$ & $0.2610^{a}$ \\
			& (0.0571) & (0.0432) & (0.0770) & (0.0374) & (0.0477) & (0.0483) & (0.0419) & (0.0488) & (0.0501) & (0.0354) & (0.0407) & (0.0448) & (0.0453) \\
			$\alpha_m$ & $0.1373^{b}$ & $0.2131^{a}$ & $0.1981^{a}$ & $0.2066^{a}$ & $0.1552^{a}$ & $0.0731$ & $0.1819^{a}$ & $0.1228^{a}$ & $0.2098^{a}$ & $0.2219^{a}$ & $0.2569^{a}$ & $0.1514^{a}$ & $0.1829^{a}$ \\
			& (0.0553) & (0.0456) & (0.0571) & (0.0374) & (0.0498) & (0.0459) & (0.0328) & (0.0416) & (0.0430) & (0.0320) & (0.0322) & (0.0486) & (0.0414) \\
			\midrule
			LogLik     & 2.03 & -536.40 & -462.58 & 177.75 & -400.88 & -422.53 & -406.38 & -567.02 & -759.92 & -510.18 & -959.57 & 78.69 & -374.60 \\
			LB 10 ($\hat\epsilon_t$)      & 0.0227 & 0.0608 & 0.0263 & 0.0004 & 0.0000 & 0.0000 & 0.0064 & 0.0082 & 0.0245 & 0.0016 & 0.0394 & 0.0005 & 0.0003 \\
			\bottomrule
		\end{tabular}
	\end{adjustbox}
	\caption{Estimates for individual stocks for the AHAR model. Estimated coefficients (robust standard errors in parenthesis) and $p$-value of the Ljung-Box statistics with 10 lags, LB 10 ($\hat\epsilon_t$).  The superscripts a, b, and c denote significance at 1\%, 5\%, and 10\% level, respectively. Estimation period: January 3, 2012 - February 4, 2020. Stocks are presented in descending order of market capitalization.}
	\label{tab:results_ahar_stacked}
\end{table}

\clearpage
\subsection{AMEM-HAR Model}
\begin{table}[!ht]
	\centering
	\setlength{\tabcolsep}{3pt}
	\renewcommand{\arraystretch}{1.1}
	\begin{adjustbox}{max width=1\linewidth,center}
		\begin{tabular}{lccccccccccccc}
			& \multicolumn{12}{c}{\textbf{Parameter Estimates (MSFT -- MCD)}} & \\
			[1.2mm]
			& MSFT & JPM & WMT & UNH & PG & JNJ & HD & KO & MRK & CVX & CRM & MCD & \\
			[0.8mm]
			$\omega$   & $0.1000^{a}$ & $0.0792^{a}$ & $0.1176^{a}$ & $0.1467^{a}$ & $0.1202^{a}$ & $0.1006^{a}$ & $0.1176^{a}$ & $0.1292^{a}$ & $0.1164^{a}$ & $0.0643^{a}$ & $0.0919^{a}$ & $0.1565^{a}$ & \\
			& (0.0160) & (0.0154) & (0.0179) & (0.0222) & (0.0219) & (0.0243) & (0.0187) & (0.0203) & (0.0204) & (0.0147) & (0.0155) & (0.0220) & \\
			$\alpha_d$ & $0.4161^{a}$ & $0.3955^{a}$ & $0.4228^{a}$ & $0.4434^{a}$ & $0.3843^{a}$ & $0.3866^{a}$ & $0.3935^{a}$ & $0.4010^{a}$ & $0.3673^{a}$ & $0.2948^{a}$ & $0.3978^{a}$ & $0.3652^{a}$ & \\
			& (0.0235) & (0.0224) & (0.0245) & (0.0241) & (0.0241) & (0.0233) & (0.0247) & (0.0231) & (0.0233) & (0.0233) & (0.0253) & (0.0242) & \\
			$\alpha_w$ & $0.3258^{a}$ & $0.3053^{a}$ & $0.3451^{a}$ & $0.2574^{a}$ & $0.3677^{a}$ & $0.4064^{a}$ & $0.3726^{a}$ & $0.2683^{a}$ & $0.2960^{a}$ & $0.4347^{a}$ & $0.3972^{a}$ & $0.3721^{a}$ & \\
			& (0.0331) & (0.0335) & (0.0323) & (0.0337) & (0.0365) & (0.0363) & (0.0340) & (0.0358) & (0.0384) & (0.0367) & (0.0314) & (0.0351) & \\
			$\alpha_m$ & $0.1539^{a}$ & $0.2127^{a}$ & $0.1156^{a}$ & $0.1579^{a}$ & $0.1213^{a}$ & $0.0980^{a}$ & $0.1215^{a}$ & $0.1997^{a}$ & $0.2257^{a}$ & $0.1952^{a}$ & $0.1106^{a}$ & $0.1085^{a}$ & \\
			& (0.0283) & (0.0291) & (0.0282) & (0.0316) & (0.0315) & (0.0296) & (0.0308) & (0.0334) & (0.0346) & (0.0298) & (0.0272) & (0.0323) & \\
			$\gamma$   & $0.0220^{a}$ & $0.0229^{a}$ & $0.0115$ & $0.0084$ & $0.0268^{a}$ & $0.0353^{a}$ & $0.0033$ & $0.0168^{b}$ & $0.0003$ & $0.0304^{a}$ & $0.0279^{a}$ & $0.0120$ & \\
			& (0.0077) & (0.0075) & (0.0082) & (0.0095) & (0.0079) & (0.0084) & (0.0087) & (0.0081) & (0.0083) & (0.0080) & (0.0105) & (0.0089) & \\
			$\theta$   & $22.1809^{a}$ & $23.8232^{a}$ & $19.1965^{a}$ & $14.7802^{a}$ & $21.0216^{a}$ & $19.6540^{a}$ & $17.3933^{a}$ & $20.3734^{a}$ & $19.7037^{a}$ & $20.6883^{a}$ & $12.0646^{a}$ & $17.2057^{a}$ & \\
			& (0.5411) & (0.5579) & (0.5597) & (0.4029) & (0.5774) & (0.5041) & (0.4606) & (0.5011) & (0.5266) & (0.5670) & (0.3273) & (0.4563) & \\
			\midrule
			LogLik     & 351.82 & 476.04 & 147.95 & -247.56 & 281.10 & 182.71 & 6.90 & 234.92 & 187.93 & 258.36 & -557.79 & -17.16 & \\
			LB 10 ($\hat\epsilon_t$) & 0.0000 & 0.0130 & 0.0248 & 0.1009 & 0.1163 & 0.0000 & 0.0433 & 0.0309 & 0.7454 & 0.0006 & 0.0000 & 0.5776 & \\
			\midrule[1.5pt]
			\\
			& \multicolumn{13}{c}{\textbf{Parameter Estimates (CSCO -- MMM)}} \\
			[1.2mm]
			& CSCO & AXP & AMGN & VZ & CAT & IBM & DIS & GS & HON & NKE & BA & INTC & MMM \\
			[0.8mm]
			$\omega$   & $0.1143^{a}$ & $0.1091^{a}$ & $0.1338^{a}$ & $0.0618^{a}$ & $0.1236^{a}$ & $0.1238^{a}$ & $0.0754^{a}$ & $0.1054^{a}$ & $0.1000^{a}$ & $0.0763^{a}$ & $0.0268^{a}$ & $0.1151^{a}$ & $0.1142^{a}$ \\
			& (0.0183) & (0.0193) & (0.0227) & (0.0147) & (0.0190) & (0.0181) & (0.0149) & (0.0175) & (0.0181) & (0.0166) & (0.0083) & (0.0179) & (0.0217) \\
			$\alpha_d$ & $0.4326^{a}$ & $0.4152^{a}$ & $0.3523^{a}$ & $0.4313^{a}$ & $0.4108^{a}$ & $0.4576^{a}$ & $0.4525^{a}$ & $0.3848^{a}$ & $0.4172^{a}$ & $0.4112^{a}$ & $0.4028^{a}$ & $0.4204^{a}$ & $0.4439^{a}$ \\
			& (0.0230) & (0.0246) & (0.0235) & (0.0238) & (0.0233) & (0.0269) & (0.0240) & (0.0237) & (0.0246) & (0.0236) & (0.0244) & (0.0238) & (0.0252) \\
			$\alpha_w$ & $0.2916^{a}$ & $0.2622^{a}$ & $0.2988^{a}$ & $0.3010^{a}$ & $0.2919^{a}$ & $0.3280^{a}$ & $0.2763^{a}$ & $0.3469^{a}$ & $0.2858^{a}$ & $0.2751^{a}$ & $0.3797^{a}$ & $0.3171^{a}$ & $0.2898^{a}$ \\
			& (0.0339) & (0.0361) & (0.0362) & (0.0364) & (0.0332) & (0.0311) & (0.0315) & (0.0329) & (0.0304) & (0.0342) & (0.0329) & (0.0328) & (0.0364) \\
			$\alpha_m$ & $0.1659^{a}$ & $0.2184^{a}$ & $0.2158^{a}$ & $0.2083^{a}$ & $0.1720^{a}$ & $0.0977^{a}$ & $0.1941^{a}$ & $0.1521^{a}$ & $0.1996^{a}$ & $0.2423^{a}$ & $0.1824^{a}$ & $0.1481^{a}$ & $0.1547^{a}$ \\
			& (0.0302) & (0.0329) & (0.0343) & (0.0309) & (0.0301) & (0.0275) & (0.0275) & (0.0286) & (0.0305) & (0.0313) & (0.0242) & (0.0300) & (0.0307) \\
			$\gamma$   & $0.0030$ & $0.0043$ & $0.0171^{c}$ & $0.0054$ & $0.0199^{b}$ & $0.0046$ & $0.0151^{c}$ & $0.0393^{a}$ & $0.0107$ & $0.0022$ & $0.0313^{a}$ & $0.0128$ & $0.0118$ \\
			& (0.0083) & (0.0097) & (0.0095) & (0.0078) & (0.0091) & (0.0091) & (0.0090) & (0.0096) & (0.0103) & (0.0094) & (0.0099) & (0.0085) & (0.0094) \\
			$\theta$   & $20.0872^{a}$ & $14.1983^{a}$ & $14.7193^{a}$ & $21.7510^{a}$ & $15.7149^{a}$ & $15.3538^{a}$ & $16.1713^{a}$ & $14.4335^{a}$ & $13.1825^{a}$ & $15.0393^{a}$ & $14.0330^{a}$ & $19.8497^{a}$ & $15.4692^{a}$ \\
			& (0.5433) & (0.3721) & (0.4007) & (0.5009) & (0.4140) & (0.4914) & (0.5175) & (0.3941) & (0.3296) & (0.4314) & (0.3843) & (0.5278) & (0.4230) \\
			\midrule
			LogLik     & 217.30 & -297.23 & -258.94 & 336.71 & -161.38 & -193.37 & -101.47 & -283.70 & -413.02 & -219.56 & -323.99 & 189.16 & -170.11 \\
			LB 10 ($\hat\epsilon_t$) & 0.0709 & 0.1809 & 0.1645 & 0.0025 & 0.0000 & 0.0000 & 0.0801 & 0.0018 & 0.0157 & 0.0015 & 0.0001 & 0.0079 & 0.0005 \\
			\bottomrule
		\end{tabular}
	\end{adjustbox}
	\caption{Estimates for individual stocks for the AMEM-HAR model. Estimated coefficients (robust standard errors in parenthesis) and $p$-value of the Ljung-Box statistics with 10 lags, LB 10 ($\hat\epsilon_t$).  The superscripts a, b, and c denote significance at 1\%, 5\%, and 10\% level, respectively. Estimation period: January 3, 2012 - February 4, 2020. Stocks are presented in descending order of market capitalization.}
	\label{tab:results_mem_ahar_individual_stocks}
\end{table}

\clearpage

\subsection{AMEM-HAR-J Model}
\begin{table}[!htbp]
	\centering
	\setlength{\tabcolsep}{3pt}
	\renewcommand{\arraystretch}{1}
	\begin{adjustbox}{max width=1\linewidth,center}
		\begin{tabular}{lccccccccccccc}
			& \multicolumn{12}{c}{\textbf{Parameter Estimates (MSFT-MCD)}}\\
			[1.2mm]
			& MSFT & JPM & WMT & UNH & PG & JNJ & HD & KO & MRK & CVX & CRM & MCD \\
			[0.8mm]
			$\omega$                & 0.0946$^{a}$   & 0.0772$^{a}$   & 0.1123$^{a}$   & 0.1445$^{a}$   & 0.1239$^{a}$   & 0.1093$^{a}$   & 0.1148$^{a}$   & 0.1292$^{a}$   & 0.1200$^{a}$   & 0.0655$^{a}$   & 0.0906$^{a}$   & 0.1573$^{a}$   &      \\
			& (0.0158) & (0.0155) & (0.0177) & (0.0223) & (0.0214) & (0.0269) & (0.0187) & (0.0203) & (0.0204) & (0.0146) & (0.0154) & (0.0219) &      \\
			$\alpha_d$              & 0.4040$^{a}$   & 0.3926$^{a}$   & 0.4159$^{a}$   & 0.4443$^{a}$   & 0.3826$^{a}$   & 0.3842$^{a}$   & 0.3893$^{a}$   & 0.4014$^{a}$   & 0.3708$^{a}$   & 0.2958$^{a}$   & 0.3925$^{a}$   & 0.3670$^{a}$   &      \\
			& (0.0228) & (0.0225) & (0.0246) & (0.0239) & (0.0241) & (0.0235) & (0.0249) & (0.0230) & (0.0232) & (0.0232) & (0.0255) & (0.0241) &      \\
			$\alpha_w$              & 0.3353$^{a}$   & 0.3145$^{a}$   & 0.3571$^{a}$   & 0.2637$^{a}$   & 0.3582$^{a}$   & 0.4017$^{a}$   & 0.3682$^{a}$   & 0.2674$^{a}$   & 0.2856$^{a}$   & 0.4323$^{a}$   & 0.3982$^{a}$   & 0.3681$^{a}$   &      \\
			& (0.0326) & (0.0334) & (0.0323) & (0.0336) & (0.0361) & (0.0374) & (0.0336) & (0.0357) & (0.0384) & (0.0366) & (0.0316) & (0.0351) &      \\
			$\alpha_m$              & 0.1635$^{a}$   & 0.2096$^{a}$   & 0.1204$^{a}$   & 0.1552$^{a}$   & 0.1299$^{a}$   & 0.0982$^{a}$   & 0.1337$^{a}$   & 0.2028$^{a}$   & 0.2298$^{a}$   & 0.1968$^{a}$   & 0.1189$^{a}$   & 0.1099$^{a}$   &      \\
			& (0.0280) & (0.0291) & (0.0278) & (0.0319) & (0.0316) & (0.0296) & (0.0310) & (0.0333) & (0.0347) & (0.0295) & (0.0272) & (0.0323) &      \\
			$\gamma$                & 0.0248$^{a}$   & 0.0231$^{a}$   & 0.0102   & 0.0070   & 0.0284$^{a}$   & 0.0342$^{a}$   & 0.0036   & 0.0153$^{c}$   & 0.0011   & 0.0291$^{a}$   & 0.0266$^{b}$   & 0.0130   &      \\
			& (0.0076) & (0.0075) & (0.0082) & (0.0095) & (0.0080) & (0.0084) & (0.0087) & (0.0081) & (0.0082) & (0.0079) & (0.0106) & (0.0088) &      \\
			$\theta$                & 24.6751$^{a}$  & 25.3525$^{a}$  & 22.1182$^{a}$  & 15.9882$^{a}$  & 22.9470$^{a}$  & 21.6986$^{a}$  & 18.4939$^{a}$  & 22.0493$^{a}$  & 21.2868$^{a}$  & 21.7231$^{a}$  & 12.9954$^{a}$  & 18.2324$^{a}$  &      \\
			& (0.6686) & (0.6371) & (0.7280) & (0.5020) & (0.7850) & (0.5897) & (0.6653) & (0.6001) & (0.5621) & (0.5761) & (0.4780) & (0.5353) &      \\
			$\zeta$                 & 6.7858$^{a}$   & 7.0844$^{a}$   & 5.7572$^{a}$   & 5.8289$^{a}$   & 8.9550$^{a}$   & 8.8129$^{a}$   & 8.3121$^{a}$   & 9.3139$^{a}$   & 6.0245$^{a}$   & 4.8328$^{c}$   & 4.9693$^{a}$   & 8.2229$^{a}$   &      \\
			& (1.0106) & (2.4675) & (0.8507) & (0.9179) & (1.1380) & (1.1748) & (1.3231) & (1.1146) & (1.4801) & (2.7985) & (0.5340) & (2.6718) &      \\
			$\kappa$                & 0.0325$^{a}$   & 0.0188$^{c}$   & 0.0416$^{a}$   & 0.0312$^{a}$   & 0.0373$^{a}$   & 0.0425$^{a}$   & 0.0297$^{a}$   & 0.0359$^{a}$   & 0.0233$^{b}$   & 0.0114   & 0.0316$^{a}$   & 0.0271$^{c}$   &      \\
			& (0.0081) & (0.0102) & (0.0098) & (0.0105) & (0.0102) & (0.0107) & (0.0114) & (0.0097) & (0.0094) & (0.0101) & (0.0107) & (0.0145) &      \\
			\midrule
			LogLik  & 367.15   & 485.99   & 171.91   & -241.06  & 286.90   & 191.44   & 9.51     & 239.37   & 199.18   & 269.57   & -553.58  & -13.14   &      \\
			LB 10($\hat\epsilon_t$) & 0.0000   & 0.0121   & 0.0290   & 0.0947   & 0.1258   & 0.0000   & 0.0500   & 0.0320   & 0.7249   & 0.0006   & 0.0000   & 0.5813   &      \\
			LB 10($\hat\xi_t$)      & 0.0000   & 0.9999   & 0.0573   & 0.0002   & 0.0000   & 0.0000   & 0.0000   & 0.4692   & 0.0000   & 0.0000   & 0.0053   & 0.3794   & \\
			\midrule[1.5pt]
			\\
			& \multicolumn{13}{c}{\textbf{Parameter Estimates (CSCO-MMM)}}\\
			[1.2mm]
			& CSCO & AXP & AMGN & VZ & CAT & IBM & DIS & GS & HON & NKE & BA & INTC & MMM \\
			[0.8mm]
			$\omega$                & 0.1133$^{a}$   & 0.1094$^{a}$   & 0.1325$^{a}$   & 0.0649$^{a}$   & 0.1237$^{a}$   & 0.1209$^{a}$   & 0.0728$^{a}$   & 0.1017$^{a}$   & 0.1$^{a}$      & 0.0738$^{a}$   & 0.0254$^{a}$   & 0.1159$^{a}$   & 0.1106$^{a}$   \\
			& (0.0182) & (0.0194) & (0.0226) & (0.0143) & ( 0.019) & (0.0178) & (0.0146) & (0.0175) & (0.0181) & (0.0165) & (0.0083) & (0.0179) & (0.0216) \\
			$\alpha_d$              & 0.4276$^{a}$   & 0.4091$^{a}$   & 0.354$^{a}$    & 0.4282$^{a}$   & 0.4086$^{a}$   & 0.4371$^{a}$   & 0.4455$^{a}$   & 0.3827$^{a}$   & 0.4172$^{a}$   & 0.402$^{a}$    & 0.4024$^{a}$   & 0.4111$^{a}$   & 0.446$^{a}$    \\
			& (0.0224) & (0.0243) & (0.0235) & (0.0229) & (0.0233) & (0.0273) & (0.0239) & (0.0238) & (0.0246) & (0.0234) & (0.0244) & ( 0.024) & (0.0252) \\
			$\alpha_w$              & 0.2882$^{a}$   & 0.2664$^{a}$   & 0.3004$^{a}$   & 0.2848$^{a}$   & 0.2963$^{a}$   & 0.3461$^{a}$   & 0.287$^{a}$    & 0.3516$^{a}$   & 0.2858$^{a}$   & 0.278$^{a}$    & 0.3835$^{a}$   & 0.324$^{a}$    & 0.2842$^{a}$   \\
			& (0.0327) & (0.0358) & ( 0.036) & (0.0358) & (0.0333) & (0.0314) & (0.0314) & (0.0324) & ( 0.034) & (0.0341) & ( 0.033) & (0.0326) & (0.0361) \\
			$\alpha_m$              & 0.1775$^{a}$   & 0.2209$^{a}$   & 0.2165$^{a}$   & 0.2244$^{a}$   & 0.1718$^{a}$   & 0.1063$^{a}$   & 0.1964$^{a}$   & 0.156$^{a}$    & 0.1996$^{a}$   & 0.2532$^{a}$   & 0.1826$^{a}$   & 0.1539$^{a}$   & 0.1631$^{a}$   \\
			& (0.0297) & (0.0329) & (0.0343) & (0.0297) & (0.0301) & (0.0272) & (0.0272) & (0.0277) & (0.0305) & (0.0308) & (0.0242) & (  0.03) & (0.0304) \\
			$\gamma$                & 0.002    & 0.0058   & 0.0154   & 0.0075   & 0.0195$^{b}$   & 0.0067   & 0.0158$^{c}$   & 0.037$^{a}$    & 0.0107   & 0.0031   & 0.0305$^{a}$   & 0.0129   & 0.0111   \\
			& (0.0081) & (0.0097) & (0.0094) & (0.0077) & (0.0091) & (0.0091) & (0.0089) & (0.0095) & (0.0103) & (0.0093) & (0.0099) & ( 0.008) & (0.0093) \\
			$\theta$                & 23.5381$^{a}$  & 15.0143$^{a}$  & 15.8335$^{a}$  & 23.9611$^{a}$  & 16.5347$^{a}$  & 18.0924$^{a}$  & 18.4376$^{a}$  & 15.5536$^{a}$  & 13.1826$^{a}$  & 16.6369$^{a}$  & 15.0761$^{a}$  & 22.5146$^{a}$  & 17.0579$^{a}$  \\
			& (0.6528) & (0.4308) & (0.4874) & (0.6756) & ( 0.579) & (0.5372) & (0.4757) & (0.4625) & (0.4044) & (0.5153) & ( 0.562) & (0.6111) & (0.5625) \\
			$\zeta$                 & 8.5373$^{a}$   & 3.9403$^{a}$   & 5.4946$^{a}$   & 7.1674$^{a}$   & 7.3633$^{a}$   & 3.8744$^{a}$   & 3.5569$^{a}$   & 4.4034$^{a}$   & 16.5954$^{a}$  & 4.4927$^{a}$   & 6.7339$^{a}$   & 6.1601$^{a}$   & 7.3746$^{a}$   \\
			& (1.0802) & (0.5517) & (0.7512) & (2.1957) & (0.9613) & (0.4866) & (0.7034) & (0.8388) & (5.7439) & (0.6372) & (1.1587) & (0.8482) & (1.2447) \\
			$\kappa$                & 0.061$^{a}$    & 0.0158$^{b}$   & 0.0277$^{a}$   & 0.0312$^{b}$   & 0.0247$^{b}$   & 0.0398$^{a}$   & 0.028$^{a}$    & 0.0227$^{a}$   & 0        & 0.0301$^{a}$   & 0.0347$^{a}$   & 0.0389$^{a}$   & 0.0447$^{a}$   \\
			& (0.0118) & (0.0065) & (0.0092) & (0.0131) & (0.0108) & (0.0087) & (0.0082) & (0.0087) & (0.0263) & (0.0087) & (0.0131) & (0.0085) & (0.0136) \\
			
			\midrule
			LogLik                  & 237.3637 & -290.449 & -253.244 & 351.8443 & -159.688 & -153.527 & -65.1614 & -273.156 & -413.025 & -204.637 & -320.496 & 207.4812 & -162.594 \\
			LB 10($\hat\epsilon_t$) & 0.0740   & 0.1876   & 0.1636   & 0.0041   & 0.0000   & 0.0000   & 0.0971   & 0.0018   & 0.0157   & 0.0016   & 0.0001   & 0.0098   & 0.0006   \\
			LB 10($\hat\xi_t$)      & 0.2350   & 0.0000   & 0.1893   & 0.0000   & 0.3390   & 0.0000   & 0.0533   & 0.0001   & 0.0131   & 0.8360   & 0.1795   & 0.0053   & 0.0000\\
			\bottomrule
		\end{tabular}
	\end{adjustbox}
	\caption{Estimates for individual stocks for the AMEM-HAR-J model. Estimated coefficients (robust standard errors in parenthesis) and $p$-value of the Ljung-Box statistics with 10 lags on model residuals, LB 10 ($\hat\epsilon_t$), and jump innovations, LB 10 ($\hat\xi_t$).  The superscripts a, b, and c denote significance at 1\%, 5\%, and 10\% level, respectively. Estimation period: January 3, 2012 - February 4, 2020. Stocks are presented in descending order of market capitalization.
		\label{tab:results_ahar_individual_stocks_b}}
\end{table}

\clearpage

\section{Model Confidence Set of Individual Stocks}\label{app:individual}
\subsection{1-step ahead}
\begin{table}[h!]
	\centering
	\setlength{\tabcolsep}{3pt}
	\renewcommand{\arraystretch}{1.1}
	\begin{adjustbox}{max width=1\linewidth,center}
		\begin{tabular}{lccccccccccccc}
			\toprule
			& (I)      & (II)      & (III)       & (IV)      & (V)      & (VI)      & (VII)      & (VIII)      & (IX)         & (X)                 & (XI)         & (XII)  & (XIII)\\
			\midrule
			MSFT & 0.0251          & 0.0575          & 0.0006          & 0.0026          & 0.0026          & 0.0097          & \textbf{0.2986*} & 0.0006          & 0.0008          & 0.0097          & \textbf{1.0000*} & 0.0008          & 0.0097          \\
			JPM  & 0.0078          & 0.0678          & 0.0678          & \textbf{0.1093*} & 0.0678          & 0.0923          & \textbf{0.1093*} & 0.0678          & \textbf{0.1093*} & \textbf{1.0000*} & \textbf{0.5804*} & 0.0678          & \textbf{0.1093*} \\
			WMT  & 0.0004          & 0.0004          & 0.0000          & 0.0000          & \textbf{0.6650*} & \textbf{0.6821*} & \textbf{0.8138*} & 0.0000          & 0.0000          & \textbf{0.3385*} & \textbf{0.9427*} & \textbf{0.9357*} & \textbf{1.0000*} \\
			UNH  & 0.0760          & 0.0760          & 0.0760          & 0.0760          & 0.0760          & 0.0760          & 0.0760          & 0.0760          & 0.0760          & \textbf{1.0000*} & \textbf{0.9781*} & \textbf{0.9781*} & 0.0760          \\
			PG   & 0.0797          & 0.0855          & 0.0021          & 0.0021          & \textbf{0.7798*} & \textbf{1.0000*} & \textbf{0.7798*} & 0.0010          & 0.0010          & 0.0797          & \textbf{0.2012*} & \textbf{0.2012*} & \textbf{0.2012*} \\
			JNJ  & 0.0994          & 0.0994          & \textbf{0.7065*} & 0.0994          & \textbf{0.7065*} & 0.0994          & \textbf{1.0000*} & \textbf{0.3777*} & 0.0994          & 0.0994          & \textbf{0.7065*} & \textbf{0.3704*} & 0.0994          \\
			HD   & 0.0755          & 0.0755          & \textbf{0.3867*} & \textbf{0.2423*} & \textbf{0.4504*} & \textbf{0.4504*} & \textbf{0.9804*} & \textbf{0.3867*} & \textbf{0.2475*} & \textbf{1.0000*} & \textbf{0.4504*} & \textbf{0.4504*} & \textbf{0.4504*} \\
			KO   & 0.0654          & 0.0580          & 0.0580          & 0.0580          & \textbf{1.0000*} & \textbf{0.3026*} & \textbf{0.8953*} & 0.0580          & 0.0580          & 0.0654          & \textbf{0.4269*} & \textbf{0.4269*} & 0.0654          \\
			MRK  & \textbf{0.1945*} & \textbf{0.1373*} & \textbf{0.1257*} & \textbf{0.1257*} & \textbf{0.2042*} & \textbf{0.2042*} & \textbf{0.2042*} & \textbf{0.1945*} & \textbf{0.1373*} & \textbf{0.2042*} & \textbf{0.8885*} & \textbf{1.0000*} & \textbf{0.2042*} \\
			CVX  & 0.0180          & 0.0180          & 0.0180          & 0.0180          & \textbf{0.1148*} & 0.0370          & \textbf{0.3098*} & 0.0180          & 0.0370          & \textbf{0.3098*} & \textbf{1.0000*} & \textbf{0.3098*} & \textbf{0.1148*} \\
			CRM  & 0.0571          & 0.0571          & \textbf{0.3501*} & \textbf{0.6426*} & 0.0571          & 0.0921          & \textbf{0.9951*} & \textbf{0.3501*} & \textbf{0.6426*} & \textbf{1.0000*} & \textbf{0.9951*} & 0.0571          & 0.0940          \\
			MCD  & 0.0314          & 0.0314          & 0.0314          & 0.0314          & \textbf{0.7464*} & 0.0314          & \textbf{1.0000*} & 0.0041          & 0.0042          & 0.0314          & 0.0314          & 0.0314          & 0.0314          \\
			CSCO & \textbf{0.4232*} & \textbf{0.3780*} & \textbf{0.4997*} & \textbf{0.4997*} & \textbf{0.5754*} & \textbf{0.4997*} & \textbf{0.8939*} & \textbf{0.5754*} & \textbf{0.4997*} & \textbf{0.8939*} & \textbf{1.0000*} & \textbf{0.5754*} & \textbf{0.4997*} \\
			AXP  & \textbf{1.0000*} & \textbf{0.7409*} & 0.0017          & 0.0017          & 0.0440          & 0.0440          & \textbf{0.1900*} & 0.0182          & 0.0131          & 0.0440          & \textbf{0.7409*} & \textbf{0.7409*} & \textbf{0.6961*} \\
			AMGN & 0.0849          & 0.0849          & 0.0849          & 0.0849          & \textbf{0.3010*} & 0.0849          & 0.0849          & 0.0849          & 0.0849          & 0.0849          & \textbf{0.3010*} & \textbf{1.0000*} & 0.0849          \\
			VZ   & 0.0032          & 0.0029          & 0.0081          & 0.0081          & 0.0800          & 0.0591          & \textbf{1.0000*} & 0.0081          & 0.0032          & 0.0081          & 0.0800          & 0.0292          & 0.0081          \\
			CAT  & \textbf{0.4758*} & \textbf{0.4758*} & 0.0023          & 0.0024          & \textbf{0.3056*} & \textbf{0.3056*} & \textbf{1.0000*} & 0.0023          & 0.0024          & 0.0358          & \textbf{0.9392*} & \textbf{0.3056*} & \textbf{0.3056*} \\
			IBM  & \textbf{0.6014*} & \textbf{0.6014*} & \textbf{0.6014*} & \textbf{0.6014*} & \textbf{0.6014*} & \textbf{0.6014*} & \textbf{1.0000*} & \textbf{0.6014*} & \textbf{0.6014*} & \textbf{0.6983*} & \textbf{0.7143*} & \textbf{0.6014*} & \textbf{0.6014*} \\
			DIS  & \textbf{0.1315*} & \textbf{0.1315*} & 0.0127          & 0.0127          & \textbf{0.7288*} & \textbf{0.7288*} & \textbf{0.9669*} & \textbf{0.1315*} & 0.0127          & \textbf{0.7288*} & \textbf{0.7288*} & \textbf{0.7288*} & \textbf{1.0000*} \\
			GS   & \textbf{0.4056*} & \textbf{0.4056*} & \textbf{0.4056*} & \textbf{0.4809*} & \textbf{0.4056*} & \textbf{0.4056*} & \textbf{0.7815*} & \textbf{0.4056*} & \textbf{0.4809*} & \textbf{1.0000*} & \textbf{0.9789*} & \textbf{0.4056*} & \textbf{0.4056*} \\
			HON  & \textbf{0.4629*} & \textbf{0.4629*} & \textbf{0.1259*} & \textbf{0.1259*} & \textbf{0.4629*} & \textbf{0.6977*} & \textbf{0.6977*} & \textbf{0.1259*} & \textbf{0.1259*} & \textbf{0.4629*} & \textbf{1.0000*} & \textbf{0.4629*} & \textbf{0.6977*} \\
			NKE  & \textbf{0.3915*} & \textbf{0.3728*} & 0.0860          & 0.0860          & \textbf{0.5043*} & \textbf{0.5043*} & \textbf{0.5168*} & \textbf{0.1960*} & 0.0860          & \textbf{0.5043*} & \textbf{1.0000*} & \textbf{0.5043*} & \textbf{0.5043*} \\
			BA   & 0.0065          & 0.0065          & 0.0003          & 0.0008          & \textbf{0.2122*} & \textbf{0.2492*} & \textbf{1.0000*} & 0.0003          & 0.0008          & 0.0065          & \textbf{0.2492*} & \textbf{0.1626*} & \textbf{0.2122*} \\
			INTC & 0.0278          & 0.0916          & 0.0253          & 0.0253          & 0.0278          & 0.0369          & \textbf{1.0000*} & 0.0253          & 0.0253          & 0.0369          & \textbf{0.5484*} & 0.0253          & 0.0278          \\
			MMM  & \textbf{0.2147*} & 0.0085          & \textbf{0.9039*} & \textbf{0.3771*} & \textbf{0.2511*} & 0.0094          & \textbf{0.2511*} & \textbf{1.0000*} & \textbf{0.3771*} & \textbf{0.2511*} & \textbf{0.3771*} & \textbf{0.2511*} & 0.0111 \\
			\bottomrule
		\end{tabular}
	\end{adjustbox}
	\caption{Forecast analysis for Individual stocks. Model Confidence Set for the 1-step ahead out-of-sample forecasting performance. Significance level 10\% (best set of models in bold and identified by an asterisk). Loss function: QLike. Estimation period: January 3, 2012 - February 4, 2020. Out-of-sample period: February 5, 2020 - January 10, 2024. Stocks are presented in descending order of market capitalization. The estimated models are (I) HAR, (II) AHAR, (III) MEM, (IV) AMEM, (V) MEM-HAR, (VI) AMEM-HAR, (VII) G-AMEM-HAR, (VIII) MEM-J, (IX) AMEM-J, (X) AMEM(2,1)-J, (XI) G-AMEM-HAR-J, (XII) MEM-HAR-J, (XIII) AMEM-HAR-J.\label{tab:mcs_1_individual} }
\end{table}
\FloatBarrier
\newpage
\subsection{5-step ahead}
\begin{table}[h!]
	\centering
	\setlength{\tabcolsep}{3pt}
	\renewcommand{\arraystretch}{1.1}
	\begin{adjustbox}{max width=1\linewidth,center}
		\begin{tabular}{lccccccccccccc}
			\toprule
			& (I)      & (II)      & (III)       & (IV)      & (V)      & (VI)      & (VII)      & (VIII)      & (IX)         & (X)                 & (XI)         & (XII)  & (XIII)\\
			\midrule                   
			MSFT & \textbf{1.0000*} & \textbf{0.3265*} & 0.0006          & 0.0006          & 0.0917          & 0.0917          & 0.0031          & 0.0002          & 0.0002          & 0.0031          & 0.0031          & 0.0912          & 0.0917          \\
			JPM  & \textbf{0.5990*} & \textbf{0.8478*} & 0.0330          & 0.0380          & \textbf{0.5169*} & \textbf{0.5169*} & \textbf{0.2671*} & \textbf{0.2199*} & \textbf{0.2671*} & \textbf{0.4297*} & \textbf{0.4297*} & \textbf{0.8478*} & \textbf{1.0000*} \\
			WMT  & 0.0489          & 0.0489          & 0.0049          & 0.0059          & \textbf{0.7297*} & \textbf{0.7297*} & 0.0359          & 0.0049          & 0.0059          & 0.0359          & 0.0359          & \textbf{0.9506*} & \textbf{1.0000*} \\
			UNH  & 0.0463          & 0.0463          & 0.0415          & 0.0463          & \textbf{0.1391*} & \textbf{0.1391*} & \textbf{0.1391*} & \textbf{0.1391*} & \textbf{0.1391*} & \textbf{0.9919*} & \textbf{0.1391*} & \textbf{0.9919*} & \textbf{1.0000*} \\
			PG   & 0.0783          & \textbf{0.2289*} & 0.0161          & 0.0363          & \textbf{0.3136*} & \textbf{1.0000*} & \textbf{0.3136*} & 0.0144          & 0.0243          & \textbf{0.2289*} & \textbf{0.2940*} & \textbf{0.3133*} & \textbf{0.3136*} \\
			JNJ  & \textbf{0.9510*} & \textbf{0.9510*} & \textbf{0.4664*} & \textbf{0.5007*} & \textbf{0.9510*} & \textbf{1.0000*} & \textbf{0.9044*} & \textbf{0.3546*} & \textbf{0.4664*} & \textbf{0.6898*} & \textbf{0.8054*} & \textbf{0.9510*} & \textbf{0.9510*} \\
			HD   & \textbf{0.6229*} & \textbf{0.7082*} & 0.0439          & 0.0600          & \textbf{0.7082*} & \textbf{1.0000*} & \textbf{0.7082*} & 0.0600          & \textbf{0.6229*} & \textbf{0.6229*} & \textbf{0.7082*} & \textbf{0.7082*} & \textbf{0.9248*} \\
			KO   & \textbf{0.2630*} & \textbf{0.2630*} & 0.0054          & 0.0054          & \textbf{1.0000*} & \textbf{0.2630*} & 0.0562          & 0.0054          & 0.0051          & 0.0149          & 0.0562          & \textbf{0.6420*} & \textbf{0.2630*} \\
			MRK  & \textbf{0.5998*} & \textbf{0.5998*} & \textbf{0.1988*} & \textbf{0.1988*} & \textbf{0.6744*} & \textbf{0.6744*} & \textbf{0.6744*} & 0.2018          & 0.2018          & \textbf{0.6744*} & \textbf{0.6744*} & \textbf{0.8048*} & \textbf{1.0000*} \\
			CVX  & 0.0681          & 0.0681          & 0.0032          & 0.0032          & \textbf{0.5200*} & \textbf{0.8554*} & 0.0379          & 0.0032          & 0.0032          & 0.0681          & 0.0323          & \textbf{0.5200*} & \textbf{1.0000*} \\
			CRM  & \textbf{0.8523*} & \textbf{1.0000*} & 0.0664          & 0.0664          & \textbf{0.7606*} & \textbf{0.8085*} & \textbf{0.8523*} & 0.0664          & 0.0664          & 0.0249          & \textbf{0.9187*} & \textbf{0.8523*} & \textbf{0.8523*} \\
			MCD  & \textbf{0.1089*} & \textbf{0.1773*} & 0.0233          & 0.0285          & \textbf{0.1773*} & \textbf{0.6806*} & \textbf{0.1773*} & 0.0243          & 0.0296          & 0.0683          & \textbf{0.5594*} & \textbf{0.1773*} & \textbf{1.0000*} \\
			CSCO & \textbf{0.2691*} & \textbf{0.2737*} & \textbf{0.1736*} & \textbf{0.1736*} & \textbf{0.2691*} & \textbf{0.2737*} & \textbf{0.2691*} & \textbf{0.1736*} & \textbf{0.2691*} & \textbf{0.2691*} & \textbf{0.2737*} & \textbf{0.2737*} & \textbf{1.0000*} \\
			AXP  & \textbf{0.4038*} & \textbf{0.4038*} & 0.0428          & 0.0461          & \textbf{0.4038*} & \textbf{0.4038*} & \textbf{0.4038*} & \textbf{0.1140*} & \textbf{0.1244*} & \textbf{0.1140*} & \textbf{0.9916*} & \textbf{0.5318*} & \textbf{1.0000*} \\
			AMGN & 0.0099          & 0.0943          & 0.0153          & 0.0943          & 0.0082          & 0.0099          & 0.0304          & 0.0304          & \textbf{1.0000*} & 0.0082          & 0.0304          & 0.0099          & 0.0099          \\
			VZ   & 0.0670          & 0.0670          & 0.0117          & 0.0118          & \textbf{1.0000*} & \textbf{0.8407*} & \textbf{0.8407*} & 0.0051          & 0.0118          & 0.0118          & \textbf{0.4330*} & \textbf{0.5665*} & \textbf{0.4330*} \\
			CAT  & \textbf{0.4355*} & \textbf{1.0000*} & 0.0013          & 0.0013          & \textbf{0.1954*} & \textbf{0.2037*} & \textbf{0.2037*} & 0.0013          & 0.0012          & 0.0013          & \textbf{0.3771*} & \textbf{0.2037*} & \textbf{0.2037*} \\
			IBM  & \textbf{0.9369*} & \textbf{0.9369*} & \textbf{0.4913*} & \textbf{0.5125*} & \textbf{0.9369*} & \textbf{1.0000*} & \textbf{0.6832*} & \textbf{0.6222*} & \textbf{0.6300*} & \textbf{0.6832*} & \textbf{0.6832*} & \textbf{0.9369*} & \textbf{0.9369*} \\
			DIS  & \textbf{0.4753*} & \textbf{1.0000*} & 0.0205          & 0.0275          & 0.0838          & 0.0838          & \textbf{0.1988*} & 0.0275          & 0.0838          & 0.0205          & \textbf{0.4753*} & \textbf{0.4753*} & \textbf{0.4753*} \\
			GS   & \textbf{0.5253*} & \textbf{0.5253*} & \textbf{0.5253*} & \textbf{0.5253*} & \textbf{0.5253*} & \textbf{0.4771*} & \textbf{0.1374*} & \textbf{1.0000*} & \textbf{0.9856*} & \textbf{0.5253*} & \textbf{0.2127*} & \textbf{0.5253*} & \textbf{0.5253*} \\
			HON  & \textbf{0.9766*} & \textbf{0.9766*} & 0.0577          & 0.0577          & \textbf{0.9766*} & \textbf{0.9766*} & \textbf{0.9766*} & \textbf{0.2161*} & \textbf{0.2161*} & 0.0577          & \textbf{1.0000*} & \textbf{0.9766*} & \textbf{0.9766*} \\
			NKE  & \textbf{0.5424*} & \textbf{0.6493*} & 0.0020          & 0.0020          & \textbf{0.5424*} & \textbf{0.5584*} & \textbf{0.3610*} & 0.0181          & 0.0189          & 0.0739          & \textbf{0.3741*} & \textbf{0.6493*} & \textbf{1.0000*} \\
			BA   & 0.0025          & 0.0026          & 0.0000          & 0.0000          & \textbf{0.4342*} & \textbf{1.0000*} & 0.0385          & 0.0000          & 0.0000          & 0.0001          & 0.0408          & \textbf{0.1065*} & \textbf{0.2056*} \\
			INTC & 0.0506          & \textbf{1.0000*} & 0.0010          & 0.0012          & 0.0447          & 0.0447          & 0.0447          & 0.0012          & 0.0013          & 0.0447          & 0.0447          & 0.0447          & 0.0506          \\
			MMM  & \textbf{0.3042*} & \textbf{0.3238*} & \textbf{0.4004*} & \textbf{0.5450*} & \textbf{0.4004*} & \textbf{0.4004*} & \textbf{0.4004*} & \textbf{0.4004*} & \textbf{1.0000*} & \textbf{0.4004*} & \textbf{0.4004*} & \textbf{0.4004*} & \textbf{0.4004*}\\
			\bottomrule
		\end{tabular}
	\end{adjustbox}
	\caption{Forecast analysis for Individual stocks. Model Confidence Set for the 5-step ahead out-of-sample forecasting performance. Significance level 10\% (best set of models in bold and identified by an asterisk). Loss function: QLike. Estimation period: January 3, 2012 - February 4, 2020. Out-of-sample period: February 5, 2020 - January 10, 2024. Stocks are presented in descending order of market capitalization. The estimated models are (I) HAR, (II) AHAR, (III) MEM, (IV) AMEM, (V) MEM-HAR, (VI) AMEM-HAR, (VII) G-AMEM-HAR, (VIII) MEM-J, (IX) AMEM-J, (X) AMEM(2,1)-J, (XI) G-AMEM-HAR-J, (XII) MEM-HAR-J, (XIII) AMEM-HAR-J.\label{tab:mcs_5_individual} }
\end{table}
\FloatBarrier
\newpage

\subsection{22-step ahead}
\begin{table}[h!]
	\centering
	\setlength{\tabcolsep}{3pt}
	\renewcommand{\arraystretch}{1.1}
	\begin{adjustbox}{max width=1\linewidth,center}
		\begin{tabular}{lccccccccccccc}
			\toprule
			& (I)      & (II)      & (III)       & (IV)      & (V)      & (VI)      & (VII)      & (VIII)      & (IX)         & (X)                 & (XI)         & (XII)  & (XIII)\\
			\midrule
			MSFT & \textbf{1.0000*} & \textbf{0.1626*} & 0.0037          & 0.0030          & 0.0290          & 0.0290          & 0.0216          & 0.0019          & 0.0013          & 0.0016          & 0.0086          & 0.0290          & 0.0290          \\
			JPM  & \textbf{0.8048*} & \textbf{0.8276*} & 0.0268          & 0.0386          & 0.0386          & 0.0386          & \textbf{0.3117*} & 0.0386          & \textbf{0.5618*} & \textbf{1.0000*} & \textbf{0.4149*} & 0.0386          & 0.0386          \\
			WMT  & \textbf{1.0000*} & \textbf{0.2750*} & 0.0946          & 0.0946          & 0.0946          & 0.0963          & 0.0946          & 0.0946          & 0.0946          & 0.0946          & 0.0946          & 0.0946          & 0.0946          \\
			UNH  & 0.0129          & 0.0129          & 0.0074          & 0.0074          & 0.0116          & 0.0116          & 0.0509          & 0.0102          & 0.0116          & 0.0116          & \textbf{1.0000*} & 0.0129          & 0.0129          \\
			PG   & 0.0448          & 0.0526          & 0.0025          & 0.0025          & \textbf{0.4152*} & \textbf{0.9457*} & \textbf{1.0000*} & 0.0025          & 0.0025          & 0.0520          & \textbf{0.9457*} & \textbf{0.4152*} & \textbf{0.9457*} \\
			JNJ  & \textbf{0.9429*} & \textbf{0.9429*} & 0.0869          & 0.0379          & \textbf{0.9429*} & \textbf{0.9429*} & \textbf{0.9429*} & 0.0869          & 0.0379          & 0.0869          & \textbf{0.9429*} & \textbf{1.0000*} & \textbf{0.9429*} \\
			HD   & \textbf{0.3013*} & \textbf{0.9353*} & 0.0135          & 0.0135          & 0.0135          & 0.0135          & \textbf{0.7278*} & 0.0135          & 0.0135          & 0.0135          & \textbf{1.0000*} & 0.0135          & 0.0135          \\
			KO   & \textbf{1.0000*} & \textbf{0.4383*} & 0.0083          & 0.0090          & \textbf{0.4383*} & \textbf{0.3771*} & \textbf{0.2499*} & 0.0090          & 0.0090          & 0.0090          & \textbf{0.3771*} & \textbf{0.4383*} & \textbf{0.3771*} \\
			MRK  & \textbf{0.3479*} & \textbf{0.3479*} & \textbf{0.1054*} & \textbf{0.1191*} & \textbf{0.3479*} & \textbf{0.3479*} & \textbf{0.3479*} & \textbf{0.2617*} & \textbf{0.2763*} & \textbf{1.0000*} & \textbf{0.9759*} & \textbf{0.3479*} & \textbf{0.3479*} \\
			CVX  & \textbf{0.1823*} & \textbf{0.2336*} & 0.0041          & 0.0041          & \textbf{0.4143*} & \textbf{0.4143*} & \textbf{0.2932*} & 0.0041          & 0.0041          & 0.0075          & \textbf{0.4143*} & \textbf{0.4143*} & \textbf{1.0000*} \\
			CRM  & \textbf{0.6819*} & \textbf{0.8438*} & 0.0848          & 0.0848          & \textbf{0.6816*} & \textbf{0.6819*} & \textbf{0.6819*} & 0.0884          & 0.0884          & \textbf{0.1348*} & \textbf{1.0000*} & \textbf{0.6819*} & \textbf{0.8438*} \\
			MCD  & 0.0053          & 0.0053          & 0.0012          & 0.0053          & 0.0284          & 0.0769          & 0.0769          & 0.0012          & 0.0053          & 0.0053          & \textbf{1.0000*} & 0.0284          & 0.0769          \\
			CSCO & \textbf{0.3887*} & \textbf{0.3887*} & 0.0501          & 0.0501          & \textbf{0.3887*} & \textbf{0.3887*} & \textbf{0.3887*} & 0.0574          & 0.0654          & \textbf{0.3887*} & \textbf{1.0000*} & \textbf{0.3887*} & \textbf{0.3887*} \\
			AXP  & 0.0227          & 0.0227          & 0.0219          & 0.0219          & 0.0227          & 0.0219          & 0.0227          & 0.0219          & 0.0219          & 0.0219          & \textbf{1.0000*} & 0.0227          & 0.0227          \\
			AMGN & 0.0109          & 0.0109          & 0.0435          & 0.0769          & 0.0105          & 0.0105          & 0.0248          & 0.0769          & \textbf{1.0000*} & 0.0248          & 0.0248          & 0.0109          & 0.0109          \\
			VZ   & \textbf{0.2405*} & \textbf{0.2405*} & 0.0138          & 0.0230          & \textbf{1.0000*} & \textbf{0.9278*} & \textbf{0.9278*} & 0.0138          & 0.0139          & 0.0952          & \textbf{0.8892*} & \textbf{0.2405*} & \textbf{0.5584*} \\
			CAT  & \textbf{0.3311*} & \textbf{1.0000*} & 0.0205          & 0.0205          & \textbf{0.3014*} & \textbf{0.3302*} & \textbf{0.3311*} & 0.0205          & 0.0205          & 0.0210          & \textbf{0.3311*} & \textbf{0.3302*} & \textbf{0.3311*} \\
			IBM  & \textbf{0.7425*} & \textbf{0.7425*} & 0.0984          & 0.0984          & 0.0984          & 0.0984          & \textbf{1.0000*} & 0.0984          & 0.0984          & 0.0984          & \textbf{0.7945*} & 0.0984          & 0.0984          \\
			DIS  & 0.0772          & \textbf{0.7515*} & 0.0160          & 0.0189          & 0.0189          & 0.0189          & 0.0210          & 0.0189          & 0.0189          & 0.0210          & \textbf{1.0000*} & 0.0210          & 0.0210          \\
			GS   & \textbf{0.1719*} & \textbf{0.1151*} & \textbf{0.1719*} & \textbf{0.1719*} & \textbf{0.1075*} & 0.0260          & \textbf{0.1151*} & \textbf{1.0000*} & \textbf{0.1719*} & \textbf{0.1719*} & \textbf{0.1151*} & \textbf{0.1075*} & 0.0260          \\
			HON  & 0.0229          & 0.0847          & 0.0006          & 0.0006          & 0.0229          & 0.0229          & \textbf{0.8057*} & 0.0006          & 0.0229          & 0.0229          & \textbf{1.0000*} & 0.0239          & 0.0239          \\
			NKE  & 0.0283          & \textbf{0.3253*} & 0.0007          & 0.0007          & 0.0253          & 0.0253          & \textbf{0.3253*} & 0.0009          & 0.0057          & 0.0253          & \textbf{1.0000*} & 0.0253          & 0.0283          \\
			BA   & 0.0005          & 0.0005          & 0.0000          & 0.0000          & \textbf{0.4967*} & \textbf{1.0000*} & \textbf{0.1120*} & 0.0000          & 0.0000          & 0.0000          & \textbf{0.2428*} & \textbf{0.2428*} & \textbf{0.4967*} \\
			INTC & \textbf{0.9690*} & \textbf{1.0000*} & 0.0226          & 0.0226          & \textbf{0.9376*} & \textbf{0.9662*} & \textbf{0.9376*} & 0.0226          & 0.0226          & \textbf{0.3620*} & \textbf{0.9662*} & \textbf{0.9690*} & \textbf{0.9690*} \\
			MMM  & 0.0002          & 0.0002          & 0.0002          & 0.0004          & 0.0002          & 0.0002          & 0.0002          & 0.0002          & \textbf{1.0000*} & 0.0004          & 0.0002          & 0.0002          & 0.0002\\
			\bottomrule
		\end{tabular}
	\end{adjustbox}
	\caption{Forecast analysis for Individual stocks. Model Confidence Set for the 22-step ahead out-of-sample forecasting performance. Significance level 10\% (best set of models in bold and identified by an asterisk). Loss function: QLike. Estimation period: January 3, 2012 - February 4, 2020. Out-of-sample period: February 5, 2020 - January 10, 2024. Stocks are presented in descending order of market capitalization. The estimated models are (I) HAR, (II) AHAR, (III) MEM, (IV) AMEM, (V) MEM-HAR, (VI) AMEM-HAR, (VII) G-AMEM-HAR, (VIII) MEM-J, (IX) AMEM-J, (X) AMEM(2,1)-J, (XI) G-AMEM-HAR-J, (XII) MEM-HAR-J, (XIII) AMEM-HAR-J.\label{tab:mcs_22_individual} }
\end{table}
\FloatBarrier

\clearpage

\section{Full-sample analysis of IlliQaR of Individual Stocks}\label{App:berkowitz_individual_full}
\FloatBarrier
\subsection{IlliQaR at 1\%}
\begin{table}[h!]
	\centering
	\setlength{\tabcolsep}{3pt}
	\renewcommand{\arraystretch}{1.1}
	\begin{adjustbox}{max width=1\linewidth,center}
		\begin{tabular}{lccccccccccccc}
			\toprule
			& (I)      & (II)      & (III)       & (IV)      & (V)      & (VI)      & (VII)      & (VIII)      & (IX)         & (X)                 & (XI)         & (XII)  & (XIII)\\
			\midrule
			MSFT & 0.0178          & 0.0157          & \textbf{0.0715} & \textbf{0.0546} & 0.0114          & \textbf{0.0515} & \textbf{0.0825} & 0.0025          & 0.0027          & 0.0052          & 0.0008          & 0.0014          & 0.0016          \\
			JPM  & 0.0303          & 0.0148          & \textbf{0.8282} & \textbf{0.6045} & \textbf{0.8444} & \textbf{0.5615} & \textbf{0.6959} & \textbf{0.3083} & \textbf{0.2829} & \textbf{0.1300} & \textbf{0.1460} & \textbf{0.1843} & \textbf{0.1434} \\
			WMT  & 0.0048          & 0.0077          & \textbf{0.8030} & \textbf{0.6447} & \textbf{0.3982} & \textbf{0.3314} & \textbf{0.3275} & 0.0127          & 0.0182          & 0.0135          & 0.0124          & 0.0133          & 0.0134          \\
			UNH  & 0.0020          & 0.0020          & \textbf{0.9606} & \textbf{0.5918} & \textbf{0.6720} & \textbf{0.9038} & \textbf{0.7919} & \textbf{0.4286} & \textbf{0.2470} & \textbf{0.4411} & \textbf{0.1739} & \textbf{0.1020} & \textbf{0.1000} \\
			PG   & 0.0007          & 0.0019          & \textbf{0.2316} & \textbf{0.2217} & \textbf{0.7494} & \textbf{0.2680} & \textbf{0.6379} & 0.0148          & 0.0253          & 0.0433          & 0.0358          & 0.0117          & 0.0355          \\
			JNJ  & 0.0015          & 0.0062          & \textbf{0.1895} & \textbf{0.2137} & \textbf{0.2824} & \textbf{0.1474} & \textbf{0.1799} & \textbf{0.3689} & \textbf{0.6062} & \textbf{0.3648} & \textbf{0.6390} & \textbf{0.3867} & \textbf{0.2461} \\
			HD   & \textbf{0.1992} & \textbf{0.2032} & \textbf{0.7226} & \textbf{0.7555} & \textbf{0.7671} & \textbf{0.7492} & \textbf{0.7420} & \textbf{0.3314} & \textbf{0.2721} & \textbf{0.3589} & \textbf{0.1907} & \textbf{0.2423} & \textbf{0.2347} \\
			KO   & \textbf{0.6910} & \textbf{0.6074} & \textbf{0.3241} & \textbf{0.2270} & \textbf{0.3202} & \textbf{0.7782} & \textbf{0.5383} & \textbf{0.0769} & \textbf{0.0809} & \textbf{0.1015} & 0.0468          & \textbf{0.0628} & \textbf{0.0521} \\
			MRK  & \textbf{0.0626} & \textbf{0.1885} & \textbf{0.4455} & \textbf{0.2099} & 0.0264          & \textbf{0.0503} & \textbf{0.0505} & \textbf{0.2504} & \textbf{0.2444} & \textbf{0.2742} & \textbf{0.1723} & \textbf{0.2444} & \textbf{0.2442} \\
			CVX  & 0.0000          & 0.0000          & 0.0099          & 0.0278          & \textbf{0.0838} & 0.0177          & 0.0093          & 0.0380          & \textbf{0.0606} & 0.0390          & \textbf{0.0648} & \textbf{0.0698} & 0.0409          \\
			CRM  & \textbf{0.7528} & \textbf{0.6970} & \textbf{0.1629} & \textbf{0.1044} & \textbf{0.1953} & \textbf{0.0660} & \textbf{0.0681} & 0.0131          & 0.0194          & 0.0162          & 0.0043          & 0.0013          & 0.0009          \\
			MCD  & 0.0010          & 0.0003          & \textbf{0.4865} & \textbf{0.5644} & \textbf{0.2745} & \textbf{0.3008} & \textbf{0.3382} & \textbf{0.8626} & \textbf{0.6662} & \textbf{0.7171} & \textbf{0.6323} & \textbf{0.5355} & \textbf{0.1701} \\
			CSCO & \textbf{0.0707} & \textbf{0.0721} & 0.0014          & 0.0008          & 0.0011          & 0.0037          & 0.0038          & \textbf{0.4274} & \textbf{0.4182} & \textbf{0.4526} & \textbf{0.3613} & \textbf{0.3991} & \textbf{0.3482} \\
			AXP  & \textbf{0.2244} & \textbf{0.1478} & \textbf{0.2520} & \textbf{0.2431} & \textbf{0.3524} & \textbf{0.3147} & \textbf{0.3145} & 0.0241          & 0.0121          & 0.0354          & \textbf{0.0846} & \textbf{0.0772} & \textbf{0.0762} \\
			AMGN & 0.0007          & 0.0007          & \textbf{0.2271} & \textbf{0.3312} & \textbf{0.4429} & \textbf{0.4242} & \textbf{0.2064} & \textbf{0.0908} & \textbf{0.0569} & \textbf{0.1055} & \textbf{0.0911} & \textbf{0.1546} & \textbf{0.1523} \\
			VZ   & 0.0207          & 0.0209          & \textbf{0.3194} & 0.0285          & 0.0264          & \textbf{0.1294} & \textbf{0.1774} & \textbf{0.5583} & \textbf{0.4580} & \textbf{0.8450} & \textbf{0.6521} & \textbf{0.5288} & \textbf{0.5786} \\
			CAT  & 0.0319          & 0.0078          & \textbf{0.2425} & \textbf{0.1684} & \textbf{0.1099} & 0.0221          & \textbf{0.1046} & \textbf{0.0675} & \textbf{0.0895} & \textbf{0.1073} & 0.0313          & 0.0152          & 0.0269          \\
			IBM  & 0.0050          & 0.0051          & \textbf{0.7851} & \textbf{0.8539} & \textbf{0.7414} & \textbf{0.7758} & \textbf{0.6702} & 0.0056          & 0.0063          & 0.0043          & 0.0017          & 0.0037          & 0.0035          \\
			DIS  & 0.0000          & 0.0000          & \textbf{0.3426} & \textbf{0.4664} & \textbf{0.3371} & \textbf{0.3289} & \textbf{0.1619} & \textbf{0.0663} & \textbf{0.0694} & \textbf{0.0539} & 0.0224          & 0.0215          & 0.0226          \\
			GS   & \textbf{0.4175} & \textbf{0.2465} & 0.0079          & 0.0153          & \textbf{0.1848} & \textbf{0.2589} & \textbf{0.3817} & 0.0213          & 0.0369          & 0.0220          & \textbf{0.0694} & \textbf{0.0836} & \textbf{0.1027} \\
			HON  & \textbf{0.1906} & \textbf{0.2056} & \textbf{0.8162} & \textbf{0.8185} & \textbf{0.7201} & \textbf{0.7159} & \textbf{0.7176} & \textbf{0.8161} & \textbf{0.8185} & \textbf{0.8128} & \textbf{0.7201} & \textbf{0.7159} & \textbf{0.8907} \\
			NKE  & 0.0063          & 0.0074          & \textbf{0.5577} & \textbf{0.5582} & \textbf{0.7147} & \textbf{0.5440} & \textbf{0.5370} & 0.0034          & 0.0015          & 0.0043          & 0.0378          & 0.0135          & 0.0127          \\
			BA   & 0.0000          & 0.0000          & \textbf{0.8823} & \textbf{0.9901} & \textbf{0.2326} & \textbf{0.8205} & \textbf{0.5584} & \textbf{0.3328} & \textbf{0.3097} & \textbf{0.3475} & \textbf{0.4155} & \textbf{0.5322} & \textbf{0.6688} \\
			INTC & \textbf{0.1537} & \textbf{0.1519} & \textbf{0.0729} & \textbf{0.0785} & \textbf{0.0684} & \textbf{0.0670} & \textbf{0.0636} & 0.0003          & 0.0007          & 0.0006          & 0.0006          & 0.0016          & 0.0012          \\
			MMM  & 0.0312          & \textbf{0.0739} & \textbf{0.3478} & \textbf{0.3218} & \textbf{0.2202} & \textbf{0.1247} & \textbf{0.1816} & \textbf{0.3829} & \textbf{0.4101} & \textbf{0.3940} & \textbf{0.4733} & \textbf{0.5872} & \textbf{0.5880}\\
			\bottomrule
		\end{tabular}
	\end{adjustbox}
	\caption{Table reports the p-values of the Berkowitz test for the full sample IlliQaR($p$) for $p=1\%$. Full sample period: January 3, 2012 - January 10, 2024. Stocks are presented in descending order of market capitalization. The estimated models are (I) HAR, (II) AHAR, (III) MEM, (IV) AMEM, (V) MEM-HAR, (VI) AMEM-HAR, (VII) G-AMEM-HAR, (VIII) MEM-J, (IX) AMEM-J, (X) AMEM(2,1)-J, (XI) G-AMEM-HAR-J, (XII) MEM-HAR-J, (XIII) AMEM-HAR-J.\label{tab:density_forecast_full_1_individual} }
\end{table}
\FloatBarrier
\clearpage

\FloatBarrier
\subsection{IlliQaR at 5\%}
\begin{table}[h!]
	\centering
	\setlength{\tabcolsep}{3pt}
	\renewcommand{\arraystretch}{1.1}
	\begin{adjustbox}{max width=1\linewidth,center}
		\begin{tabular}{lccccccccccccc}
			\toprule
			& (I)      & (II)      & (III)       & (IV)      & (V)      & (VI)      & (VII)      & (VIII)      & (IX)         & (X)                 & (XI)         & (XII)  & (XIII)\\
			\midrule
			MSFT & 0.0022          & 0.0060          & 0.0323          & 0.0375          & 0.0430          & \textbf{0.0676} & \textbf{0.1062} & 0.0027          & 0.0034          & 0.0063          & 0.0040          & 0.0103          & 0.0182          \\
			JPM  & 0.0053          & 0.0028          & \textbf{0.2386} & \textbf{0.2536} & \textbf{0.3004} & \textbf{0.2227} & \textbf{0.3968} & \textbf{0.6097} & \textbf{0.5232} & \textbf{0.6887} & \textbf{0.2923} & \textbf{0.3644} & \textbf{0.3734} \\
			WMT  & 0.0005          & 0.0002          & 0.0137          & 0.0245          & \textbf{0.1161} & \textbf{0.0846} & \textbf{0.1392} & 0.0354          & 0.0376          & 0.0423          & 0.0202          & 0.0228          & 0.0207          \\
			UNH  & 0.0000          & 0.0000          & \textbf{0.8722} & \textbf{0.9430} & \textbf{0.4311} & \textbf{0.2669} & \textbf{0.5872} & \textbf{0.8596} & \textbf{0.7917} & \textbf{0.8858} & \textbf{0.7795} & \textbf{0.7222} & \textbf{0.7127} \\
			PG   & 0.0000          & 0.0000          & \textbf{0.4413} & \textbf{0.2666} & \textbf{0.4742} & \textbf{0.1002} & \textbf{0.3067} & \textbf{0.6220} & \textbf{0.4764} & \textbf{0.8933} & \textbf{0.5941} & \textbf{0.4735} & \textbf{0.6247} \\
			JNJ  & 0.0005          & 0.0002          & \textbf{0.0771} & \textbf{0.2098} & \textbf{0.2380} & 0.0251          & \textbf{0.1484} & \textbf{0.9300} & \textbf{0.9110} & \textbf{0.9301} & \textbf{0.9027} & \textbf{0.6675} & \textbf{0.8632} \\
			HD   & \textbf{0.0665} & \textbf{0.0672} & \textbf{0.1651} & \textbf{0.2310} & \textbf{0.1752} & \textbf{0.1057} & \textbf{0.1055} & \textbf{0.1747} & \textbf{0.2561} & \textbf{0.2819} & \textbf{0.2576} & \textbf{0.2538} & \textbf{0.2344} \\
			KO   & 0.0200          & 0.0282          & \textbf{0.3677} & \textbf{0.3927} & \textbf{0.4612} & \textbf{0.3840} & \textbf{0.4764} & 0.0435          & \textbf{0.0641} & \textbf{0.0669} & \textbf{0.0816} & \textbf{0.1761} & \textbf{0.1992} \\
			MRK  & 0.0012          & 0.0018          & 0.0426          & \textbf{0.0721} & \textbf{0.1894} & \textbf{0.1600} & \textbf{0.1601} & \textbf{0.2059} & \textbf{0.2051} & \textbf{0.2400} & \textbf{0.2178} & \textbf{0.2284} & \textbf{0.2241} \\
			CVX  & 0.0000          & 0.0000          & \textbf{0.0781} & \textbf{0.1215} & \textbf{0.1061} & \textbf{0.2057} & \textbf{0.1000} & 0.0082          & \textbf{0.0586} & 0.0114          & \textbf{0.0872} & \textbf{0.0760} & \textbf{0.0981} \\
			CRM  & 0.0087          & 0.0169          & \textbf{0.5437} & \textbf{0.5971} & \textbf{0.2688} & \textbf{0.4873} & \textbf{0.4960} & \textbf{0.2616} & \textbf{0.3234} & \textbf{0.3187} & \textbf{0.1835} & \textbf{0.2819} & \textbf{0.3130} \\
			MCD  & 0.0104          & 0.0060          & \textbf{0.6857} & \textbf{0.8178} & \textbf{0.6681} & \textbf{0.6411} & \textbf{0.6513} & \textbf{0.8370} & \textbf{0.8161} & \textbf{0.7399} & \textbf{0.9096} & \textbf{0.9686} & \textbf{0.5284} \\
			CSCO & 0.0052          & 0.0053          & 0.0034          & 0.0124          & 0.0066          & 0.0007          & 0.0004          & \textbf{0.6337} & \textbf{0.6285} & \textbf{0.7221} & \textbf{0.6127} & \textbf{0.6688} & \textbf{0.6810} \\
			AXP  & 0.0133          & 0.0159          & \textbf{0.5286} & \textbf{0.6025} & \textbf{0.3849} & \textbf{0.3882} & \textbf{0.3803} & \textbf{0.2999} & \textbf{0.2743} & \textbf{0.2993} & \textbf{0.3201} & \textbf{0.2889} & \textbf{0.2406} \\
			AMGN & 0.0013          & 0.0009          & \textbf{0.2882} & \textbf{0.2667} & \textbf{0.3760} & \textbf{0.4849} & \textbf{0.4952} & 0.0408          & 0.0386          & \textbf{0.0647} & \textbf{0.1488} & \textbf{0.1443} & \textbf{0.1748} \\
			VZ   & 0.0019          & 0.0019          & \textbf{0.2277} & \textbf{0.3334} & \textbf{0.4591} & 0.0129          & 0.0124          & \textbf{0.5693} & \textbf{0.4710} & \textbf{0.7365} & \textbf{0.3276} & \textbf{0.1620} & \textbf{0.1812} \\
			CAT  & 0.0186          & 0.0155          & 0.0409          & 0.0475          & 0.0087          & 0.0048          & 0.0078          & 0.0124          & 0.0156          & 0.0213          & 0.0025          & 0.0013          & 0.0019          \\
			IBM  & 0.0007          & 0.0011          & \textbf{0.1345} & \textbf{0.1128} & 0.0248          & 0.0307          & 0.0392          & \textbf{0.1076} & \textbf{0.0934} & \textbf{0.1274} & \textbf{0.0630} & \textbf{0.0592} & 0.0478          \\
			DIS  & 0.0000          & 0.0000          & 0.0280          & 0.0455          & 0.0272          & 0.0236          & 0.0146          & \textbf{0.0729} & \textbf{0.0735} & \textbf{0.0651} & 0.0240          & 0.0191          & 0.0165          \\
			GS   & 0.0318          & 0.0380          & \textbf{0.1427} & \textbf{0.2072} & \textbf{0.1641} & \textbf{0.1074} & \textbf{0.2605} & 0.0156          & 0.0107          & 0.0185          & 0.0366          & \textbf{0.0565} & \textbf{0.0761} \\
			HON  & 0.0000          & 0.0000          & \textbf{0.5794} & \textbf{0.5300} & \textbf{0.5043} & \textbf{0.5371} & \textbf{0.4747} & \textbf{0.5839} & \textbf{0.5351} & \textbf{0.4896} & \textbf{0.5043} & \textbf{0.5371} & \textbf{0.4902} \\
			NKE  & 0.0000          & 0.0000          & \textbf{0.5629} & \textbf{0.6288} & \textbf{0.1864} & \textbf{0.3477} & \textbf{0.3480} & \textbf{0.7826} & \textbf{0.7817} & \textbf{0.8363} & \textbf{0.2595} & \textbf{0.2545} & \textbf{0.2547} \\
			BA   & 0.0000          & 0.0000          & \textbf{0.6208} & \textbf{0.8599} & \textbf{0.6107} & \textbf{0.3783} & \textbf{0.1553} & \textbf{0.4206} & \textbf{0.3673} & \textbf{0.4952} & \textbf{0.2129} & \textbf{0.3794} & \textbf{0.3382} \\
			INTC & \textbf{0.0587} & \textbf{0.0563} & 0.0010          & 0.0013          & 0.0009          & 0.0010          & 0.0024          & 0.0008          & 0.0010          & 0.0021          & 0.0005          & 0.0007          & 0.0008          \\
			MMM  & 0.0002          & 0.0001          & \textbf{0.2147} & \textbf{0.3784} & \textbf{0.0876} & \textbf{0.1414} & \textbf{0.1805} & \textbf{0.8492} & \textbf{0.8770} & \textbf{0.8720} & \textbf{0.7772} & \textbf{0.8076} & \textbf{0.7999}\\
			\bottomrule
		\end{tabular}
	\end{adjustbox}
	\caption{Table reports the p-values of the Berkowitz test for the full sample IlliQaR($p$) at $p=5\%$. Full sample period: January 3, 2012 - January 10, 2024. Stocks are presented in descending order of market capitalization. The estimated models are (I) HAR, (II) AHAR, (III) MEM, (IV) AMEM, (V) MEM-HAR, (VI) AMEM-HAR, (VII) G-AMEM-HAR, (VIII) MEM-J, (IX) AMEM-J, (X) AMEM(2,1)-J, (XI) G-AMEM-HAR-J, (XII) MEM-HAR-J, (XIII) AMEM-HAR-J.\label{tab:density_forecast_full_5_individual} }
\end{table}
\cleardoublepage

\FloatBarrier
\subsection{IlliQaR at 10\%}
\begin{table}[h!]
	\centering
	\setlength{\tabcolsep}{3pt}
	\renewcommand{\arraystretch}{1.1}
	\begin{adjustbox}{max width=1\linewidth,center}
		\begin{tabular}{lccccccccccccc}
			\toprule
			& (I)      & (II)      & (III)       & (IV)      & (V)      & (VI)      & (VII)      & (VIII)      & (IX)         & (X)                 & (XI)         & (XII)  & (XIII)\\
			\midrule
			MSFT & 0.0016          & 0.0023          & \textbf{0.0630} & \textbf{0.0808} & \textbf{0.0845} & \textbf{0.1079} & \textbf{0.1169} & 0.0311          & 0.0397          & \textbf{0.0657} & 0.0430          & 0.0277          & 0.0368          \\
			JPM  & 0.0000          & 0.0000          & \textbf{0.3766} & \textbf{0.3031} & \textbf{0.1693} & \textbf{0.2062} & \textbf{0.1093} & \textbf{0.6358} & \textbf{0.5222} & \textbf{0.7448} & \textbf{0.3193} & \textbf{0.3126} & \textbf{0.2088} \\
			WMT  & 0.0002          & 0.0000          & 0.0399          & \textbf{0.0768} & \textbf{0.0511} & \textbf{0.0884} & \textbf{0.0880} & 0.0308          & 0.0419          & 0.0357          & \textbf{0.0517} & \textbf{0.0584} & \textbf{0.0678} \\
			UNH  & 0.0000          & 0.0000          & \textbf{0.6480} & \textbf{0.8641} & \textbf{0.1414} & \textbf{0.2323} & \textbf{0.2116} & \textbf{0.9099} & \textbf{0.8969} & \textbf{0.9946} & \textbf{0.6123} & \textbf{0.5870} & \textbf{0.4979} \\
			PG   & 0.0000          & 0.0000          & \textbf{0.1201} & \textbf{0.0869} & \textbf{0.2256} & \textbf{0.4237} & \textbf{0.1845} & \textbf{0.3563} & \textbf{0.2440} & \textbf{0.6019} & \textbf{0.4476} & \textbf{0.5577} & \textbf{0.4322} \\
			JNJ  & 0.0001          & 0.0002          & \textbf{0.3459} & \textbf{0.1163} & \textbf{0.2868} & \textbf{0.2449} & \textbf{0.1733} & \textbf{0.7623} & \textbf{0.8828} & \textbf{0.7826} & \textbf{0.9146} & \textbf{0.8775} & \textbf{0.8900} \\
			HD   & 0.0315          & 0.0440          & \textbf{0.2206} & \textbf{0.2234} & \textbf{0.3519} & \textbf{0.4052} & \textbf{0.4586} & \textbf{0.2523} & \textbf{0.2536} & \textbf{0.4591} & \textbf{0.2156} & \textbf{0.2563} & \textbf{0.2529} \\
			KO   & 0.0042          & 0.0069          & \textbf{0.2056} & \textbf{0.2510} & \textbf{0.1230} & \textbf{0.2304} & \textbf{0.2876} & \textbf{0.1687} & \textbf{0.1712} & \textbf{0.2566} & \textbf{0.1415} & \textbf{0.1945} & \textbf{0.1951} \\
			MRK  & 0.0034          & 0.0023          & \textbf{0.4593} & \textbf{0.4481} & \textbf{0.3751} & \textbf{0.2495} & \textbf{0.2302} & \textbf{0.2150} & \textbf{0.1117} & \textbf{0.1562} & \textbf{0.2198} & \textbf{0.2325} & \textbf{0.2217} \\
			CVX  & 0.0000          & 0.0000          & \textbf{0.1743} & \textbf{0.1722} & \textbf{0.0662} & \textbf{0.2070} & \textbf{0.1799} & \textbf{0.1453} & \textbf{0.1772} & \textbf{0.2161} & 0.0144          & \textbf{0.1164} & \textbf{0.0726} \\
			CRM  & 0.0320          & 0.0123          & \textbf{0.2839} & \textbf{0.3459} & \textbf{0.3430} & \textbf{0.3098} & \textbf{0.3369} & \textbf{0.3377} & \textbf{0.3973} & \textbf{0.4267} & \textbf{0.2690} & \textbf{0.3398} & \textbf{0.3260} \\
			MCD  & 0.0011          & 0.0008          & \textbf{0.7363} & \textbf{0.6816} & \textbf{0.2567} & \textbf{0.2297} & \textbf{0.2986} & \textbf{0.9748} & \textbf{0.9834} & \textbf{0.9339} & \textbf{0.8778} & \textbf{0.9076} & \textbf{0.8591} \\
			CSCO & 0.0000          & 0.0000          & 0.0199          & 0.0127          & 0.0270          & 0.0132          & 0.0104          & \textbf{0.6097} & \textbf{0.6106} & \textbf{0.6678} & \textbf{0.5191} & \textbf{0.6227} & \textbf{0.6421} \\
			AXP  & 0.0062          & 0.0060          & \textbf{0.5137} & \textbf{0.5597} & \textbf{0.3541} & \textbf{0.3306} & \textbf{0.3333} & \textbf{0.3572} & \textbf{0.3534} & \textbf{0.3978} & \textbf{0.3580} & \textbf{0.2987} & \textbf{0.3197} \\
			AMGN & 0.0005          & 0.0010          & \textbf{0.3209} & \textbf{0.3086} & \textbf{0.3151} & \textbf{0.4926} & \textbf{0.4203} & 0.0466          & \textbf{0.0759} & \textbf{0.0587} & \textbf{0.0894} & \textbf{0.1504} & \textbf{0.1787} \\
			VZ   & 0.0000          & 0.0000          & \textbf{0.1612} & \textbf{0.1858} & \textbf{0.1119} & \textbf{0.1719} & \textbf{0.1325} & \textbf{0.5405} & \textbf{0.5450} & \textbf{0.8680} & \textbf{0.2992} & \textbf{0.2718} & \textbf{0.2796} \\
			CAT  & 0.0004          & 0.0003          & 0.0364          & 0.0414          & 0.0055          & 0.0050          & 0.0073          & 0.0130          & 0.0103          & 0.0179          & 0.0014          & 0.0014          & 0.0023          \\
			IBM  & 0.0000          & 0.0000          & \textbf{0.0561} & 0.0284          & 0.0166          & 0.0078          & 0.0092          & \textbf{0.1247} & \textbf{0.1044} & \textbf{0.1584} & \textbf{0.0540} & \textbf{0.0603} & \textbf{0.0562} \\
			DIS  & 0.0000          & 0.0000          & \textbf{0.0677} & 0.0349          & 0.0121          & 0.0111          & 0.0084          & 0.0402          & 0.0340          & 0.0373          & 0.0338          & 0.0336          & 0.0254          \\
			GS   & 0.0037          & 0.0040          & \textbf{0.1264} & \textbf{0.2231} & \textbf{0.0840} & \textbf{0.0775} & \textbf{0.1885} & 0.0064          & 0.0115          & 0.0107          & 0.0143          & 0.0057          & 0.0347          \\
			HON  & 0.0000          & 0.0000          & \textbf{0.6604} & \textbf{0.6485} & \textbf{0.5830} & \textbf{0.2781} & \textbf{0.3918} & \textbf{0.6668} & \textbf{0.6426} & \textbf{0.5476} & \textbf{0.5830} & \textbf{0.2781} & \textbf{0.4034} \\
			NKE  & 0.0000          & 0.0000          & \textbf{0.4283} & \textbf{0.2858} & \textbf{0.3534} & \textbf{0.3765} & \textbf{0.3780} & \textbf{0.6714} & \textbf{0.6770} & \textbf{0.7343} & \textbf{0.1791} & \textbf{0.1550} & \textbf{0.1564} \\
			BA   & 0.0000          & 0.0000          & \textbf{0.1534} & \textbf{0.3843} & \textbf{0.1902} & \textbf{0.3828} & \textbf{0.3621} & \textbf{0.3154} & \textbf{0.4765} & \textbf{0.3543} & \textbf{0.3255} & \textbf{0.3439} & \textbf{0.3693} \\
			INTC & \textbf{0.0721} & \textbf{0.0847} & 0.0006          & 0.0005          & 0.0002          & 0.0004          & 0.0003          & 0.0005          & 0.0006          & 0.0015          & 0.0002          & 0.0002          & 0.0002          \\
			MMM  & 0.0004          & 0.0017          & \textbf{0.5004} & \textbf{0.4956} & \textbf{0.4049} & \textbf{0.3590} & \textbf{0.4123} & \textbf{0.9387} & \textbf{0.8337} & \textbf{0.9565} & \textbf{0.6397} & \textbf{0.8430} & \textbf{0.7349}\\
			\bottomrule
		\end{tabular}
	\end{adjustbox}
	\caption{Table reports the p-values of the Berkowitz test for the full sample  IlliQaR($p$) at $p=10\%$. Full sample period: January 3, 2012 - January 10, 2024.  Stocks are presented in descending order of market capitalization. The estimated models are (I) HAR, (II) AHAR, (III) MEM, (IV) AMEM, (V) MEM-HAR, (VI) AMEM-HAR, (VII) G-AMEM-HAR, (VIII) MEM-J, (IX) AMEM-J, (X) AMEM(2,1)-J, (XI) G-AMEM-HAR-J, (XII) MEM-HAR-J, (XIII) AMEM-HAR-J.\label{tab:density_forecast_full_10_individual} }
\end{table}
\FloatBarrier
\newpage

\section{Out-of-sample analysis of IlliQaR of Individual Stocks}\label{App:berkowitz_individual_out}
\subsection{IlliQaR at 1\%}
\begin{table}[h!]
	\centering
	\setlength{\tabcolsep}{3pt}
	\renewcommand{\arraystretch}{1.1}
	\begin{adjustbox}{max width=1\linewidth,center}
		\begin{tabular}{lccccccccccccc}
			\toprule
			& (I)      & (II)      & (III)       & (IV)      & (V)      & (VI)      & (VII)      & (VIII)      & (IX)         & (X)                 & (XI)         & (XII)  & (XIII)\\
			\midrule
			MSFT & \textbf{0.4504} & \textbf{0.5908} & 0.0003          & 0.0014          & 0.0037          & \textbf{0.0527} & 0.0423          & 0.0000          & 0.0000          & 0.0005          & 0.0003          & 0.0083          & 0.0050          \\
			JPM  & 0.0007          & 0.0004          & \textbf{0.1990} & \textbf{0.4552} & \textbf{0.1263} & \textbf{0.1428} & \textbf{0.2401} & \textbf{0.0985} & \textbf{0.4553} & \textbf{0.7643} & \textbf{0.1384} & \textbf{0.0814} & \textbf{0.1650} \\
			WMT  & 0.0109          & 0.0057          & \textbf{0.5014} & \textbf{0.2428} & \textbf{0.4580} & \textbf{0.6072} & \textbf{0.5972} & \textbf{0.3661} & \textbf{0.3876} & \textbf{0.4637} & \textbf{0.3456} & \textbf{0.2773} & \textbf{0.2800} \\
			UNH  & 0.0006          & 0.0004          & \textbf{0.7292} & \textbf{0.6567} & \textbf{0.3722} & \textbf{0.9449} & \textbf{0.6676} & \textbf{0.4455} & \textbf{0.2248} & \textbf{0.4379} & \textbf{0.2501} & \textbf{0.6265} & \textbf{0.3809} \\
			PG   & 0.0032          & 0.0007          & \textbf{0.2959} & \textbf{0.4671} & \textbf{0.9973} & \textbf{0.5924} & \textbf{0.7633} & \textbf{0.1076} & \textbf{0.1988} & \textbf{0.2796} & \textbf{0.4233} & \textbf{0.0766} & \textbf{0.2817} \\
			JNJ  & 0.0023          & 0.0028          & \textbf{0.4890} & \textbf{0.3750} & \textbf{0.2628} & \textbf{0.2088} & \textbf{0.1374} & \textbf{0.9235} & \textbf{0.7517} & \textbf{0.4379} & \textbf{0.5812} & \textbf{0.7101} & \textbf{0.5129} \\
			HD   & 0.0027          & 0.0022          & \textbf{0.1495} & \textbf{0.1448} & \textbf{0.2067} & \textbf{0.4118} & \textbf{0.3919} & \textbf{0.0780} & \textbf{0.0696} & \textbf{0.1754} & \textbf{0.1097} & \textbf{0.1407} & \textbf{0.1131} \\
			KO   & \textbf{0.0572} & \textbf{0.2733} & 0.0071          & 0.0070          & 0.0021          & 0.0113          & \textbf{0.1925} & 0.0227          & 0.0204          & 0.0263          & 0.0062          & 0.0334          & \textbf{0.1030} \\
			MRK  & 0.0097          & 0.0094          & \textbf{0.6090} & \textbf{0.6123} & 0.0294          & 0.0296          & 0.0285          & \textbf{0.5472} & \textbf{0.7886} & \textbf{0.9208} & \textbf{0.5931} & \textbf{0.5947} & \textbf{0.5781} \\
			CVX  & 0.0000          & 0.0000          & \textbf{0.0995} & 0.0300          & \textbf{0.1048} & \textbf{0.0899} & 0.0196          & \textbf{0.5755} & \textbf{0.1714} & \textbf{0.4957} & \textbf{0.4166} & \textbf{0.7346} & \textbf{0.2258} \\
			CRM  & \textbf{0.1174} & \textbf{0.1260} & \textbf{0.8365} & \textbf{0.8385} & \textbf{0.6675} & \textbf{0.2721} & \textbf{0.6398} & \textbf{0.7440} & \textbf{0.7631} & \textbf{0.5956} & \textbf{0.5155} & \textbf{0.3927} & \textbf{0.3845} \\
			MCD  & 0.0032          & 0.0014          & \textbf{0.0796} & \textbf{0.0914} & \textbf{0.0622} & \textbf{0.0628} & \textbf{0.0631} & \textbf{0.2530} & \textbf{0.2703} & \textbf{0.3249} & \textbf{0.3871} & \textbf{0.3082} & \textbf{0.3212} \\
			CSCO & 0.0007          & 0.0006          & \textbf{0.0562} & \textbf{0.0547} & 0.0194          & 0.0109          & 0.0126          & \textbf{0.0917} & \textbf{0.0929} & \textbf{0.1432} & \textbf{0.1788} & \textbf{0.2010} & \textbf{0.1955} \\
			AXP  & 0.0287          & 0.0297          & \textbf{0.1867} & \textbf{0.2709} & \textbf{0.2348} & \textbf{0.9364} & \textbf{0.7854} & \textbf{0.2967} & \textbf{0.1409} & \textbf{0.1649} & \textbf{0.1588} & \textbf{0.6918} & \textbf{0.6639} \\
			AMGN & \textbf{0.0624} & 0.0241          & \textbf{0.6004} & \textbf{0.8079} & \textbf{0.6132} & \textbf{0.5725} & \textbf{0.3173} & \textbf{0.5472} & \textbf{0.4991} & \textbf{0.5270} & \textbf{0.5634} & \textbf{0.5928} & \textbf{0.4206} \\
			VZ   & \textbf{0.3654} & \textbf{0.3536} & \textbf{0.1734} & \textbf{0.1867} & \textbf{0.1054} & \textbf{0.1051} & \textbf{0.1031} & \textbf{0.0992} & \textbf{0.1124} & 0.0369          & 0.0378          & \textbf{0.0817} & 0.0377          \\
			CAT  & 0.0000          & 0.0000          & \textbf{0.3579} & \textbf{0.3450} & \textbf{0.4428} & \textbf{0.5825} & \textbf{0.8167} & \textbf{0.5644} & \textbf{0.5365} & \textbf{0.7100} & \textbf{0.5849} & \textbf{0.7209} & \textbf{0.9205} \\
			IBM  & 0.0022          & 0.0021          & \textbf{0.4485} & \textbf{0.5325} & \textbf{0.8460} & \textbf{0.6751} & \textbf{0.5553} & \textbf{0.4379} & \textbf{0.4719} & \textbf{0.3962} & \textbf{0.1794} & \textbf{0.4761} & \textbf{0.4989} \\
			DIS  & 0.0000          & 0.0000          & \textbf{0.2629} & \textbf{0.1492} & \textbf{0.8975} & \textbf{0.5949} & \textbf{0.7923} & \textbf{0.5587} & \textbf{0.0968} & \textbf{0.9107} & \textbf{0.8701} & \textbf{0.2765} & \textbf{0.5217} \\
			GS   & 0.0007          & 0.0004          & 0.0250          & 0.0171          & \textbf{0.0606} & \textbf{0.0941} & \textbf{0.0679} & \textbf{0.4071} & \textbf{0.3268} & \textbf{0.1797} & \textbf{0.5119} & \textbf{0.4345} & \textbf{0.3710} \\
			HON  & \textbf{0.2218} & \textbf{0.2338} & \textbf{0.2579} & \textbf{0.2579} & \textbf{0.6019} & \textbf{0.2443} & \textbf{0.2392} & \textbf{0.4194} & \textbf{0.4193} & 0.0377          & \textbf{0.5343} & \textbf{0.2048} & \textbf{0.1994} \\
			NKE  & 0.0000          & 0.0000          & \textbf{0.7334} & \textbf{0.7644} & \textbf{0.1200} & 0.0104          & 0.0107          & \textbf{0.4405} & \textbf{0.4815} & \textbf{0.1415} & 0.0196          & 0.0232          & 0.0103          \\
			BA   & 0.0000          & 0.0000          & \textbf{0.1420} & \textbf{0.0620} & \textbf{0.3564} & \textbf{0.4440} & \textbf{0.5413} & \textbf{0.3789} & \textbf{0.1801} & \textbf{0.4107} & \textbf{0.3785} & \textbf{0.8294} & \textbf{0.6141} \\
			INTC & \textbf{0.5176} & \textbf{0.5072} & \textbf{0.3267} & \textbf{0.2262} & \textbf{0.5488} & \textbf{0.7500} & \textbf{0.7426} & \textbf{0.1092} & \textbf{0.0926} & \textbf{0.1531} & \textbf{0.2460} & \textbf{0.3822} & \textbf{0.3784} \\
			MMM  & 0.0000          & 0.0000          & 0.0046          & 0.0012          & 0.0004          & 0.0016          & 0.0001          & 0.0144          & 0.0063          & 0.0029          & 0.0013          & 0.0074          & 0.0015\\
			\bottomrule
		\end{tabular}
	\end{adjustbox}
	\caption{Table reports the p-values of the Berkowitz test for the out-of-sample IlliQaR at 1\%. Estimation period: January 3, 2012 - February 4, 2020. Out-of-sample period: February 5, 2020 - January 10, 2024. Stocks are presented in descending order of market capitalization. The estimated models are (I) HAR, (II) AHAR, (III) MEM, (IV) AMEM, (V) MEM-HAR, (VI) AMEM-HAR, (VII) G-AMEM-HAR, (VIII) MEM-J, (IX) AMEM-J, (X) AMEM(2,1)-J, (XI) G-AMEM-HAR-J, (XII) MEM-HAR-J, (XIII) AMEM-HAR-J.\label{tab:density_forecast_oos_1_individual} }
\end{table}

\clearpage
\FloatBarrier
\subsection{IlliQaR at 5\%}
\begin{table}[h!]
	\centering
	\setlength{\tabcolsep}{3pt}
	\renewcommand{\arraystretch}{1.1}
	\begin{adjustbox}{max width=1\linewidth,center}
		\begin{tabular}{lccccccccccccc}
			\toprule
			& (I)      & (II)      & (III)       & (IV)      & (V)      & (VI)      & (VII)      & (VIII)      & (IX)         & (X)                 & (XI)         & (XII)  & (XIII)\\
			\midrule
			MSFT & 0.0002          & 0.0001          & 0.0069          & 0.0041          & 0.0017          & 0.0148          & 0.0086          & 0.0050          & 0.0030          & 0.0010          & 0.0006          & 0.0055          & 0.0039          \\
			JPM  & 0.0004          & 0.0003          & \textbf{0.3930} & \textbf{0.4999} & \textbf{0.3910} & \textbf{0.2876} & \textbf{0.3358} & \textbf{0.4595} & \textbf{0.4889} & \textbf{0.1181} & \textbf{0.4470} & \textbf{0.3100} & \textbf{0.3259} \\
			WMT  & 0.0074          & 0.0203          & 0.0473          & 0.0082          & 0.0244          & 0.0100          & 0.0164          & \textbf{0.0721} & 0.0426          & 0.0254          & 0.0390          & 0.0242          & 0.0389          \\
			UNH  & 0.0009          & 0.0025          & \textbf{0.5660} & \textbf{0.5012} & \textbf{0.9246} & \textbf{0.7445} & \textbf{0.8709} & \textbf{0.2376} & \textbf{0.2002} & \textbf{0.4824} & \textbf{0.8360} & \textbf{0.7820} & \textbf{0.5919} \\
			PG   & 0.0002          & 0.0002          & \textbf{0.7658} & \textbf{0.8980} & \textbf{0.9612} & \textbf{0.6885} & \textbf{0.7421} & \textbf{0.4674} & \textbf{0.6879} & \textbf{0.8788} & \textbf{0.6424} & \textbf{0.8364} & \textbf{0.8519} \\
			JNJ  & 0.0006          & 0.0025          & \textbf{0.2907} & \textbf{0.5922} & \textbf{0.0980} & \textbf{0.1015} & \textbf{0.0875} & \textbf{0.7344} & \textbf{0.8332} & \textbf{0.6369} & \textbf{0.4036} & \textbf{0.4746} & \textbf{0.3476} \\
			HD   & 0.0067          & 0.0058          & \textbf{0.5113} & \textbf{0.3439} & \textbf{0.6327} & \textbf{0.6194} & \textbf{0.6236} & \textbf{0.4953} & \textbf{0.3524} & \textbf{0.5797} & \textbf{0.7462} & \textbf{0.7520} & \textbf{0.7498} \\
			KO   & 0.0152          & 0.0052          & 0.0001          & 0.0002          & 0.0000          & 0.0000          & 0.0000          & 0.0001          & 0.0002          & 0.0000          & 0.0000          & 0.0000          & 0.0000          \\
			MRK  & 0.0005          & 0.0005          & \textbf{0.1015} & \textbf{0.1469} & 0.0442          & 0.0459          & 0.0437          & \textbf{0.3904} & \textbf{0.6084} & \textbf{0.3762} & \textbf{0.3049} & \textbf{0.3144} & \textbf{0.3021} \\
			CVX  & 0.0000          & 0.0000          & \textbf{0.2530} & \textbf{0.0732} & \textbf{0.1001} & \textbf{0.0922} & 0.0060          & \textbf{0.5295} & \textbf{0.1354} & 0.0487          & \textbf{0.2841} & \textbf{0.3299} & \textbf{0.1053} \\
			CRM  & \textbf{0.0869} & \textbf{0.1599} & \textbf{0.5191} & \textbf{0.4121} & \textbf{0.7671} & \textbf{0.8290} & \textbf{0.7408} & \textbf{0.6057} & \textbf{0.4919} & \textbf{0.5585} & \textbf{0.7500} & \textbf{0.9073} & \textbf{0.7734} \\
			MCD  & 0.0299          & 0.0194          & \textbf{0.2400} & \textbf{0.1500} & \textbf{0.0607} & 0.0431          & \textbf{0.0505} & \textbf{0.4384} & \textbf{0.3873} & \textbf{0.4087} & \textbf{0.3025} & \textbf{0.2273} & \textbf{0.2042} \\
			CSCO & 0.0156          & 0.0256          & \textbf{0.0547} & \textbf{0.0802} & 0.0160          & 0.0071          & 0.0132          & 0.0273          & 0.0321          & 0.0265          & 0.0188          & 0.0198          & 0.0204          \\
			AXP  & 0.0021          & 0.0015          & 0.0114          & 0.0062          & 0.0438          & 0.0433          & 0.0431          & 0.0025          & 0.0025          & 0.0007          & 0.0153          & 0.0147          & 0.0062          \\
			AMGN & \textbf{0.2747} & \textbf{0.1605} & \textbf{0.5824} & \textbf{0.3179} & \textbf{0.6303} & \textbf{0.6026} & \textbf{0.6081} & \textbf{0.2884} & \textbf{0.0903} & \textbf{0.3226} & \textbf{0.4849} & \textbf{0.4132} & \textbf{0.4105} \\
			VZ   & 0.0011          & 0.0010          & 0.0000          & 0.0000          & 0.0000          & 0.0000          & 0.0000          & 0.0000          & 0.0000          & 0.0000          & 0.0003          & 0.0001          & 0.0001          \\
			CAT  & 0.0000          & 0.0000          & \textbf{0.1596} & \textbf{0.0911} & \textbf{0.5149} & \textbf{0.1001} & \textbf{0.2788} & \textbf{0.1457} & \textbf{0.0804} & \textbf{0.0573} & \textbf{0.4710} & \textbf{0.0767} & \textbf{0.2375} \\
			IBM  & 0.0004          & 0.0007          & \textbf{0.7065} & \textbf{0.7063} & \textbf{0.5984} & \textbf{0.9419} & \textbf{0.9422} & \textbf{0.4391} & \textbf{0.5673} & \textbf{0.7007} & \textbf{0.8376} & \textbf{0.3546} & \textbf{0.2767} \\
			DIS  & 0.0000          & 0.0000          & 0.0069          & 0.0094          & 0.0305          & 0.0144          & 0.0218          & 0.0061          & 0.0093          & 0.0058          & 0.0109          & 0.0112          & 0.0114          \\
			GS   & 0.0003          & 0.0005          & 0.0268          & 0.0238          & 0.0416          & \textbf{0.1003} & 0.0424          & 0.0268          & 0.0196          & \textbf{0.0631} & \textbf{0.0510} & \textbf{0.2337} & \textbf{0.0848} \\
			HON  & \textbf{0.0770} & \textbf{0.0788} & \textbf{0.2206} & \textbf{0.2206} & \textbf{0.4910} & \textbf{0.8022} & \textbf{0.8028} & \textbf{0.2247} & \textbf{0.2247} & \textbf{0.4987} & \textbf{0.4783} & \textbf{0.8502} & \textbf{0.8154} \\
			NKE  & 0.0000          & 0.0000          & \textbf{0.9556} & \textbf{0.9486} & \textbf{0.2528} & \textbf{0.1843} & \textbf{0.2213} & \textbf{0.9582} & \textbf{0.9661} & \textbf{0.8399} & \textbf{0.1636} & \textbf{0.1293} & \textbf{0.1605} \\
			BA   & 0.0000          & 0.0000          & \textbf{0.6783} & \textbf{0.3740} & \textbf{0.6282} & \textbf{0.8041} & \textbf{0.7273} & \textbf{0.3886} & \textbf{0.0863} & \textbf{0.3966} & \textbf{0.3149} & \textbf{0.6832} & \textbf{0.6609} \\
			INTC & \textbf{0.1270} & \textbf{0.1211} & 0.0400          & 0.0479          & 0.0114          & \textbf{0.0599} & 0.0422          & 0.0361          & 0.0394          & \textbf{0.0616} & 0.0421          & 0.0498          & 0.0374          \\
			MMM  & 0.0000          & 0.0000          & 0.0010          & 0.0006          & 0.0004          & 0.0005          & 0.0002          & 0.0053          & 0.0025          & 0.0008          & 0.0038          & 0.0042          & 0.0013 \\
			\bottomrule
		\end{tabular}
	\end{adjustbox}
	\caption{Table reports the p-values of the Berkowitz test for the out-of-sample IlliQaR at 5\%. Estimation period: January 3, 2012 - February 4, 2020.  Out-of-sample period: February 5, 2020 - January 10, 2024. Stocks are presented in descending order of market capitalization. The estimated models are (I) HAR, (II) AHAR, (III) MEM, (IV) AMEM, (V) MEM-HAR, (VI) AMEM-HAR, (VII) G-AMEM-HAR, (VIII) MEM-J, (IX) AMEM-J, (X) AMEM(2,1)-J, (XI) G-AMEM-HAR-J, (XII) MEM-HAR-J, (XIII) AMEM-HAR-J.\label{tab:density_forecast_oos_5_individual}  }
\end{table}

\clearpage
\FloatBarrier
\subsection{IlliQaR at 10\%}
\begin{table}[h!]
	\centering
	\setlength{\tabcolsep}{3pt}
	\renewcommand{\arraystretch}{1.1}
	\begin{adjustbox}{max width=1\linewidth,center}
		\begin{tabular}{lccccccccccccc}
			\toprule
			& (I)      & (II)      & (III)       & (IV)      & (V)      & (VI)      & (VII)      & (VIII)      & (IX)         & (X)                 & (XI)         & (XII)  & (XIII)\\
			\midrule
			MSFT & 0.0026          & 0.0006          & 0.0013          & 0.0010          & 0.0015          & 0.0055          & 0.0050          & 0.0028          & 0.0017          & 0.0021          & 0.0026          & 0.0101          & 0.0084          \\
			JPM  & 0.0003          & 0.0000          & \textbf{0.0994} & \textbf{0.2858} & \textbf{0.4038} & \textbf{0.2247} & \textbf{0.3459} & \textbf{0.0742} & \textbf{0.1382} & \textbf{0.1408} & \textbf{0.4326} & \textbf{0.3017} & \textbf{0.3641} \\
			WMT  & 0.0455          & \textbf{0.0855} & \textbf{0.1694} & \textbf{0.2475} & \textbf{0.1100} & \textbf{0.0782} & \textbf{0.0811} & \textbf{0.1241} & \textbf{0.0716} & \textbf{0.0632} & 0.0490          & 0.0315          & 0.0458          \\
			UNH  & 0.0071          & 0.0114          & \textbf{0.2922} & \textbf{0.1721} & \textbf{0.7775} & \textbf{0.9491} & \textbf{0.7807} & \textbf{0.0960} & \textbf{0.0633} & \textbf{0.2271} & \textbf{0.5567} & \textbf{0.7219} & \textbf{0.4651} \\
			PG   & 0.0299          & 0.0108          & \textbf{0.1924} & \textbf{0.1819} & \textbf{0.4815} & \textbf{0.2407} & \textbf{0.3926} & \textbf{0.5938} & \textbf{0.5099} & \textbf{0.3209} & \textbf{0.5603} & \textbf{0.4978} & \textbf{0.7578} \\
			JNJ  & 0.0013          & 0.0006          & \textbf{0.2147} & \textbf{0.0837} & \textbf{0.4128} & \textbf{0.4733} & \textbf{0.5049} & \textbf{0.3676} & \textbf{0.2719} & \textbf{0.2956} & \textbf{0.6450} & \textbf{0.8552} & \textbf{0.9080} \\
			HD   & 0.0052          & 0.0071          & \textbf{0.1295} & \textbf{0.1293} & \textbf{0.5507} & \textbf{0.6380} & \textbf{0.6486} & \textbf{0.1242} & \textbf{0.1247} & \textbf{0.4089} & \textbf{0.2057} & \textbf{0.7192} & \textbf{0.5826} \\
			KO   & 0.0003          & 0.0003          & 0.0000          & 0.0000          & 0.0000          & 0.0001          & 0.0000          & 0.0000          & 0.0000          & 0.0000          & 0.0000          & 0.0000          & 0.0000          \\
			MRK  & 0.0276          & 0.0177          & \textbf{0.8889} & \textbf{0.8345} & \textbf{0.1275} & \textbf{0.1316} & \textbf{0.1259} & \textbf{0.9482} & \textbf{0.9818} & \textbf{0.9893} & \textbf{0.4577} & \textbf{0.5435} & \textbf{0.4530} \\
			CVX  & 0.0000          & 0.0000          & \textbf{0.4409} & \textbf{0.4950} & \textbf{0.5161} & \textbf{0.5494} & \textbf{0.2337} & \textbf{0.4955} & \textbf{0.5005} & \textbf{0.5404} & \textbf{0.4360} & \textbf{0.5819} & \textbf{0.3237} \\
			CRM  & \textbf{0.2676} & \textbf{0.2273} & \textbf{0.7853} & \textbf{0.7836} & \textbf{0.8358} & \textbf{0.6019} & \textbf{0.6797} & \textbf{0.9012} & \textbf{0.8625} & \textbf{0.7947} & \textbf{0.7804} & \textbf{0.7469} & \textbf{0.6923} \\
			MCD  & 0.0179          & 0.0286          & \textbf{0.1659} & \textbf{0.2734} & \textbf{0.5151} & \textbf{0.4656} & \textbf{0.4354} & \textbf{0.1202} & \textbf{0.2572} & \textbf{0.4089} & \textbf{0.7809} & \textbf{0.7479} & \textbf{0.7477} \\
			CSCO & 0.0029          & 0.0028          & \textbf{0.0934} & \textbf{0.0809} & 0.0336          & 0.0245          & 0.0329          & 0.0161          & 0.0219          & 0.0198          & 0.0294          & 0.0262          & 0.0296          \\
			AXP  & 0.0020          & 0.0014          & 0.0035          & 0.0017          & \textbf{0.0894} & \textbf{0.0553} & 0.0307          & 0.0007          & 0.0002          & 0.0044          & \textbf{0.0649} & 0.0300          & 0.0229          \\
			AMGN & \textbf{0.0515} & \textbf{0.0978} & \textbf{0.6324} & \textbf{0.6475} & \textbf{0.5837} & \textbf{0.4063} & \textbf{0.6524} & \textbf{0.4395} & \textbf{0.4630} & \textbf{0.1785} & \textbf{0.2885} & \textbf{0.1844} & \textbf{0.5439} \\
			VZ   & 0.0002          & 0.0002          & 0.0000          & 0.0000          & 0.0000          & 0.0000          & 0.0000          & 0.0000          & 0.0000          & 0.0000          & 0.0000          & 0.0000          & 0.0000          \\
			CAT  & 0.0000          & 0.0000          & \textbf{0.2766} & \textbf{0.2143} & \textbf{0.4772} & \textbf{0.5153} & \textbf{0.3968} & \textbf{0.3120} & \textbf{0.2092} & \textbf{0.0971} & \textbf{0.3953} & \textbf{0.4940} & \textbf{0.3812} \\
			IBM  & 0.0008          & 0.0021          & \textbf{0.2577} & \textbf{0.3673} & \textbf{0.3276} & \textbf{0.1725} & \textbf{0.1727} & \textbf{0.8120} & \textbf{0.8051} & \textbf{0.9104} & \textbf{0.7681} & \textbf{0.6891} & \textbf{0.6395} \\
			DIS  & 0.0000          & 0.0000          & 0.0042          & 0.0037          & \textbf{0.0706} & 0.0307          & 0.0233          & 0.0008          & 0.0006          & 0.0122          & 0.0346          & 0.0164          & 0.0178          \\
			GS   & 0.0003          & 0.0004          & 0.0020          & 0.0007          & 0.0045          & 0.0083          & 0.0071          & 0.0012          & 0.0004          & 0.0011          & 0.0033          & 0.0052          & 0.0072          \\
			HON  & \textbf{0.3323} & \textbf{0.2854} & \textbf{0.0785} & \textbf{0.0785} & \textbf{0.3973} & \textbf{0.6089} & \textbf{0.6096} & \textbf{0.0794} & \textbf{0.0794} & \textbf{0.1753} & \textbf{0.3936} & \textbf{0.6153} & \textbf{0.5675} \\
			NKE  & 0.0000          & 0.0000          & \textbf{0.1301} & \textbf{0.1062} & \textbf{0.0530} & \textbf{0.0707} & \textbf{0.0742} & 0.0227          & 0.0460          & 0.0487          & 0.0098          & 0.0470          & 0.0375          \\
			BA   & 0.0000          & 0.0000          & \textbf{0.9802} & \textbf{0.9819} & \textbf{0.8180} & \textbf{0.6837} & \textbf{0.7473} & \textbf{0.8850} & \textbf{0.8974} & \textbf{0.5977} & \textbf{0.7060} & \textbf{0.4931} & \textbf{0.4361} \\
			INTC & \textbf{0.3053} & \textbf{0.2416} & 0.0036          & 0.0076          & 0.0013          & 0.0101          & 0.0057          & 0.0057          & 0.0071          & 0.0155          & 0.0026          & 0.0189          & 0.0157          \\
			MMM  & 0.0000          & 0.0000          & 0.0008          & 0.0007          & 0.0003          & 0.0006          & 0.0002          & 0.0022          & 0.0019          & 0.0007          & 0.0037          & 0.0034          & 0.0021 \\
			\bottomrule
		\end{tabular}
	\end{adjustbox}
	\caption{Table reports the p-values of the Berkowitz test for the full sample and out-of-sample IlliQaR at 10\%. Estimation period: January 3, 2012 - February 4, 2020.  Out-of-sample period: February 5, 2020 - January 10, 2024. Stocks are presented in descending order of market capitalization. The estimated models are (I) HAR, (II) AHAR, (III) MEM, (IV) AMEM, (V) MEM-HAR, (VI) AMEM-HAR, (VII) G-AMEM-HAR, (VIII) MEM-J, (IX) AMEM-J, (X) AMEM(2,1)-J, (XI) G-AMEM-HAR-J, (XII) MEM-HAR-J, (XIII) AMEM-HAR-J.\label{tab:density_forecast_oos_10_individual}  }
\end{table}
\FloatBarrier

\end{appendix}
\end{document}